\documentclass[
  aps,
  prx,
  twocolumn,
  superscriptaddress,
  longbibliography,
  floatfix
]{revtex4-2}

\usepackage[T1]{fontenc}
\usepackage[utf8]{inputenc}
\usepackage{amsmath,amssymb,amsfonts,amsthm,mathtools,bm}
\usepackage{braket}
\usepackage{graphicx}
\usepackage{xcolor}
\usepackage{hyperref}
\hypersetup{colorlinks=true,linkcolor=blue,citecolor=blue,urlcolor=blue}
\usepackage{booktabs,array,longtable}
\usepackage[nameinlink,capitalize,noabbrev]{cleveref}
\usepackage{tikz}
\usepackage{algorithm}
\usepackage{algpseudocode}

\makeatletter
\edef\ftype@algorithm{\number\numexpr\value{float@type}/2\relax}
\makeatother

\algrenewcommand\algorithmicrequire{\textbf{Input:}}
\algrenewcommand\algorithmicensure{\textbf{Output:}}

\newtheorem{theorem}{Theorem}
\newtheorem{proposition}{Proposition}

\newtheorem{corollary}{Corollary}
\theoremstyle{definition}

\begin{document}

\title{From Syndromes to Computation: A General Framework for Logical Operations in Quantum Error-Correcting Codes}

\author{Hunter Nelson}
\email{huntertn@vt.edu}
\author{Arshag Danageozian}
\author{ChunJun Cao}
\author{Sophia E. Economou}
\author{Edwin Barnes}
\email{efbarnes@vt.edu}
\affiliation{Department of Physics, Virginia Tech, Blacksburg, VA 24061, USA}
\affiliation{Virginia Tech Center for Quantum Information Science and Engineering, Blacksburg, VA 24061, USA}

\date{\today}

\begin{abstract}
Fault-tolerant quantum computation requires error-correcting codes that not only protect quantum information but also support useful logical operations with practical physical realizations. Here we develop a general framework in which the stabilizer syndromes of commuting Pauli rotations directly organize logical-gate design. We derive an exact codespace-preservation condition valid at arbitrary rotation angles, providing a common stabilizer-level description for CSS and non-CSS codes and for both Clifford and non-Clifford operations. A particularly useful equal-syndrome specialization reduces Clifford gate design to stabilizer factorization, yields a general generator weight bound of $\lceil d/2\rceil$ for distance-$d$ codes, and systematically extends to logical rotations higher in the Clifford hierarchy. Applied to bivariate-bicycle (BB) and non-CSS Mirror qLDPC codes, the framework gives explicit low-weight Clifford and controlled-phase gates.
Our implementations yield new low-weight gates, including an addressed $T$ gate on the $[[16,4,4]]$ BB code, $T$-type rotations on the $[[64,8,8]]$ BB code, and CCZ-, CS-, and $T$-type operations on the $[[36,6,6]]$ Mirror code. We also develop analytic constructions using Clifford algebra bases, which turn the syndrome constraints into explicit algebraic relations and provide a systematic route to logical addressability. We use this approach to analytically construct  addressed CNOT, SWAP, controlled, and CCZ-type gates on recursive hypercube codes. An independent construction on middle quantum Reed--Muller codes yields C${}^r$Z gates that saturate the equal-syndrome weight bound. These results establish stabilizer syndromes as a constructive link between the stabilizer structure of a quantum code, the logical computations it can directly support, and the fault-tolerance properties of their physical realizations.

\end{abstract}

\maketitle

% ============================================================
\section{Introduction}
\label{sec:introduction}
% ============================================================

Quantum error correction is essential to scalable quantum computation,
both for protecting quantum information against noise and for enabling
long-depth logical processing
\cite{shor1995scheme,steane1996multiple,gottesman1997stabilizer,
terhal2015quantum,campbell2017roads}. Over the past three decades, the
field has developed increasingly sophisticated code families with
improved trade-offs among distance, rate, locality, and overhead,
including concatenated, topological, color-code, and quantum
low-density-parity-check (qLDPC) constructions
\cite{knill1996concatenated,dennis2002topological,bombin2006topological,kovalev2013quantum,
tillich2014quantum,breuckmann2021balanced,panteleev2022asymptotically,
leverrier2022quantum,bravyi2024highthreshold, wang2026algebra}. For such codes to form useful computational architectures, however,
error protection is only one part of the overall design. One must also understand
how to perform a sufficiently rich set of logical operations while
preserving the advantages of the encoding \cite{campbell2017roads,bombin2023logicalblocks,
GotoManyHypercube2024}. The design of logical gates
should therefore be intertwined with the design of the
codes themselves.

A number of complementary paradigms have been developed for
implementing logical gates fault-tolerantly. Transversal gates are
especially attractive because their restricted support strongly limits
the propagation of errors within a code block
\cite{gottesman1998theory,zeng2011transversality}. Their computational
power, however, is fundamentally limited by the Eastin--Knill theorem, which rules out universal transversal gate sets
\cite{eastin2009restrictions}, while locality-preserving logical gates in \(D\)-dimensional topological stabilizer codes are restricted by the spatial dimension, with constant-depth geometrically local gates lying within the \(D\)th level of the Clifford hierarchy
\cite{bravyi2013classification}. More recently,
Ref.~\cite{chakraborty2026no} showed that no stabilizer code encoding more than one logical qubit can realize the full logical Clifford group using Clifford transversal gadgets. It also established analogous no-go results for code-automorphism constructions and showed that \(k\) logical qubits require at least \(k\)-fold transversal Clifford gadgets. These limitations have motivated
broader approaches to logical-gate design. Fold-transversal operations
and code automorphisms relax the physical circuit structure, while
lattice surgery, code deformation, code switching, and gauge fixing
modify or couple encoded representations
\cite{grassl2013automorphisms,sayginel2025fault,
breuckmann2024fold,bombin2009deformation,horsman2012lattice,
litinski2019game,paetznick2013universal,anderson2014conversion,
bombin2015gauge,kubica2015universal,vuillot2019code}. A particularly striking example is provided by phantom codes, where
physical-qubit permutations alone can induce entangling logical
operations \cite{koh2026entangling}. Tensor-network constructions provide another route, using code blocks with useful symmetries to build larger codes with addressable transversal Clifford and non-Clifford gates \cite{cao2026quantum}. 
Algebraic, homological, and geometric constructions have further
revealed rich logical gate sets in qLDPC and related codes
\cite{breuckmann2026cups,chainmap2026hierarchy,
zhu2025higher,lin2024sheaf,golowich2025algebraic}.
Together, these developments make it natural to seek structured
logical-gate constructions beyond strict transversality.

Relaxing transversality, however, presents a distinct challenge. Given a stabilizer code and a target logical operation, how should one choose a physically structured ansatz for the implementing unitary, and how can one determine whether that ansatz preserves the codespace and produces the intended logical transformation? Ideally, the answer would follow directly from the stabilizer description of the code itself. Such a viewpoint would be especially useful for codes encoding many logical qubits, where identifying Pauli frames with nice properties, and determining which physical operations implement useful logical gates can become difficult. It would also provide a common starting point for CSS and non-CSS stabilizer codes.

A second challenge arises beyond the Clifford group. Clifford gates are relatively more amenable to systematic analysis because conjugation by a Clifford unitary maps Pauli operators into Pauli operators. Code
preservation and logical action can therefore be expressed directly in
the binary symplectic language of the stabilizer formalism, enabling
general logical-Clifford synthesis, automorphism searches, and
structural classifications of code-preserving Clifford circuits
\cite{rengaswamy2020logical,sayginel2025fault,
albert2026beyond,dasu2025classification,kuehnke2025hardware}. For non-Clifford gates this
closure is lost, and the same stabilizer machinery no longer directly
provides a general codespace-preservation criterion. Systematic results
are known for restricted physical ans\"atze, including transversal
$T$ gates and more general transversal diagonal Clifford-hierarchy
operations
\cite{rengaswamy2020optimality,webster2023diagonal,bauer2026diagonal}. Recent constructions for qLDPC and related high-rate quantum codes have gone substantially further, realizing transversal
or constant-depth non-Clifford gates through intersection forms,
multiplication properties, cup products, higher symmetries, and
special sheaf, product, or manifold structure
\cite{scruby2024rainbow,zhu2025higher,breuckmann2026cups,
tiew2026copycup,lin2024sheaf,golowich2025algebraic,
golowich2025asymptoticallygood,elishco2026ccz,
zhu2025topological,menon2026tricycles,li2026almostgood}. These results demonstrate that non-Clifford computation is compatible with high-rate and
qLDPC codes, but the known constructions are typically tied to
additional algebraic or geometric structure beyond the stabilizer
commutation relations themselves. These properties are not guaranteed for an arbitrary stabilizer code and must first be established before the corresponding gate construction can be applied. 

Recent work has begun to narrow this gap. Ref.~\cite{yoshioka2025syndrome} develops a systematic approach to weakly transversal logical gates in general CSS codes, including a procedure for constructing physical Pauli rotations whose logical action is realized probabilistically through syndrome measurement and conditional feedforward. The very recently introduced chain-map hierarchy provides a complementary homological framework for systematically constructing Pauli, Clifford, and non-Clifford logical operations, including constant-depth Clifford circuits and non-Clifford controlled-phase gates~\cite{chainmap2026hierarchy}. Additionally, another very recent contribution constructs CSS codes supporting universal logical gate sets by coupling lower-distance seed codes admitting non-Clifford gates to auxiliary kernel codes through entangling circuits. The resulting logical gates are implemented by depth-one layers of disjoint few-qubit gates, potentially using different physical-qubit partitions~\cite{dua2026qubitcss}.

However, these approaches never answer a more direct question in the stabilizer language itself: Given only the stabilizers of a code, how can we systematically construct unitary circuits that preserve the codespace and implement nontrivial logical gates? Such a general approach would apply to both CSS and non-CSS codes, handle Clifford and non-Clifford rotations under the same preservation condition, and avoid requiring syndrome measurement, or an additional chain-complex or geometric construction. It would also offer a direct starting point for systematic gate searches on existing codes and analytic constructions across multiple levels of the Clifford hierarchy.

In this work, we develop such a drastically simplifying general framework in which stabilizer
syndromes become an active design principle for logical gate design. We consider logical gate ans\"atze (LGAs) built from mutually commuting Pauli rotations and derive an
exact codespace-preservation condition valid at arbitrary rotation
angles. The stabilizer syndrome of each generator records
which stabilizer checks select (i.e., are altered by) that rotation under conjugation, so the
syndrome pattern identifies the doubled-angle operation that each
stabilizer check selects. Codespace preservation requires every such
operation to act as the identity on the codespace. Grouping the generators by these patterns
gives what we call a \emph{syndrome partition}. Because the condition is
formulated directly in the stabilizer and Pauli-operator language, it applies
uniformly to CSS and non-CSS codes and at both Clifford and
non-Clifford rotation angles. It therefore provides a common
stabilizer-level starting point for constructing logical gates beyond
both strict transversality and the Clifford setting.

A useful realization of this principle occurs when all Pauli generators share the same nonzero syndrome. In this setting, pairwise products of the Pauli generators have zero syndrome, so the LGA can be organized directly
by its relative logical Pauli content. At the Clifford angle $\pi/4$,
gate design reduces to factoring a stabilizer into commuting
equal-syndrome Pauli generators and choosing physical representatives
with favorable weight and support. We prove that any nontrivial
construction of this form on a distance-$d$ code must contain a
Pauli generator of weight at least $\lceil d/2\rceil$. The same viewpoint also
organizes and simplifies the construction of gates higher in the
Clifford hierarchy. In particular, equal-syndrome pairs provide a
systematic way to construct rotations about a chosen logical Pauli
axis, yielding logical $\pi/4$ phase rotations and $\pi/8$ $T$-type
rotations. More generally, syndrome relations between lower-level
gates reveal how they can be combined to obtain higher-level
operations. Thus the same
stabilizer-syndrome framework connects Clifford and non-Clifford gate
design within a single construction principle.

We first apply the framework to existing qLDPC codes, including
self-dual bivariate-bicycle codes and the non-CSS Mirror code. The
resulting searches produce large catalogs of logical Clifford
operations together with several direct non-Clifford constructions. Our low-weight
implementations include an addressed $T$ gate on the $[[16,4,4]]$ BB
code, weight-four $T$-type rotations on the $[[64,8,8]]$ BB code, and
CCZ-, CS-, and $T$-type operations on the $[[36,6,6]]$ Mirror code, which to the best of our knowledge have not been previously found.
We also obtain low-depth CCZ-type implementations on several BB
instances. These qLDPC examples demonstrate that the method can uncover both Clifford and non-Clifford operations in CSS and non-CSS codes, including gates that are not apparent from transversality or code-automorphism structure alone.

A second major contribution is an analytic route to constructing
solutions of the syndrome-partition conditions. We introduce Clifford
algebra bases (CABs) as a coordinate system in which stabilizers, logical
Paulis, and Pauli operators carrying nonzero syndrome share a common index
structure. Pauli multiplication and stabilizer syndromes then reduce to
simple relations among these indices, making equal-syndrome
constructions accessible analytically. We demonstrate this on recursive
hypercube codes, where simple CAB-index patterns systematically produce
addressed logical CNOT and SWAP gates, more general controlled
operations, and higher-level gates such as CCZ, all with structured
physical realizations. In a separate
syndrome based analytic construction, we show that middle quantum Reed--Muller codes
can support logical C${}^r$Z gates whose generators have disjoint support and weight exactly $d/2$, saturating the general equal-syndrome bound.

The low-weight and low-depth Pauli-rotation realizations found
throughout this work are also favorable from a fault-tolerance
perspective. Many of the gates use few-qubit Pauli generators arranged
into one or a small number of Pauli-rotation layers with disjoint
support within each layer. Small generator supports limit the initial
spread of an error, while low Pauli-rotation depth provides natural
points at which intermediate error correction can be inserted when
successive layers overlap. Beyond these geometric properties, the syndrome information used to
construct the gates also contains fault-tolerance information.
Relations among the generator syndromes can identify small collections
of Pauli-rotation supports that can carry logical operators, while the
rank of the generator-syndrome span determines how many independent
stabilizer combinations remain unchanged throughout the gate.

Together, these results establish syndrome partitions as a general
design principle for direct logical gates. This places CSS and
non-CSS codes, as well as Clifford and non-Clifford gates, within the
same construction framework. Importantly, the same syndrome structure
used to construct a logical gate also constrains its fault-tolerance
properties, including how many Pauli-rotation supports can combine to
carry a logical operator and which stabilizer combinations remain
available during the gate. Syndrome partitions therefore provide a
systematic route not only to uncovering direct logical operations in
existing codes, but also to searching for implementations with
favorable fault-propagation structure and to designing new codes
together with their logical gate sets. More broadly, this suggests a
route toward co-designing error correction, logical computation, and
fault-tolerant implementation within a common stabilizer-syndrome
framework.

The remainder of the paper is organized as follows.
Section~\ref{sec:gate_design_workflows} develops the general
syndrome-partition condition, its equal-syndrome specialization, and
the constructions connecting different levels of the Clifford
hierarchy.
Section~\ref{sec:qldpc_searches} applies these ideas to existing CSS
and non-CSS qLDPC codes and presents the resulting Clifford and
non-Clifford gate constructions.
Section~\ref{sec:analytic_constructions} develops analytic
syndrome-partition constructions using CAB coordinates and recursive
hypercube codes, followed by the middle Reed--Muller hierarchy.
Section~\ref{sec:fault_diagnostics} examines fault propagation,
relates syndrome relations to Pauli-rotation partition distance, and
identifies stabilizer checks that remain available for intermediate
error correction in higher-depth constructions. Technical proofs and search details are given
in the appendices.

% ============================================================
\section{Logical-gate design from syndrome partitions}
\label{sec:gate_design_workflows}
% ============================================================

Logical gates in stabilizer codes have been studied from many different perspectives. Transversal constructions require the logical gate to factor into physical operations that act independently across corresponding physical qubits of different code blocks, so that a single physical fault cannot spread within a code block. One then asks which logical operations are compatible with this restricted form
\cite{gottesman1998theory,zeng2011transversality,
eastin2009restrictions,chen2022automated,holmes2026quantumlogic}.
Code-automorphism methods exploit physical symmetries, such as qubit permutations or local Clifford transformations, that preserve the stabilizer group. The corresponding action on the logical Pauli operators then determines the induced logical Clifford gate
\cite{grassl2013automorphisms,sayginel2025fault}. Other algebraic and homological approaches represent the code through symplectic, chain-complex, cohomological, or related structures, and construct logical gates as transformations that preserve those structures while acting nontrivially on the encoded degrees of freedom
\cite{albert2026beyond,chainmap2026hierarchy}. In this work, instead of relying on a predetermined transversal pattern, a known code symmetry, or a specific homological construction, we focus on commuting physical Pauli rotations as our fundamental building blocks of logical gates. The main challenge is to then find sets of Pauli generators whose stabilizer syndromes are compatible with codespace preservation.

Our goal is to go beyond strictly transversal gates while retaining
physical structure that is favorable for fault tolerance. In
particular, we allow the physical Pauli rotations to act on several
qubits within the same code block, which considerably enlarges the
space of possible logical gates. At the same time, we seek
implementations built from low-weight rotations that, where possible,
act on disjoint sets of physical qubits or can be divided into only a small number of layers. Disjointness limits how far an error can spread during
a single layer, while a small number of layers leaves open the
possibility of performing error correction between them. These
properties do not by themselves guarantee a fault-tolerant
implementation, but they provide useful physical criteria for seeking fault-tolerant
logical gates beyond transversality.

To introduce these concepts pedagogically, we begin with an example on the $[[4,2,2]]$ code, with stabilizers
\begin{align}
S_{\mathrm A}
&=
XXXX,
&
S_{\mathrm B}
&=
ZZZZ.
\end{align}
A transversal layer of root-$Y$ gates acting on all four qubits uses the four generators
$G_j=Y_j$.  All four have the same nonzero syndrome, since every
$Y_j$ anticommutes with both stabilizer checks. It also holds that
\begin{align}
G_1G_2G_3G_4
&=
YYYY
=
S_{\mathrm A}S_{\mathrm B}.
\end{align}
Therefore, at angle $\pi/4$, conjugation multiplies $S_{\mathrm{A}}$ and $S_{\mathrm{B}}$ by the product $S_{\mathrm A}S_{\mathrm B}$, which exchanges the two stabilizer generators,
\begin{align}
S_{\mathrm A}
&\mapsto
S_{\mathrm B},
&
S_{\mathrm B}
&\mapsto
S_{\mathrm A}.
\end{align}
In an appropriately chosen logical Pauli frame, the resulting logical operation is a SWAP of the two encoded qubits. This example captures a basic mechanism in which several physical commuting Pauli generators share one nonzero syndrome and their product lies in the stabilizer group.

More generally the Pauli generators will not have equal syndromes. For example, consider the transversal phase gate of the $[[7,1,3]]$ Steane code, which is comprised of phase gates on all seven qubits. A convenient stabilizer generating set is
\begin{align}
S_{X,1}&=X_1X_2X_3X_4,
&
S_{Z,1}&=Z_1Z_2Z_3Z_4,
\nonumber\\
S_{X,2}&=X_1X_2X_5X_6,
&
S_{Z,2}&=Z_1Z_2Z_5Z_6,
\nonumber\\
S_{X,3}&=X_1X_3X_5X_7,
&
S_{Z,3}&=Z_1Z_3Z_5Z_7.
\end{align}
The Pauli generators $Z_j$ of the phase gate generally have different stabilizer syndromes because different physical qubits participate in different $X$-type stabilizer checks. A given $X$ stabilizer therefore does not select the entire transversal layer; it selects only the $Z_j$ rotations acting on qubits in its support. Nevertheless, the product of those selected rotations is exactly the $Z$-type stabilizer corresponding to the $X$-type stabilizer acted on. Consequently,
\begin{align}
S_a^X
&\mapsto
S_a^X S_a^Z,
&
S_a^Z
&\mapsto
S_a^Z.
\end{align}
Thus codespace preservation does not require all Pauli generators to share one syndrome. Different stabilizer checks may select different subsets of the same LGA, provided that each selected subset produces a trivial action on the codespace.

The same viewpoint can also be applied to physical
circuits that are not a single product of commuting Pauli rotations.
For example, a Clifford automorphism may combine several circuit layers
with qubit permutations. The individual layers need not preserve the
original codespace: a stabilizer check may be mapped outside the
stabilizer group after one layer and returned to it by a later layer.
The phase-gate automorphism of the $[[5,1,3]]$ code provides such an
example, where a layer of phase gates on all five qubits combines with
two SWAP gates to produce a logical phase gate
\cite{sayginel2025fault}. The bookkeeping is the same as in the
$[[4,2,2]]$ and Steane examples above. At each commuting
Pauli-rotation step, the syndrome of the physical Pauli generators
determines which rotations are selected by a given stabilizer check,
while a permutation relabels the resulting Pauli operator. The
successive contributions need only combine so that, after the complete
circuit, every stabilizer check is mapped back into the stabilizer group
with the correct sign. For non-Clifford gates, transversal $T/T^\dagger$ patterns acting on each $X$-type stabilizer select a product of doubled-angle Clifford operations, and divisibility or triorthogonality conditions then ensure codespace preservation
\cite{bravyi2012magic,rengaswamy2020optimality,
scruby2024rainbow,zhu2025higher}. Related qLDPC controlled-phase constructions enforce the same type of lower-level cancellation through cup products, intersection forms, sheaf structure, or algebraic multiplication rules
\cite{breuckmann2026cups,lin2024sheaf,golowich2025algebraic,
zhu2025topological,menon2026tricycles,li2026almostgood,
golowich2026improved}.

These algebraic mechanisms all differ, but the central preservation question remains the same. For every stabilizer check, the physical rotations selected by anticommutation must
combine to an operation that is trivial on the codespace. The key question is how to systematically leverage the syndrome information of the physical gates to aid the search for logical operations.

% ------------------------------------------------------------
\subsection{Syndrome-partition theorem}
\label{subsec:syndrome_principle}
% ------------------------------------------------------------

Let $\mathcal C$ be an $[[n,k,d]]$ stabilizer code with independent
stabilizer generators $\boldsymbol S=(S_1,\ldots,S_r)$, where $r=n-k$,
and stabilizer group $\mathcal S$. For a Pauli operator $P$, let
$\boldsymbol{\sigma}_{\boldsymbol S}(P)\in\mathbb F_2^r$ denote its
binary stabilizer syndrome, with
$(\boldsymbol{\sigma}_{\boldsymbol S}(P))_a=1$ exactly when $P$
anticommutes with $S_a$. In this work we consider logical gate ans\"atze (LGAs) built from mutually commuting Pauli rotations of the form

\begin{align}
U(\boldsymbol\theta,\boldsymbol\epsilon)
&=
\prod_{j=1}^{q}
\exp\left(
i\epsilon_j\theta_jG_j
\right),
&
\epsilon_j
&\in
\{\pm1\},
\label{eq:general_arbitrary_angle_layer}
\end{align}
where the $G_j$ are pairwise-commuting Hermitian Pauli strings, and the
$\theta_j\geq0$ specify the rotation magnitudes. We call each $G_j$ a
\emph{Pauli generator}. Grouping the Pauli generators by their
stabilizer-syndrome vectors defines the LGA's \emph{syndrome partition}.
If $G_j$ commutes with $S_a$, its rotation leaves that check unchanged;
if it anticommutes, conjugation contributes a doubled-angle rotation
$e^{2i\epsilon_j\theta_jG_j}$. Thus, for each $S_a$, let
$\mathcal J_a$ denote the set of indices $j$ for which $G_j$ anticommutes with $S_a$ and define
\begin{align}
V_a
&=
\prod_{j\in\mathcal J_a}
\exp\left(
2i\epsilon_j\theta_jG_j
\right).
\label{eq:check_conditioned_unitary}
\end{align}
The conjugated stabilizer is then
$U^\dagger S_aU=S_aV_a$.
We refer to $V_a$ as the \emph{check-conditioned unitary}: it is the
part of the LGA selected by the stabilizer check $S_a$.

\begin{theorem}[Arbitrary-angle syndrome-partition theorem]
\label{thm:arbitrary_angle_syndrome_partition}
The LGA in Eq.~\eqref{eq:general_arbitrary_angle_layer} preserves the
$+1$ codespace $\mathcal C$ if and only if
\begin{align}
V_a|\psi\rangle
&=
|\psi\rangle
\qquad
\forall\,|\psi\rangle\in\mathcal C,
\quad
\forall a.
\label{eq:general_arbitrary_angle_preservation}
\end{align}
\end{theorem}

The theorem gives the common check-by-check condition underlying the
examples above. Each stabilizer selects the physical rotations whose
generators anticommute with it, and the resulting doubled-angle
operation must act trivially on the codespace. A proof is given in
Appendix~\ref{app:desired_syndrome_proof}.

The $[[4,2,2]]$ and Steane examples illustrate the two extremes of the same preservation problem. In the $[[4,2,2]]$ construction, all Pauli generators have the same syndrome, so every affected stabilizer check selects the same set of rotations. In the Steane construction, by contrast, the Pauli generators have different syndromes, and each stabilizer check selects a different subset of the generators. The exact preservation condition accommodates both situations. More general circuits outside the single-LGA ansatz can be treated by
composing the same check-by-check propagation rule across successive
operations. For instance, a qubit permutation such as those used in the [[5,1,3]] example above simply relabels the current Pauli check, while each subsequent commuting Pauli-rotation in the LGA contributes the corresponding check-conditioned transformation. Codespace preservation is then determined only after the complete sequence of operations has been applied. Our LGA could be extended to include such permutations, but we do not develop this extension in the present work.

In general, the preservation condition can be difficult to solve directly, because different stabilizer checks may select different subsets of the rotations in Eq.~\eqref{eq:general_arbitrary_angle_layer}, producing many coupled constraints on a candidate LGA. We therefore focus on three constructions that are easier to analyze: equal-syndrome LGAs, collections of equal-syndrome pairs, and lifted constructions that combine lower-level Clifford-hierarchy operations to produce higher-level logical gates. We begin with the equal-syndrome case.

% ------------------------------------------------------------
\subsection{Equal-syndrome LGAs}
\label{subsec:equal_syndrome_general}
% ------------------------------------------------------------

Suppose that all Pauli generators in the LGA carry the same nonzero stabilizer syndrome,
\begin{align}
\boldsymbol{\sigma}_{\boldsymbol S}(G_j)
&=
\boldsymbol v
\neq
\boldsymbol 0
\qquad
\forall j.
\label{eq:equal_syndrome_condition}
\end{align}
A stabilizer check then either commutes with every Pauli generator or
anticommutes with every Pauli generator. Consequently, every affected stabilizer check selects the entire LGA $U$. Its check-conditioned unitary is therefore just the doubled-angle operation $U^2$, reducing the many check-by-check constraints of the general theorem to a single preservation condition.

More importantly, if two Pauli generators have the same syndrome, their product has zero syndrome and therefore lies in the Pauli normalizer $N(\mathcal S)$ of the stabilizer group. Choose one Pauli generator $E=G_1$ as a reference and define $L_j=EG_j$.
Because $E$ and $G_j$ have the same syndrome, $L_j$ has zero syndrome
and therefore lies in the Pauli normalizer. Since the original Pauli generators
commute, $E$ commutes with every $L_j$, and the $L_j$ commute pairwise. Thus,
on the codespace they represent a commuting family of logical Pauli
operators. Since
$E^2=I$, we may equivalently write $G_j=EL_j$. Every physical Pauli generator can therefore be viewed as the same
syndrome-carrying Pauli $E$ multiplied by a different logical Pauli
operator $L_j$.

This separation is the key simplification provided by equal syndromes.
Substituting $G_j=EL_j$ into the LGA allows us to write the check-conditioned unitary as
\begin{align}
U^2=
\cos\left(2
\sum_{j=1}^{q}
\epsilon_j\theta_jL_j
\right)
+
iE
\sin\left(2
\sum_{j=1}^{q}
\epsilon_j\theta_jL_j
\right).
\label{eq:equal_syndrome_operator_decomposition}
\end{align}
The first term remains in the codespace, whereas the second carries the
state into the nonzero syndrome sector associated with $E$. Codespace preservation therefore requires the sine term to vanish on the codespace and the cosine term to act as the identity. The important consequence is that, once the common factor $E$ has been
separated, the gate-design problem involves only the
commuting logical Pauli operators $L_j$, their rotation magnitudes, and
their rotation directions. The common factor $E$ no longer determines
which logical gate is produced; instead, it determines how that logical
construction is realized physically.

To express the preservation condition as a scalar, we restrict to the codespace. Because the $L_j$ commute, the $2^k$ logical basis states can be chosen as simultaneous eigenstates of all $L_j$. We label these states by $\lambda$, and denote the eigenvalue of $L_j$ on logical basis state $\lambda$ by $\ell_j^{(\lambda)}=\pm1$. Define
\begin{align}
\alpha_\lambda
=
\sum_{j=1}^{q}
\epsilon_j\theta_j\ell^{(\lambda)}_j.
\label{eq:logical_sector_angle}
\end{align}
The sine and cosine in
Eq.~\eqref{eq:equal_syndrome_operator_decomposition} therefore have
argument $2\alpha_\lambda$ on logical basis state $\lambda$.
Codespace preservation requires the resulting check-conditioned
unitary to be the identity for every logical basis state $\lambda$. Hence:

\begin{corollary}[Equal-syndrome arbitrary-angle condition]
\label{cor:equal_syndrome_arbitrary_angle}
An equal-syndrome LGA preserves the codespace if and only if
\begin{align}
\sum_{j=1}^{q}
\epsilon_j\theta_j\ell_j^{(\lambda)}
&\in
\pi\mathbb Z
\label{eq:equal_syndrome_angle_condition}
\end{align}
for every logical basis state $\lambda$.
\end{corollary}

When this condition holds, the check-conditioned unitary
is the identity on every logical basis state. The original LGA $U$
can nevertheless retain a nontrivial logical phase. Thus equal syndromes cleanly
separate the logical and physical parts of the construction. The
logical operators $L_j$, rotation angle, and rotation directions
determine the encoded phase, while the choice of physical
representatives and the common syndrome-carrying Pauli $E$ determines
the actual Pauli generators $G_j=EL_j$.

Once the logical action of each zero-syndrome product $EG_j$ is fixed, the Pauli strings used for $E$ and $G_j$ can be changed. The encoded gate is unchanged as long as each new product $EG_j$ induces the same logical Pauli on the codespace and the resulting Pauli generators retain the commutation relations required by the construction. This freedom can be used to reduce both Pauli generator weight and physical-support overlap. 

For example, consider the $[[4,2,2]]$ code. A logical SWAP can be written using the generators $IYYY$, $YIYY$, $YYIY$, and $YYYI$. These generators have weight three and every pair overlaps on two qubits. Since the stabilizer $S_Y=YYYY$ acts as the identity on the codespace, multiplying each of these generators by $S_Y$ preserves its action on every encoded state and gives $Y_1,Y_2,Y_3,Y_4$, respectively. The resulting gate therefore implements the same logical SWAP using four disjoint weight-one rotations.

A \emph{Pauli-rotation schedule} for an LGA $U$ is a partition of its
constituent rotations into \emph{Pauli-rotation layers}, such that the
rotations within each layer have pairwise-disjoint physical support and
may therefore be executed in parallel when each multiqubit Pauli
rotation is treated as an elementary operation. We define the
\emph{Pauli-rotation depth} $D$ as the minimum number of layers in such
a schedule. Thus, an LGA whose Pauli generators have pairwise-disjoint
support has Pauli-rotation depth one. This quantity is distinct from
the circuit depth obtained after compiling each multiqubit Pauli
rotation into a hardware-native gate set.

Equal-syndrome LGAs also obey a weight bound.
Every pair $G_iG_j$ has zero syndrome. For the LGA to produce a
nontrivial logical operation, at least one such pair must represent a
nontrivial logical Pauli. Indeed, if every pair represented the trivial logical Pauli class,
then with $E=G_1$ one would have $L_j=G_1G_j\in\pm\mathcal S$
for every $j$. Each $L_j$ would therefore act as a scalar $\pm1$
on the codespace. The zero-syndrome part of the LGA would then be
scalar, while codespace preservation forces its nonzero-syndrome part
to vanish. The encoded operation would consequently be a global phase. Writing $w_{\max}=\max_j\operatorname{wt}(G_j)$, a nontrivial
logical Pauli has weight at least the code distance $d$, whereas a
product of two generators has weight at most $2w_{\max}$. We therefore
obtain
\begin{corollary}[Equal-syndrome weight bound]
\label{cor:equal_syndrome_weight_bound}
Any nontrivial equal-syndrome LGA satisfies
\begin{align}
w_{max}
&\geq
\left\lceil
\frac{d}{2}
\right\rceil.
\label{eq:equal_syndrome_weight_bound}
\end{align}
\end{corollary}
Further details are given in
Appendix~\ref{app:desired_syndrome_proof}.

\begin{figure*}[t]
\centering
\includegraphics[width=0.95\textwidth]{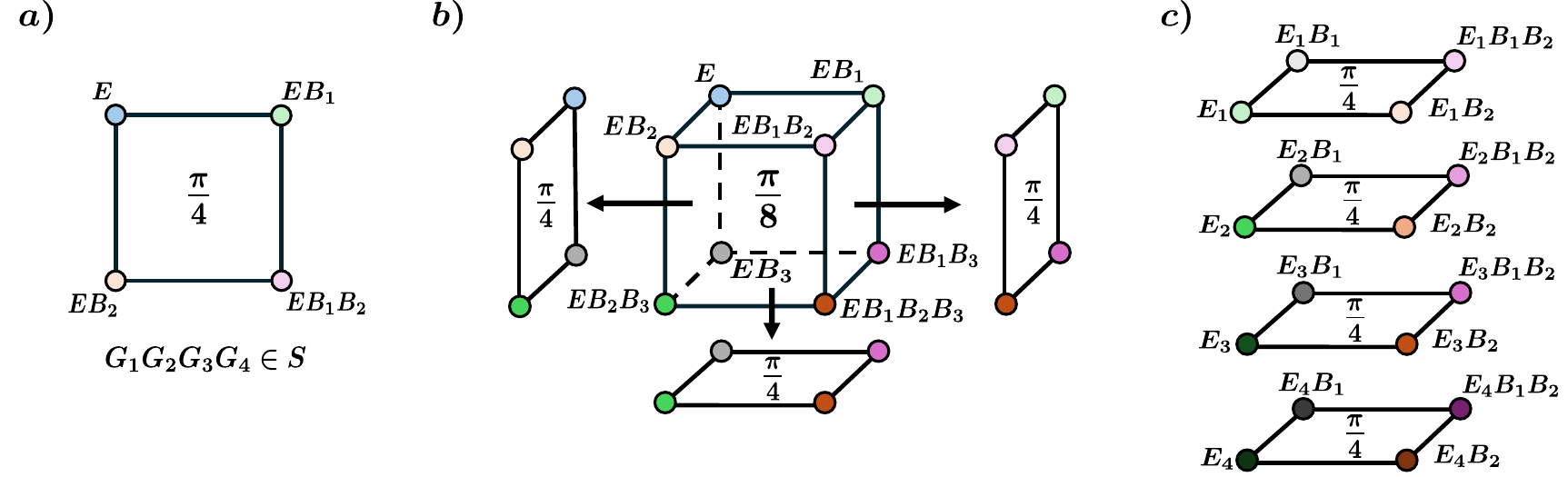}
\caption{
Geometric organization of equal- and unequal-syndrome logical gates.
Each vertex represents one Pauli-rotation generator, labeled by its
logical Pauli content relative to the common syndrome-carrying reference
$E$ (or $E_\mu$ for each partition in panel c). Here
$B_1,B_2,\ldots$ denote independent generators for commuting logical Pauli operators,
whose products label the vertices of the logical square or cube.
$\boldsymbol{a)}$ Four equal-syndrome Pauli generators form a rank-two logical square
$\{E,EB_1,EB_2,EB_1B_2\}$.  At angle $\pi/4$, the four Pauli generators
multiply to a stabilizer up to a sign fixed by the rotation directions and implement a logical Clifford operation.
$\boldsymbol{b)}$ Eight equal-syndrome Pauli generators form a rank-three logical cube at
angle $\pi/8$.  A logical-Pauli derivative selects one four-vertex
face of the cube and doubles its rotation angle to $\pi/4$, exposing
a lower-level Clifford gate; the three independent derivative
directions give the three pairs of opposite square faces.
$\boldsymbol{c)}$ Conversely, several rank-two Clifford partitions may have the
same order-two logical Clifford action but different common syndrome carriers
$E_\mu$. When their syndrome vectors satisfy
$\bigoplus_\mu \boldsymbol s_\mu=0$, the doubled-angle actions cancel check by check, giving a
code-preserving half-angle LGA; the encoded action determines
whether the resulting gate is non-Clifford.
}
\label{fig:syndrome_partition_hierarchy}
\end{figure*}
% ------------------------------------------------------------
\subsection{Equal-syndrome Clifford constructions}
\label{subsec:pi4_syndrome_partition}
% ------------------------------------------------------------

We first specialize this structure to the Clifford angle. Let every
rotation have angle $\pi/4$, while retaining independent directions
$\epsilon_j=\pm1$. At this angle $U^2$ is
proportional to the product of its generators, so every affected stabilizer check collects the same Pauli product
\begin{align}
R
&=
G_1G_2\cdots G_q.
\label{eq:equal_syndrome_stabilizer_multiplier}
\end{align}
The equal-syndrome preservation condition therefore reduces to a
stabilizer-factorization rule: The product $R$ must lie in
$\pm\mathcal S$, with the rotation directions fixing the required
$+1$-eigenspace sign.

\begin{corollary}[Equal-syndrome $\pi/4$ condition]
\label{thm:desired_syndrome_partition}
An equal-syndrome $\pi/4$ LGA whose Pauli generators $G_1,\ldots,G_q$ have a
common nonzero syndrome preserves the $+1$ codespace if and only if
\begin{align}
i^q
\left(\prod_{j=1}^{q}\epsilon_j\right)
\left(\prod_{j=1}^{q}G_j\right)
&\in
\mathcal S.
\label{eq:general_syndrome_partition_condition}
\end{align}
Equivalently, $q$ is even and the Pauli product
$R=\prod_jG_j$ satisfies $R=\rho S$ for some
$S\in\mathcal S$ and $\rho\in\{+1,-1\}$, with
$i^q\rho\prod_j\epsilon_j=1$.
\end{corollary}

This is precisely the stabilizer-factorization picture encountered in
the $[[4,2,2]]$ example, and it turns the search for code-preserving
Clifford LGAs into a particularly concrete design problem. Starting
from a stabilizer $R$, we ask whether its support can be broken into
commuting Pauli generators
\begin{align}
R
&=
G_1G_2\cdots G_q
\end{align}
that all carry the same nonzero syndrome. Whenever such a factorization
exists, the corresponding $\pi/4$ rotations preserve the codespace
after the rotation directions are chosen to satisfy the stabilizer-sign
constraint $i^q\prod_{j=1}^{q}\epsilon_j
=
1$.

The physical support of the Pauli generators then determines the Pauli-rotation depth.
In particular, if the $G_j$ have pairwise-disjoint support, all rotations
occupy one Pauli-rotation layer. Thus the problem of finding shallow constructions
in this Pauli-rotation model can be recast as the problem of finding disjoint
equal-syndrome factorizations of stabilizers. This viewpoint underlies
the direct searches performed below, where we enumerate such
factorizations on existing stabilizer codes and determine the logical
Clifford operations they induce.

% ------------------------------------------------------------
\subsection{Equal-syndrome gates higher in the Clifford hierarchy}
\label{subsec:equal_angle_hierarchy}
% ------------------------------------------------------------

At $\pi/4$, codespace preservation can be determined from the total product of the physical Pauli generators. At smaller angles, this simplification no longer occurs. As Eq.~\eqref{eq:equal_syndrome_operator_decomposition} shows, the zero-syndrome logical operators associated with the Pauli generators appear inside an operator-valued sine and cosine. Codespace preservation therefore depends on how these logical Pauli contributions combine to ensure the sine term vanishes.

Consider equal rotation angles $\theta_j=\theta$. Again let $\lambda$ label a logical basis state that is a simultaneous eigenstate of all $L_j$, and let $\ell_j^{(\lambda)}=\pm1$ be the eigenvalue of $L_j$ on that state. The angle entering the preservation condition for logical basis state $\lambda$ is then
\begin{align}
\alpha_\lambda
&=
\theta A_\lambda,
&
A_\lambda
&=
\sum_{j=1}^{q}
\epsilon_j\ell_j^{(\lambda)}.
\end{align}
Codespace preservation requires $\theta A_\lambda$ to be an integer multiple of $\pi$ for every logical basis state $\lambda$. For $\theta=\pi/4,\pi/8,\pi/16,\ldots$, the corresponding values of $A_\lambda$ must therefore be divisible by $4,8,16,\ldots$, respectively. This immediately places a lower bound on the number of Pauli generators needed to obtain a nontrivial logical phase. Since $A_\lambda$ is a sum of $q$ numbers equal to $\pm1$, producing $A_\lambda=\pm 2^m$ at angle $\theta=\pi/2^m$ requires at least $2^m$ Pauli generators. Thus a $\pi/8$ construction requires at least eight Pauli generators, a $\pi/16$ construction requires at least sixteen, and so forth.

We now use this condition to construct a specific class of logical gates. On the logical subsystem to be acted on, choose a basis state $|\lambda_0\rangle$ and define the phase-flip gate
\begin{align}
U_{\lambda_0}
&=
I-2\Pi_{\lambda_0},
&
\Pi_{\lambda_0}
&=
|\lambda_0\rangle\langle \lambda_0|.
\end{align}
This gate multiplies $|\lambda_0\rangle$ by $-1$ and leaves every other logical basis state unchanged. Controlled-$Z$ gates are examples of this form where choosing $|\lambda_0\rangle=|11\rangle$ gives CZ, choosing $|\lambda_0\rangle=|111\rangle$ gives CCZ, and more generally this pattern continues to give C${}^r$Z gates.

For $\theta=\pi/2^m$, this phase pattern can be generated from $m$ independent commuting logical Pauli operators $B_i$ with $i=1,\ldots,m$. Their $2^m$ possible products provide the $2^m$ operators $L_j$ required by the bound above. The rotation signs $\epsilon_j$ are chosen so that, on the target logical basis state $|\lambda_0\rangle$, all $2^m$ terms contribute with the same sign, giving $A_{\lambda_0}=\pm2^m$. On every other logical basis state, half of the terms contribute $+1$ and half contribute $-1$, giving $A_{\lambda\ne\lambda_0}=0$. The target state therefore acquires the phase $e^{\pm i\pi}=-1$, while every other logical basis state acquires phase $+1$. This choice of rotation signs is a Walsh-sign pattern. The general sign construction is given in Appendix~\ref{app:desired_syndrome_proof}.

This construction has a simple geometric representation, shown in Fig.~\ref{fig:syndrome_partition_hierarchy}. Here the rank is the number of independent commuting logical Pauli
operators that generate the product family. Thus two independent
logical Paulis generate the four vertices of a rank-two square, while
three generate the eight vertices of a rank-three cube. The square represents the four logical products used in the $\pi/4$ construction, while the cube represents the eight logical products used in the $\pi/8$ construction.

The discussion above determines which logical Pauli action each $L_j$ must have, but it does not uniquely determine the physical Pauli string used to realize that action. If $S$ is any stabilizer, then $L_j$ and $L_jS$ act identically on every encoded state because $S|\psi\rangle=|\psi\rangle$ throughout the codespace. We may therefore multiply the chosen $L_j$ by stabilizers without changing any of the eigenvalues $\ell_j^{(\lambda)}$, the Walsh-sign condition, or the resulting logical gate. These substitutions can change which physical qubits appear in $L_j$ and can reduce its Pauli weight.

After choosing physical representatives for the $L_j$, we choose a Pauli $E$ with nonzero syndrome that commutes with all of them and define $G_j=EL_j$. Every $G_j$ then has the syndrome of $E$, since each $L_j$ has zero syndrome. The stabilizer-equivalent choices of the $L_j$ and the allowed choices of $E$ can be searched over to reduce the weight and support overlap of the physical Pauli generators $G_j$ while keeping the logical gate fixed. The existence argument is given in Appendix~\ref{app:logical_sign_construction}. In the constructions below, $w_{\max}=\lceil d/2\rceil$ attains the lower bound of
Corollary~\ref{cor:equal_syndrome_weight_bound}.

% ------------------------------------------------------------
\subsection{Syndrome-partition lifting and logical derivatives}
\label{subsec:syndrome_module_lifting}
% ------------------------------------------------------------

We now show how syndrome partitions can connect different levels of the Clifford hierarchy. We first use known code-preserving Clifford gates and run them at half-angle to construct non-Clifford gates, and then show how logical Pauli commutators recover lower-level gates from higher-level ones.

An interesting possibility presents itself once several code-preserving
equal-syndrome LGAs are known. Suppose there are $M$ such doubled-angle
LGAs $R_\mu$, with $\mu=1,\ldots,M$, and write
\begin{align}
R_\mu
&=
\prod_{j=1}^{q_\mu}
\exp\left(
2i\epsilon_{\mu j}\theta_{\mu j}G_{\mu j}
\right).
\end{align}
Within each $R_\mu$, all Pauli generators $G_{\mu j}$ share one common
nonzero syndrome,
\begin{align}
\boldsymbol{\sigma}_{\boldsymbol S}(G_{\mu j})
&=
\boldsymbol s_\mu,
\qquad
j=1,\ldots,q_\mu,
\end{align}
while $\boldsymbol s_\mu$ may differ between the different LGAs.

We define the corresponding half-angle LGA
\begin{align}
W_\mu
&=
\prod_{j=1}^{q_\mu}
\exp\left(
i\epsilon_{\mu j}\theta_{\mu j}G_{\mu j}
\right),
\end{align}
so that $W_\mu^2=R_\mu$. Because all Pauli generators within
$W_\mu$ have syndrome $\boldsymbol s_\mu$, a stabilizer check $S_a$
selects the entire doubled-angle operation $R_\mu$ when
$(\boldsymbol s_\mu)_a=1$ and selects none of it when
$(\boldsymbol s_\mu)_a=0$.

\begin{theorem}[Syndrome-partition lifting]
\label{thm:syndrome_module_lifting}
Assume that the $W_\mu$ commute and that each $R_\mu$
preserves the codespace with logical action $\overline R_\mu$.  Then
\begin{align}
W
&=
\prod_\mu W_\mu
\end{align}
preserves the codespace if and only if
\begin{align}
\prod_\mu
\overline R_\mu^{\,(\boldsymbol s_\mu)_a}=\overline{I}
\qquad
\forall a .
\label{eq:module_logical_cancellation}
\end{align}
\end{theorem}

For a fixed stabilizer check $S_a$, the combined LGA selects exactly
those component operations $R_\mu$ for which
$(\boldsymbol s_\mu)_a=1$. Equation~\eqref{eq:module_logical_cancellation}
therefore requires the logical actions of those selected operations to
multiply to the identity.

A particularly simple sufficient condition occurs when every $R_\mu$
has the same order-two logical action,
$\overline R_\mu=\overline C$, with
$\overline C^2=\overline I$. If
\begin{align}
\bigoplus_{\mu=1}^{M}\boldsymbol s_\mu
&=
\boldsymbol 0,
\label{eq:module_zero_syndrome_relation}
\end{align}
then every stabilizer check selects an even number of the component
LGAs, so their doubled-angle logical actions multiply to the identity.
The total number $M$ of component LGAs need not itself be even: two,
three, four, five, or more LGAs may participate as long as their
syndrome vectors satisfy the zero-sum relation. For example, three
components with
$\boldsymbol s_3=\boldsymbol s_1\oplus\boldsymbol s_2$
already satisfy the condition, because every stabilizer check selects
either zero or two of them. 

This construction is illustrated in
Fig.~\ref{fig:syndrome_partition_hierarchy}(c).  Each square represents
an equal-syndrome Clifford partition with the same logical structure
but a different common syndrome.  When the syndrome vectors satisfy
the required zero-sum relation, the partitions can be run at half
angle and combined so that their doubled-angle Clifford actions cancel
check by check, leaving a code-preserving half-angle LGA. Its encoded
action is then evaluated to determine whether it is non-Clifford.

There is also a complementary operation that works in the opposite
direction. Let $L$ be a physical Pauli representative of a logical Pauli
operator and let $U$ be a code-preserving LGA. Consider
\begin{align}
D_L(U)=L U^\dagger L U.
\label{eq:logical_pauli_derivative}
\end{align}
We call this group commutator $D_L(U)$ the \emph{logical derivative} of $U$ with respect to
$L$. Conjugation by $L$ leaves rotations generated by Paulis that commute with $L$ unchanged and reverses the sign of those that anticommute with it. Consequently, the commuting terms cancel while the anticommuting terms are doubled,
\begin{align}
D_L(U)
&=
\prod_{j:\{L,G_j\}=0}
\exp\left(
2i\epsilon_j\theta_jG_j
\right).
\label{eq:logical_derivative_physical}
\end{align}
Thus a logical Pauli plays the same selecting role as a stabilizer check in that it picks out exactly the Pauli generators that anticommute with it. The difference here is that a stabilizer check must select an operation that is trivial on the codespace, whereas a logical Pauli can select a nontrivial logical operation. On the codespace,
\begin{align}
\overline{D_L(U)}
&=
\overline L\,
\overline U^\dagger\,
\overline L\,
\overline U,
\label{eq:logical_derivative_encoded}
\end{align}
which is the encoded Pauli commutator of $\overline U$ with $\overline L$. In this way we can start from a logical gate $U$ at some level of the Clifford hierarchy and produce a logical gate $D_L(U)$ that lies in a lower level of the hierarchy.

The use of logical derivatives to descend the Clifford hierarchy is well
established in the theory of fault-tolerant logical gates
\cite{bravyi2013classification,pastawski2015fault,vasmer2019surface,
webster2023diagonal,haruna2026homological,williamson2026fast}, and related
constructions obtain lower-level gates from specific transversal or
topological higher-level gates. Here the same descent is expressed
for a general LGA and, unlike existing literature, does not require a
CSS structure, a transversal diagonal gate, or a particular geometry.
Equation~\eqref{eq:logical_derivative_physical} identifies exactly which
Pauli generators anticommute with the chosen logical Pauli and doubles their angles.
When the Pauli generators form a rank-$r$ logical cube, a nontrivial logical derivative
selects a rank-$(r-1)$ subcube.

% ------------------------------------------------------------
\subsection{Pauli roots from equal-syndrome pairs}
\label{subsec:syndrome_embedded_roots}
% ------------------------------------------------------------
The equal-syndrome construction is powerful, but its defining preservation condition is that every affected check sees the same doubled-angle operation $U^2$, which must act trivially on the codespace. Consequently, the logical gates obtained from the logical-square and logical-cube constructions are naturally order-two operations, such as C$^r$Z type gates. That does not directly give the other higher-order operations we would like to construct such as logical roots of $X$, $Y$, $Z$, or $T$ gates.

This motivates a second structured reduction of the general syndrome condition. This construction begins by using a rotation identity for commuting Pauli operators. Let $P_1,\ldots,P_r$ be independent commuting Hermitian Pauli
operators, and let
\begin{align}
\mathcal A
&=
\langle P_1,\ldots,P_r\rangle
=
\left\{
P_1^{v_1}\cdots P_r^{v_r}:
\boldsymbol v\in\mathbb F_2^r
\right\}
\end{align}
be the abelian Pauli subgroup they generate.  It contains $2^r$
elements, one for each binary choice of the generators.  At the angle $\theta_r=\frac{\pi}{2^{r-1}}$ rotating about every nonidentity element of $\mathcal A$ gives only a
global phase,
\begin{align}
\prod_{P\in\mathcal A\setminus\{I\}}
\exp\left(i\theta_r P\right)
&\doteq I.
\label{eq:complete_pauli_root_identity}
\end{align}
where $\doteq$ denotes equality up to a global phase. This commuting-Pauli identity has been used in parity-phase circuit synthesis, non-Clifford resource reduction, and constructions of diagonal logical operators
\cite{amy2019tcount,debeaudrap2020spider,kissinger2024spider,
webster2023diagonal}. Here we use it to construct roots of a logical Pauli $L$ from physical Paulis with controlled syndrome structure.

Set $P_1=L$, where $L$ is a physical Pauli with zero stabilizer
syndrome that acts as the desired logical Pauli on the codespace.
Omitting the rotation about $L$ from
Eq.~\eqref{eq:complete_pauli_root_identity} leaves $2^r-2$
rotations. Since the complete product is proportional to the identity,
these remaining rotations implement
$\exp(-i\theta_r L)$ up to a global phase.

For $r=3$, choose two additional commuting Pauli operators $P_2$ and
$P_3$, both commuting with $L$, with independent nonzero syndromes
$\boldsymbol a$ and $\boldsymbol b$. The six operators left after
removing $I$ and $L$ are
\begin{align}
\{
P_2,LP_2,
P_3,LP_3,
P_2P_3,LP_2P_3
\}.\label{eq:Clifford_root_construction}
\end{align}
Choosing $L$ as one of the independent generators has an additional
consequence. Every
remaining Pauli $Q$ in the subgroup appears together with $LQ$.
Because $L$ has zero stabilizer syndrome, $Q$ and $LQ$ automatically
have the same syndrome, while their product is the same logical Pauli
$L$. The Pauli-root identity therefore organizes the physical
generators into equal-syndrome pairs of the form $(Q,LQ)$.

For $r=3$, taking $P_2$ and $P_3$ to have independent nonzero
syndromes $\boldsymbol a$ and $\boldsymbol b$ gives the three pairs
\begin{align}
&(P_2,LP_2)
&&\text{with syndrome }\boldsymbol a,
\nonumber\\
&(P_3,LP_3)
&&\text{with syndrome }\boldsymbol b,
\nonumber\\
&(P_2P_3,LP_2P_3)
&&\text{with syndrome }\boldsymbol a+\boldsymbol b.
\label{eq:syndrome_plane_root_pairs}
\end{align}
Thus the three pair syndromes are precisely the three nonzero vectors
of the rank-two syndrome subspace generated by
$\boldsymbol a$ and $\boldsymbol b$. This provides a direct search rule in which one
looks for three equal-syndrome pairs whose pair products represent the
same logical Pauli $L$ and whose syndrome labels are
$\boldsymbol a$, $\boldsymbol b$, and
$\boldsymbol a+\boldsymbol b$.

The same syndrome relation also makes codespace preservation automatic.
For any stabilizer check $S_a$,
$(\boldsymbol a+\boldsymbol b)_a
=(\boldsymbol a)_a+(\boldsymbol b)_a \pmod 2$, so the check selects
either none or exactly two of the three pairs. At physical angle
$\pi/4$, a selected pair with a common rotation direction contributes
the doubled-angle logical operation $-L$. A check that selects two
pairs therefore sees $(-L)^2=I$, while a check that selects no pairs
sees the identity. Hence the check-conditioned unitary is trivial for
every stabilizer check, as required by
Theorem~\ref{thm:arbitrary_angle_syndrome_partition}.

The logical action of the full six-rotation LGA follows independently
from Eq.~\eqref{eq:complete_pauli_root_identity}. These six operators
are exactly the terms that remain after the rotations about $I$ and
$L$ are removed, so applying a $\pi/4$ rotation about each gives
\begin{align}
R_L(\pi/4)
&\equiv
\exp\left(-i\frac{\pi}{4}L\right)
\end{align}
up to a global phase. If $L=\overline Z_i$, this is the logical phase
gate $S_i$. More generally, we call
$R_L(\theta)=e^{-i\theta L}$ a logical Pauli root when an integer power
of the rotation equals $L$ up to a global phase. Thus $R_L(\pi/4)$ is a
square root of $L$, while $R_L(\pi/8)$ is a fourth root.

The same construction at $r=4$ gives $\pi/8$ rotations. Choose a third commuting Pauli $P_4$, with syndrome
$\boldsymbol c$ independent of $\boldsymbol a$ and
$\boldsymbol b$. The seven nonzero syndrome values are
\begin{align}
\{
\boldsymbol a,
\boldsymbol b,
\boldsymbol c,
\boldsymbol a+\boldsymbol b,
\boldsymbol a+\boldsymbol c,
\boldsymbol b+\boldsymbol c,
\boldsymbol a+\boldsymbol b+\boldsymbol c
\}.
\label{eq:seven_root_syndromes}
\end{align}
Each syndrome labels an equal-syndrome pair whose two members differ by
the same logical Pauli $L$. The seven pairs therefore contain fourteen
Pauli generators $G_j$, exactly the $2^4-2$ operators remaining in
the $r=4$ Pauli-root identity. Their $\pi/8$ rotations realize
\begin{align}
R_L(\pi/8)
&\equiv
\exp\left(-i\frac{\pi}{8}L\right),
\end{align}
up to a global phase. For $L=\overline Z_i$, this is the logical
$T_i$ gate.

Codespace preservation follows directly from this syndrome structure for any $r$. The $r-1$ independent syndrome vectors generate $2^{r-1}$ syndrome combinations, and any stabilizer check that is affected by the construction selects exactly half of them. It therefore
selects $2^{r-2}$ equal-syndrome pairs, or $2^{r-1}$ Pauli generators
in total. Relative to any one selected generator $E$, these generators
can be written as $E\mathcal B$, where $\mathcal B$ is a commuting
Pauli group of size $2^{r-1}$. The sum over such a group is $\sum_{B\in\mathcal B} B=2^{r-1}\Pi$, where $\Pi$ projects onto its common $+1$ eigenspace. Since $E$ commutes with $\Pi$, $E^2=I$, and $\theta_r=\pi/2^{r-1}$, the check-conditioned unitary is $V_a=\exp(i2\pi E\Pi)=I$. A stabilizer check that is unaffected by the construction selects no
rotations and also sees the identity. Thus every stabilizer check
satisfies the codespace-preservation condition.

\begin{figure}[t]
\centering
\includegraphics[width=0.95\columnwidth]{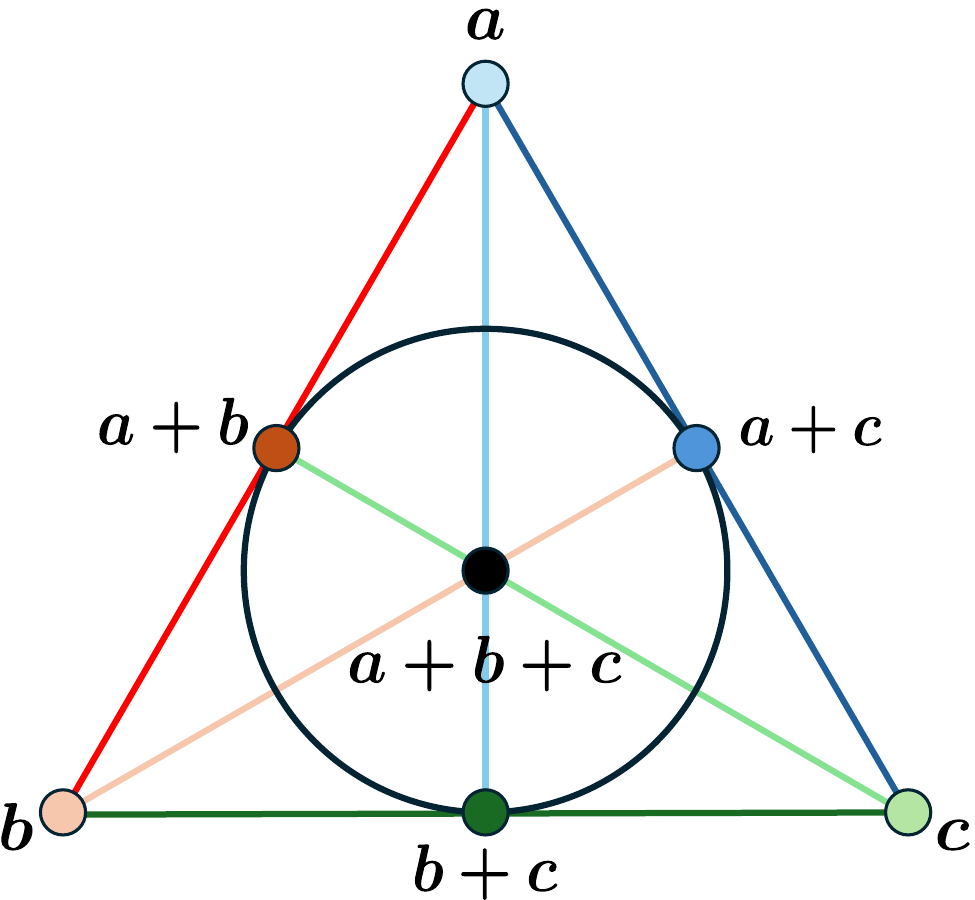}
\caption{
Representation of the Pauli-root construction of a logical $T$-type gate as a Fano plane.
Each point represents one equal-syndrome pair whose product is the
same logical Pauli $L$. The seven points are the seven nonzero
syndromes in a rank-three syndrome space. Each line contains a triple
$\{\boldsymbol u,\boldsymbol v,\boldsymbol u+\boldsymbol v\}$ and hence
the three equal-syndrome pairs of one six-rotation phase gate $S$ or
$S^\dagger$ construction. All seven points together give the seven
pairs, or fourteen physical rotations, of the $T$-gate construction.
For the direct $T$-gate identity all fourteen rotations may be taken with
the same sign. If instead the gate is reconstructed from the seven
overlapping $S/S^\dagger$ constructions, the three line
contributions meeting at each point are chosen with signs
$+,+,-$, leaving one net positive copy of that syndrome pair.
}
\label{fig:Fano_plane}
\end{figure}

Figure~\ref{fig:Fano_plane} organizes the relation between the $\pi/4$ and $\pi/8$ constructions. The seven nonzero syndrome vectors can be interpreted as the seven points of the Fano plane associated with this rank-three syndrome space. Any three points whose labels sum to
zero are the three nonzero vectors of a rank-two syndrome subspace and hence one of the three-pair
$\pi/4$ constructions above. For example, the line:
$\{
\boldsymbol a,\,
\boldsymbol b,\,
\boldsymbol a+\boldsymbol b
\}$
defines one such Clifford root about $L$. The seven lines of the Fano
plane enumerate the seven overlapping three-pair Clifford constructions
contained in the same seven-syndrome construction.

Similar to the lifting procedure of Sec.~\ref{subsec:syndrome_module_lifting}, this geometry provides another setting in which Clifford gates can be used as a resource for higher-level constructions. Each syndrome point belongs to
three Fano lines, so each equal-syndrome pair occurs in three different
Clifford gates. Simply running all seven Clifford constructions at half angle would therefore count each equal-syndrome pair three times. The rotation signs must instead be chosen compatibly so that the three half-angle contributions meeting at each Fano-plane point combine to one net $\pi/8$ rotation on that pair. The signs are assigned to each occurrence of an equal-syndrome pair
within a three-pair Clifford construction. They are not a single
uniform sign attached to an entire Fano line. Reversing both Pauli generators
of one pair preserves the root construction, with its direction
verified from the resulting six-generator operator. One can choose the sign combination,
\begin{align}
+1+1-1=+1,
\label{eq:fano_sign_rule}
\end{align}
up to permutation of the three lines associated with a given vertex. Thus two positive
half-angle contributions and one negative contribution leave exactly one
net $\pi/8$ rotation on each member of the pair. When such a compatible sign assignment is made across all seven lines,
the seven overlapping Clifford gates reproduce the fourteen-rotation logical gate $R_L(\pi/8)$.

% ===========================================================
\section{Syndrome-partition searches on existing qLDPC codes}
\label{sec:qldpc_searches}
% ============================================================

The constructions of the previous section turn codespace preservation into algebraic conditions on the syndromes and relative logical actions of the physical Pauli generators making up Eq.~(\ref{eq:general_arbitrary_angle_layer}). We now use those conditions as search rules for direct logical gates on existing qLDPC codes. In addition to finding nontrivial logical operations, we look for physically simple realizations with low Pauli generator weight and low Pauli-rotation depth, so that the resulting Pauli rotations act on small and, where possible, disjoint sets of physical qubits. The accompanying JSON file contains the signed stabilizer generators, fixed logical Pauli frames, and complete Pauli-rotation generators for the code-preserving LGAs found in this work.

To begin, we use the syndrome framework to construct catalogs of logical Clifford gates on existing qLDPC codes. First, we factor stabilizers into disjoint, low-weight Pauli generators with a common nonzero syndrome, yielding code-preserving Clifford LGAs. Second, we combine equal-syndrome pairs whose products represent the same logical Pauli $L$ so that three suitably related pairs realize a logical root $R_L(\pi/4)$. These two constructions generate complementary catalogs of logical Clifford gates. We then use these Clifford gates to construct gates one level higher in the Clifford hierarchy using the methods in Secs.~\ref{subsec:syndrome_module_lifting} and ~\ref{subsec:syndrome_embedded_roots}. We also search directly for non-Clifford gates in both settings. Rank-three common-syndrome cubes yield CCZ-type gates, while complete seven-pair structures yield logical Pauli roots at angle $\pi/8$.

We apply these searches primarily to the self-dual bivariate-bicycle (BB) codes \cite{liang2026self}, whose previously studied logical-gate mechanisms are predominantly Clifford \cite{eberhardt2407logical,sayginel2025fault,liang2026self,albert2026beyond}. This makes them a useful setting for demonstrating how the syndrome constructions above can produce additional low-weight Clifford gates and operations higher in the Clifford hierarchy. We also study the non-CSS $[[36,6,6]]$ Mirror code of Ref.~\cite{khesin2026mirror} to test the applicability of our constructions beyond the CSS setting.

% ------------------------------------------------------------
\subsection{Clifford gates from direct stabilizer search}
\label{subsec:qldpc_clifford_search}
% ------------------------------------------------------------

We begin with the equal-syndrome Clifford construction of
Sec.~\ref{subsec:pi4_syndrome_partition}.  In the qLDPC searches below we restrict attention to
four-generator LGAs with Pauli-rotation depth one.  At physical angle $\pi/4$,
such an LGA preserves the code when the four commuting Pauli generators share
a common nonzero syndrome and, with the appropriate rotation signs,
their product is a $+1$ stabilizer.

We restrict the direct Clifford searches to four Pauli generators because
this is the smallest nontrivial equal-syndrome construction at angle
$\pi/4$. If all $q$ generators have the same nonzero syndrome
$\boldsymbol v$, then their product has syndrome
$q\boldsymbol v$ modulo two. Since the $\pi/4$ preservation condition
requires this product to be a stabilizer, its syndrome must vanish, so
$q$ must be even. For $q=2$, the product $G_1G_2$ is itself a
stabilizer, so the two generators differ only by a stabilizer and the
resulting LGA has no nontrivial logical action. The first nontrivial case
is therefore $q=4$.

This approach provides a direct way to search for codespace-preserving Clifford LGAs.  For
the BB codes, we begin with a stabilizer and ask if it can be split
into four low-weight, pairwise-disjoint Pauli generators with the same
nonzero syndrome.  A proposed factorization is retained only after the
equal-syndrome and stabilizer-product conditions are verified and the
rotation signs are chosen so that the specified codespace is preserved.
We then compute the induced action in a fixed logical Pauli frame. Alternatively, one can first enumerate low-weight Pauli generators,
group them by their stabilizer syndrome, and look for four mutually
disjoint Pauli generators whose product is a stabilizer.  This is the strategy
used for part of the Mirror search. Throughout the searches in this work, a weight cutoff
$w_{\max}$ on Pauli rotation generators is applied independently
$\operatorname{wt}(G_j)\leq w_{\max}$, so the search includes all mixed
Pauli-generator weight profiles within this bound and does not require
the generators in an LGA to have equal weight.

The finite search domain depends on the code size. For the $[[16,4,4]]$ BB code, we enumerate the full stabilizer group and search for four-generator disjoint factorizations with Pauli generator weight at most four. For the larger code instances, the BB and Mirror constructions are defined on periodic lattices with translationally related stabilizer checks which we can use to organize the search space. For the $[[64,8,8]]$ BB code, we search products of at most three translated stabilizer checks, with stabilizer weight at most $24$ and Pauli generator weight at most six. For the Mirror code, we use translation symmetry to identify translation-related Pauli generators and search one representative from each symmetry orbit, both at weight four and in a mixed-weight search through weight six. Full search details are given in Appendix~\ref{app:qldpc_clifford_search_details}.

Since the binary symplectic action of a Clifford gate is unchanged under multiplication by a Pauli string, two LGAs are identified when they
induce the same binary symplectic action on the fixed logical Pauli
frame. LGAs that differ by a logical Pauli string are therefore
counted together. For the self-dual BB codes, the $[[16,4,4]]$ search exhausts the full search domain and finds $52{,}056$ code-preserving LGAs. Evaluating their action in a fixed logical Pauli frame gives $291$ distinct logical Clifford operations. Under the bounded stabilizer domain, the $[[64,8,8]]$ search finds $3{,}072$ code-preserving LGAs realizing $112$ distinct logical Clifford operations. Finally, for the non-CSS $[[36,6,6]]$ Mirror code, the bounded searches yield $47$ translation-inequivalent orbits that realize $25$ distinct logical Clifford operations. In the chosen logical Pauli frames, the catalogs
include addressed CNOT- and SWAP-type operations as well as more correlated
multiqubit Cliffords. Importantly, we retain the physically distinct realizations of the
same logical action as their stabilizer syndromes can differ,
providing different building blocks for non-Clifford gates.

Every Clifford gate found consists of four pairwise
support-disjoint Pauli generators with one common nonzero syndrome. For
example, the $[[16,4,4]]$ BB catalog contains the LGA with Pauli-rotation depth one, generated by $G_1=X_1X_3$, $G_2=X_6X_{10}$, $G_3=X_9X_{11}$, and
$G_4=X_2X_{16}$. The four weight-two Pauli generators multiply to a stabilizer,
so this construction attains the equal-syndrome bound
$w_{\max}=d/2=2$. In the bounded $[[64,8,8]]$ BB search, the successful
weight profiles of the four generators $G_j$ are $(2,6,6,6)$ and $(6,6,6,6)$. The Mirror catalog
contains both uniform $(4,4,4,4)$ partitions and mixed-weight solutions,
including $(2,4,6,6)$. Representative signed Pauli strings, full
logical actions, and the code-specific search domains are given in
Appendix~\ref{app:qldpc_clifford_search_details}.

\begin{table}[t]
\centering
\begin{tabular*}{\columnwidth}{@{\extracolsep{\fill}}ccc@{}}
\hline\hline
Code & Code type & Distinct logical gates \\
\hline
$[[16,4,4]]$ BB      & CSS     & $291$ \\
$[[64,8,8]]$ BB      & CSS     & $112$ \\
$[[36,6,6]]$ Mirror  & non-CSS & $25$ \\
\hline\hline
\end{tabular*}
\caption{Direct equal-syndrome Clifford search summary. The counts give distinct fixed-frame logical Clifford actions modulo
logical Pauli operators and global phases, after identifying repeated
physical realizations. The $[[16,4,4]]$ count exhausts the finite domain defined in
Appendix~\ref{app:qldpc_clifford_search_details}; the $[[64,8,8]]$ and Mirror
counts use the bounded domains defined there.}
\label{tab:qldpc_clifford_summary}
\end{table}

% ------------------------------------------------------------
\subsection{Building higher-level gates from Clifford partitions}
\label{subsec:qldpc_module_lifting}
% ------------------------------------------------------------

We next use the catalogs of Clifford gates found in the previous section as building
blocks for gates one level higher in the Clifford hierarchy. Consider a
small collection of equal-syndrome $\pi/4$ Clifford LGAs. The LGAs
may have different nonzero syndrome vectors $\boldsymbol s_\mu$, but we
require them to implement the same order-two logical Clifford
$\overline C$ on the codespace. Here $\boldsymbol s_\mu$ is the common stabilizer syndrome of the
physical Pauli generators in the $\mu$th LGA, and $\overline C$ denotes the exact operator acting on the codespace, including its overall sign. This distinction matters because two LGAs that induce the same Clifford action but contribute $\overline C$ and $-\overline C$ would combine to $-I$, rather than $+I$, when selected by the same stabilizer check.

Now halve every physical rotation angle from $\pi/4$ to $\pi/8$. For a
given stabilizer check, the doubled-angle operation contributed by each
selected Clifford LGA is again $\overline C$. If their syndromes satisfy
\begin{align}
\bigoplus_\mu \boldsymbol s_\mu
&=
\boldsymbol 0,
\end{align}
then each stabilizer check selects an even number of the LGAs used in the construction. Since every selected LGA contributes the same order-two logical operation $\overline C$, the
doubled-angle contributions cancel pairwise for every check, and the total
number of LGAs used in the construction may be odd or even. Since $\overline C^2=I$ with the correct codespace phase,
these doubled-angle contributions cancel for every check. The complete
half-angle LGA can nevertheless act nontrivially on the logical
qubits, producing a gate one level higher in the hierarchy.

This is the lifting mechanism of
Sec.~\ref{subsec:syndrome_module_lifting}. In the Clifford catalogs we search for small sets of Clifford LGAs
whose common-syndrome vectors sum to zero modulo two. The lifting theorem allows any number of Clifford LGAs to be used in the non-Clifford gate construction; the
three-LGA and four-LGA restriction here is only a finite-search choice. The BB16 and BB64 constructions reported
below use four Clifford LGAs, whereas the Mirror construction uses
three. After a relation is found, the corresponding physical rotations
can be run at half angle to give a new code-preserving LGA. We then
compute its encoded action to check whether it corresponds to a non-Clifford logical gate.

The $[[16,4,4]]$ BB code gives the simplest example. Four distinct equal-syndrome Clifford LGAs, each containing four disjoint weight-two $X$-type Pauli generators, implement the same exact order-two logical Clifford at $\pi/4$. Their four syndrome vectors have
zero binary sum. Combining these sixteen Pauli generators together into one LGA, each with angle $\pi/8$, provides a code-preserving unequal-syndrome gate with $w_{\max}=2$ and
Pauli-rotation depth two, with eight Pauli generators in each Pauli-rotation layer. Its
logical action is of CCZ-type. For instance in the chosen logical Pauli frame it implements $\overline{Z}_3\overline{CZ}_{13}\overline{CZ}_{23}\overline{CCZ}_{123}$.

The same mechanism appears independently in the larger BB and non-CSS
Mirror catalogs. On the $[[64,8,8]]$ BB code, four Clifford LGAs
produce a verified non-Clifford lift with $w_{\max}=6$ and
Pauli-rotation depth two, again with eight Pauli generators per Pauli-rotation layer; its logical action is also of CCZ-type. On the $[[36,6,6]]$ Mirror
code, three Clifford LGAs give a construction with
Pauli-rotation depth three and $w_{\max}=6$, producing a CS-type
logical action. Additional search details are provided in
Appendix~\ref{app:unequal_lifting_examples}. These lifting results are summarized in the first three rows of
Table~\ref{tab:qldpc_nonclifford_results}.

% ------------------------------------------------------------
\subsection{Direct rank-three non-Clifford partitions}
\label{subsec:bb_nonclifford}
% ------------------------------------------------------------

A second route to non-Clifford gates is to search directly. Candidate Pauli generators are grouped by a common
nonzero syndrome, and their relative logical Paulis are tested for the
complete rank-three commuting family described in
Sec.~\ref{subsec:equal_angle_hierarchy}. Eight Pauli generators realizing this logical cube can implement a CCZ-type gate. All direct rank-three constructions in this subsection use eight mutually
commuting Pauli rotations. We separately record
whether their physical supports are mutually disjoint or can be
divided into a small number of Pauli-rotation layers.

\begin{table}[t]
\centering
\begin{tabular*}{\columnwidth}{@{\extracolsep{\fill}}cccc@{}}
\hline\hline
Code & Syndrome structure & $w_{max}$ & Schedule \\
\hline
$[[16,4,4]]$ BB
& four-partition lift & $2$ & $8+8$ \\
$[[64,8,8]]$ BB
& four-partition lift & $6$ & $8+8$ \\
$[[36,6,6]]$ Mirror
& three-partition lift & $6$ & $4+4+4$ \\
\hline
$[[16,4,4]]$ BB
& equal rank three & $3$ & $4+4$ \\
$[[64,8,8]]$ BB
& equal rank three & $6$ & $4+4$ \\
$[[64,8,8]]$ BB
& equal rank three & $8$ & $8$ disjoint \\
$[[96,12,8]]$ BB
& equal rank three & $8$ & $8$ disjoint \\
$[[128,16,8]]$ BB
& equal rank three & $8$ & $8$ disjoint \\
$[[36,6,6]]$ Mirror
& equal rank three & $5$ & $4+4$ \\
\hline\hline
\end{tabular*}
\caption{Representative higher-level syndrome-partition constructions
on qLDPC codes. The first three rows are unequal-syndrome lifts of
lower-level Clifford gates: the two BB lifts consist of CCZ-type gates, whereas the Mirror lift is a CS-type operation. The
remaining rows are equal-syndrome rank-three CCZ-type constructions. A schedule entry $p+q+\cdots$ gives the number of Pauli generators in
each Pauli-rotation layer, while ``$8$ disjoint'' denotes
Pauli-rotation depth one.}
\label{tab:qldpc_nonclifford_results}
\end{table}

On the $[[16,4,4]]$ BB code, we exhaust all nonidentity Pauli generators through weight three and test the resulting rank-three families at Pauli-rotation depths one and two. No family with Pauli-rotation depth one occurs in this domain. A verified weight-three construction occurs at Pauli-rotation depth two, with four rotations in each Pauli-rotation layer.

On the $[[64,8,8]]$ BB code, we exhaust the pure-$Z$ Pauli generator pool cumulatively through weight six. No rank-three family appears through $w_{\max}=5$; at $w_{\max}=6$ the search finds a construction at Pauli-rotation depth two. A Pauli-rotation depth one solution can be found with eight mutually disjoint weight-eight Pauli generators. We also search on the larger $[[96,12,8]]$ and $[[128,16,8]]$ BB codes, with constructions built from two four-generator translation orbits. Each orbit is generated by repeatedly translating a Pauli generator $G_j$ under an order-four lattice translation. We test all such order-four translations and increase the Pauli generator weights through $w_{\max}=8$. No rank-three logical cube is found through weight seven, while weight-eight Pauli generators produce fully support-disjoint CCZ-type constructions. The complete search domain is described in Appendix~\ref{app:large_bb_nonclifford_search}.

For the non-CSS $[[36,6,6]]$ Mirror code, we search mixed-Pauli generators using the code's four verified uniform translations, as detailed in Appendix~\ref{app:mirror_nonclifford_search}. This search contains no rank-three family through weight four. At $w_{\max}=5$ it finds two families, each with eight commuting Pauli generators, one common nonzero syndrome, and Pauli-rotation depth two yielding a CCZ-type gate. 

The resulting constructions are summarized in
Table~\ref{tab:qldpc_nonclifford_results}. Once a non-Clifford gate has
been found directly, the logical derivative provides an avenue to move back down the Clifford hierarchy and recover lower-level
logical gates.  Applying the derivative
$D_L(U)$ of Eq.~\eqref{eq:logical_derivative_physical} with respect to
a suitable logical Pauli $L$ selects the physical Pauli generators that
anticommute with $L$ and doubles their rotation angles. For a rank-three eight-generator cube at angle $\pi/8$, choosing $L$ to
anticommute with one logical cube direction and commute with the other
two selects four of the eight Pauli generators and doubles their angles to
$\pi/4$.  These four Pauli generators form a square face of the logical cube,
as illustrated in Fig.~\ref{fig:syndrome_partition_hierarchy}(b), and
implement a code-preserving logical Clifford gate.

\begin{table*}[t]
\centering
\small

\begin{tabular*}{\textwidth}{@{\extracolsep{\fill}}lccccc@{}}
\hline\hline
\multicolumn{6}{c}{
(a) six-generator logical $\pi/4$ roots
}
\\
\hline
Code
& Pauli generators
& Search domain
& Physical realizations
& Logical axes
& Pauli-rotation depth
\\
\hline

$[[16,4,4]]$ BB
& $Z_{\mathcal A}$
& $w=2$
& $72$
& $9$
& all $D=2$
\\

$[[16,4,4]]$ BB
& $X_{\mathcal A}$
& $w=2$
& $72$
& $9$
& all $D=2$
\\

$[[16,4,4]]$ BB
& $Z_{\mathcal A}$
& $w\le 3$
& $18{,}072$
& $9$
& all $D=2$
\\

$[[16,4,4]]$ BB
& $X_{\mathcal A}$
& $w\le 3$
& $18{,}072$
& $9$
& all $D=2$
\\

$[[64,8,8]]$ BB
& $Z_{\mathcal A}$
& $w=4$
& $1{,}360$
& $12$
& $144$ at $D=1$; $1{,}216$ at $D=2$
\\

$[[64,8,8]]$ BB
& $X_{\mathcal A}$
& $w=4$
& $1{,}360$
& $12$
& $144$ at $D=1$; $1{,}216$ at $D=2$
\\

\hline
\multicolumn{6}{c}{
(b) fourteen-generator logical $\pi/8$ constructions
}
\\
\hline
Code
& Pauli generators
& Logical family
& Physical realizations
& $(w_{\max},D)$
& Construction route
\\
\hline

$[[16,4,4]]$ BB
& $Z_{\mathcal A}$
& $9$ $Z$-type axes
& $2{,}304$
& $(3,3)$
& Clifford-root 
\\

$[[16,4,4]]$ BB
& $X_{\mathcal A}$
& $9$ $X$-type axes
& $2{,}304$
& $(3,3)$
& Clifford-root 
\\

$[[64,8,8]]$ BB
& $Z_{\mathcal A}$
& $4$ $Z$-type axes
& $32$
& $(4,5)$
& Clifford-root 
\\

$[[64,8,8]]$ BB
& $X_{\mathcal A}$
& $4$ $X$-type axes
& $32$
& $(4,5)$
& Clifford-root
\\

$[[36,6,6]]$ Mirror
& $X_{\mathcal A}$
& $3$ logical axes
& $270$ 
& $(4,7)$
& Direct construction
\\

\hline\hline
\end{tabular*}

\caption{
Logical $\pi/4$ roots and $\pi/8$ constructions from equal-syndrome
Pauli pairs. Panel (a) counts distinct physical six-generator roots and
the logical Pauli axes they realize in the stated finite search domains.
The exact-weight-two BB16 catalogs already contain roots on nine axes in
each pure-$X$ and pure-$Z$ search, but contain no compatible overlapping
root pairs under the completion rule. Extending the BB16 search
through weight three produces $18{,}072$ roots in each sector, with
$2{,}008$ physical realizations on each of the same nine axes. Panel (b)
counts distinct fourteen-generator physical realizations. The BB16 and
BB64 families are obtained by completing compatible overlapping
$\pi/4$ roots, while the Mirror gates are directly constructed.
}
\label{tab:qldpc_pauli_root_results}
\end{table*}

% ------------------------------------------------------------
\subsection{From logical \texorpdfstring{$\pi/4$}{pi/4} roots to
\texorpdfstring{$T$}{T}-type gates}
\label{subsec:qldpc_pauli_roots}
% ------------------------------------------------------------

The final search route uses the Pauli-root construction of
Sec.~\ref{subsec:syndrome_embedded_roots}. As in the lifting search
above, we first build a catalog of logical Clifford gates and then use
that catalog as a set of building blocks for gates one level higher in
the Clifford hierarchy. Here the lower-level gates are logical
$\pi/4$ rotations about Pauli axes. Each such Clifford root is formed
from three equal-syndrome pairs whose products represent the same
logical Pauli $L$, as in Eq.~\eqref{eq:Clifford_root_construction}. We then look for compatible Clifford roots that can
be combined to complete the seven-pair structure required for
$R_L(\pi/8)$.

For the self-dual BB codes, we perform separate searches using only $Z$ and $X$-type Pauli generators. These two Pauli families simplify the search as all Pauli generators commute, and equal-syndrome pairs remain within the same Pauli family. This gives two simple families of physical Pauli generators,
\begin{align}
Z_{\mathcal A}
&=
\prod_{j\in\mathcal A} Z_j,
\qquad
X_{\mathcal A}
=
\prod_{j\in\mathcal A} X_j,
\end{align}
where $\mathcal A$ denotes the physical support. For each Pauli type, we enumerate the allowed low-weight Pauli generators and group them by stabilizer syndrome. Two Pauli generators with the same syndrome have a zero-syndrome product, and we retain the pair when that product represents a nontrivial logical Pauli $L$. We then combine three such pairs with syndrome labels
$\boldsymbol a$, $\boldsymbol b$, and
$\boldsymbol a+\boldsymbol b$, requiring all three pair products to represent the same $L$. We also check the multiplicative relations among the chosen representatives
modulo stabilizers and retain the construction only after its complete
six-generator logical operator is verified, as additional zero-syndrome logical factors can exist between the three chosen syndrome carrying Pauli generators.
An accepted construction realizes
\begin{align}
R_L(\pi/4)
&\equiv
\exp\left(-i\frac{\pi}{4}L\right)
\end{align}
or its inverse, up to a global phase.

The searched root catalogs are summarized in
Table~\ref{tab:qldpc_pauli_root_results}(a). For the $[[16,4,4]]$ BB
code, in each pure-Pauli sector, the exact-weight-two search yields $72$
code-preserving roots about nine distinct logical Pauli axes, all
with Pauli-rotation depth two.
These catalogs include individually addressed phase rotations,
for example about $\overline Z_2$, $\overline Z_3$, and $\overline Z_4$
in the $Z$-type search and about $\overline X_1$, $\overline X_2$, and
$\overline X_4$ in the $X$-type search. However, no two same-axis roots in
either $72$-root catalog share a common equal-syndrome pair that is required for a non-Clifford gate. 

For the $[[64,8,8]]$ BB
code, we perform a search on weight-four pure-$Z$ and pure-$X$ Pauli generators. Each search gives $1{,}360$ code-preserving root gates about $12$ distinct logical axes; $144$ have Pauli-rotation depth one and $1{,}216$ have
Pauli-rotation depth two. For the Mirror code, we search only for roots of Pauli-rotation depth at most two. The search covers pure-$X$ and pure-$Z$ Pauli generators at exact weights three and four, together with mixed-Pauli generators through weight five whose syndromes are invariant under the code's four verified uniform translations. None of these three finite searches yields a solution.

We next ask whether the BB Clifford-root catalogs can be combined to
produce logical $\pi/8$ rotations. Suppose two verified $\pi/4$ roots act about the same logical
Pauli $L$, share one equal-syndrome pair of Pauli generators, and form two
compatible rank-two faces of a rank-three syndrome space. Modulo
stabilizers, their pair structure may be written schematically as
\begin{align}
\mathcal C_{ab}
&=
\left\{
(A,LA),
(B,LB),
(AB,LAB)
\right\},
\nonumber\\
\mathcal C_{ac}
&=
\left\{
(A,LA),
(C,LC),
(AC,LAC)
\right\}.
\label{eq:overlapping_phase_gate_completion}
\end{align}
The two roots already provide five of the seven nonzero syndrome classes,
\begin{align}
\boldsymbol a,
\quad
\boldsymbol b,
\quad
\boldsymbol c,
\quad
\boldsymbol a+\boldsymbol b,
\quad
\boldsymbol a+\boldsymbol c.
\end{align}
The two missing syndrome values are therefore
$\boldsymbol b+\boldsymbol c$ and
$\boldsymbol a+\boldsymbol b+\boldsymbol c$. At the level of Pauli
multiplication they are represented by $BC$ and $ABC$, so the complete
seven-pair structure requires pairs equivalent to $(BC,LBC)$ and
$(ABC,LABC)$. The five pairs inherited from the two Clifford-root gates already come with physical Pauli generators that satisfy the source weight, commutation, and Pauli-rotation depth constraints. The products $BC$ and $ABC$ formed from the existing physical
representatives, however, need not satisfy the corresponding constraints for the completed construction. We therefore use the freedom to multiply these products by stabilizers, which preserves their syndrome and logical Pauli class while changing their physical weight and overlap with the existing Pauli generators. We then retain only those representatives that satisfy the required weight, mutual-commutation, and Pauli-rotation depth bounds.

The resulting seven equal-syndrome pairs occupy the seven nonzero
syndromes of a rank-three space:
\begin{align}
\{&\boldsymbol a,
\boldsymbol b,
\boldsymbol c,
\boldsymbol a+\boldsymbol b,
\boldsymbol a+\boldsymbol c,
\boldsymbol b+\boldsymbol c,
\boldsymbol a+\boldsymbol b+\boldsymbol c\}.
\end{align}
Every completed fourteen-generator candidate is then subjected to exact codespace verification, including its logical phase. Accepted candidates realize the gate:
\begin{align}
R_L(\pi/8)
&\equiv
\exp\!\left(-i\frac{\pi}{8}L\right),
\end{align}
up to a global phase.

It turns out the weight-two BB16 catalogs are not sufficient for this
construction as no two gates about the same logical axis share the physical
equal-syndrome pair required to form the two overlapping faces above.
We therefore extend the pure-$Z$ and pure-$X$ searches through
Pauli generator weight three. Each enlarged search contains
$18{,}072$ code-preserving $\pi/4$ root gates of Pauli-rotation depth two, again about the same $9$ logical axes, with $2{,}008$ gates per axis.

In the context of this enlarged Clifford catalog we obtain
$859{,}464$ compatible overlapping gate pairs in each Pauli axis type.
Completing all such overlaps with Pauli generators through weight three and
requiring $w_{\max}\le3$ and final Pauli-rotation depth $D\le3$ yields
$2{,}304$ distinct physical $\pi/8$ constructions in each search.
All $9$ logical axes admit solutions, with exactly $256$ per logical axis. Every solution has Pauli-rotation depth three and consists of six weight-two and eight weight-three
Pauli generators.

One $Z$-type construction acts about
$L=\overline Z_2$ and therefore implements the addressed logical gate
$T_2$. Its seven equal-syndrome pairs are
\begin{align}
\mathcal P_1
&=
\left(
Z_{14}Z_{16},
Z_7Z_{11}
\right),
\nonumber\\
\mathcal P_2
&=
\left(
Z_2Z_5,
Z_4Z_9
\right),
\nonumber\\
\mathcal P_3
&=
\left(
Z_4Z_5,
Z_2Z_9
\right),
\nonumber\\
\mathcal P_4
&=
\left(
Z_1 Z_6 Z_7,
Z_7 Z_8Z_{15}
\right),
\nonumber\\
\mathcal P_5
&=
\left(
Z_8Z_{11}Z_{15},
Z_1Z_6Z_{11}
\right),
\nonumber\\
\mathcal P_6
&=
\left(
Z_3Z_{10}Z_{16},
Z_{12}Z_{13}Z_{16}
\right),
\nonumber\\
\mathcal P_7
&=
\left(
Z_3Z_{10}Z_{14},
Z_{12}Z_{13}Z_{14}
\right).
\label{eq:bb16_t2_physical_pairs}
\end{align}
The two Pauli generators in every pair have the same nonzero syndrome, every
pair product represents $\overline Z_2$, and the seven pair syndromes
are the seven nonzero vectors of a rank-three syndrome subspace.
Applying all fourteen rotations with angle $+\pi/8$ gives
\begin{align}
\prod_{j=1}^{7}
\prod_{G\in\mathcal P_j}
\exp\!\left(i\frac{\pi}{8}G\right)
&\doteq
T_2.
\label{eq:bb16_t2_physical_gate}
\end{align}
This gate is one of the $256$ solutions of the
$\overline Z_2$ axis in the full BB16 catalog.

For the $[[64,8,8]]$ BB code, the weight-four catalogs are
already sufficiently rich to yield non-Clifford gates. In each of the
pure-$Z$ and pure-$X$ catalogs, the $1{,}360$ code-preserving LGAs contain
$8{,}800$ compatible overlapping pairs. We exhaust the allowed
ways of combining these roots and enumerate the weight-four
representatives of the two missing equal-syndrome pairs. Requiring final
Pauli-rotation depth at most five yields exactly $32$ distinct physical constructions in each pure-$X$
and pure-$Z$ search. These realize four logical Pauli axes, with eight
physical realizations per axis. They realize four distinct logical axes, with eight code-preserving LGAs per axis, and every retained construction has
Pauli-rotation depth five. For example, one solution of $Z$-type acts about
\begin{align}
L_{\mathrm{BB64}}
=
\overline Z_3\overline Z_4\overline Z_7\overline Z_8
\end{align}
and realizes $R_{L_{\mathrm{BB64}}}(\pi/8)$. Every Pauli generator has weight
four, so the equal-syndrome pairs saturate the lower
bound $w_{\max}=d/2=4$.

For the non-CSS Mirror code, the bounded Clifford-root searches considered above do not produce the $\pi/4$ roots needed to construct a non-Clifford gate. We therefore search directly for the full seven-pair structure required for a logical $\pi/8$ root gate. We search over pure-$X$, weight-four Pauli strings finding $270$ distinct fourteen-generator constructions consisting of three logical Pauli axes,
\begin{align}
\overline Y_1\overline Z_3,\qquad
\overline Z_3\overline Z_6,\qquad
\overline Y_1\overline Z_6,
\end{align}
with $90$ physical realizations per axis. In every construction, the fourteen weight-four Pauli generators are arranged into seven equal-syndrome pairs, and the product of the two generators in each pair represents the corresponding logical Pauli axis. Applying all fourteen rotations with angle $\pi/8$ realizes the associated logical root $R_L(\pi/8)$. Every construction has Pauli-rotation depth seven.

% ============================================================
\section{Analytic syndrome-partition constructions}
\label{sec:analytic_constructions}
% ============================================================
The direct stabilizer searches of
Sec.~\ref{subsec:qldpc_clifford_search} choose physical Pauli
generators from syndrome, commutation, and support constraints without
prescribing a target logical gate. A logical Pauli frame is used
afterward to determine the action of each accepted LGA. For analytic
gate design, we instead choose the logical frame and desired gate
structure upfront and use a \emph{Clifford algebra basis} (CAB) presentation to relate them directly
to physical Pauli operators. This makes logical addressability an input to the construction.

\begin{figure*}[t]
\centering
\includegraphics[width=0.95\textwidth]{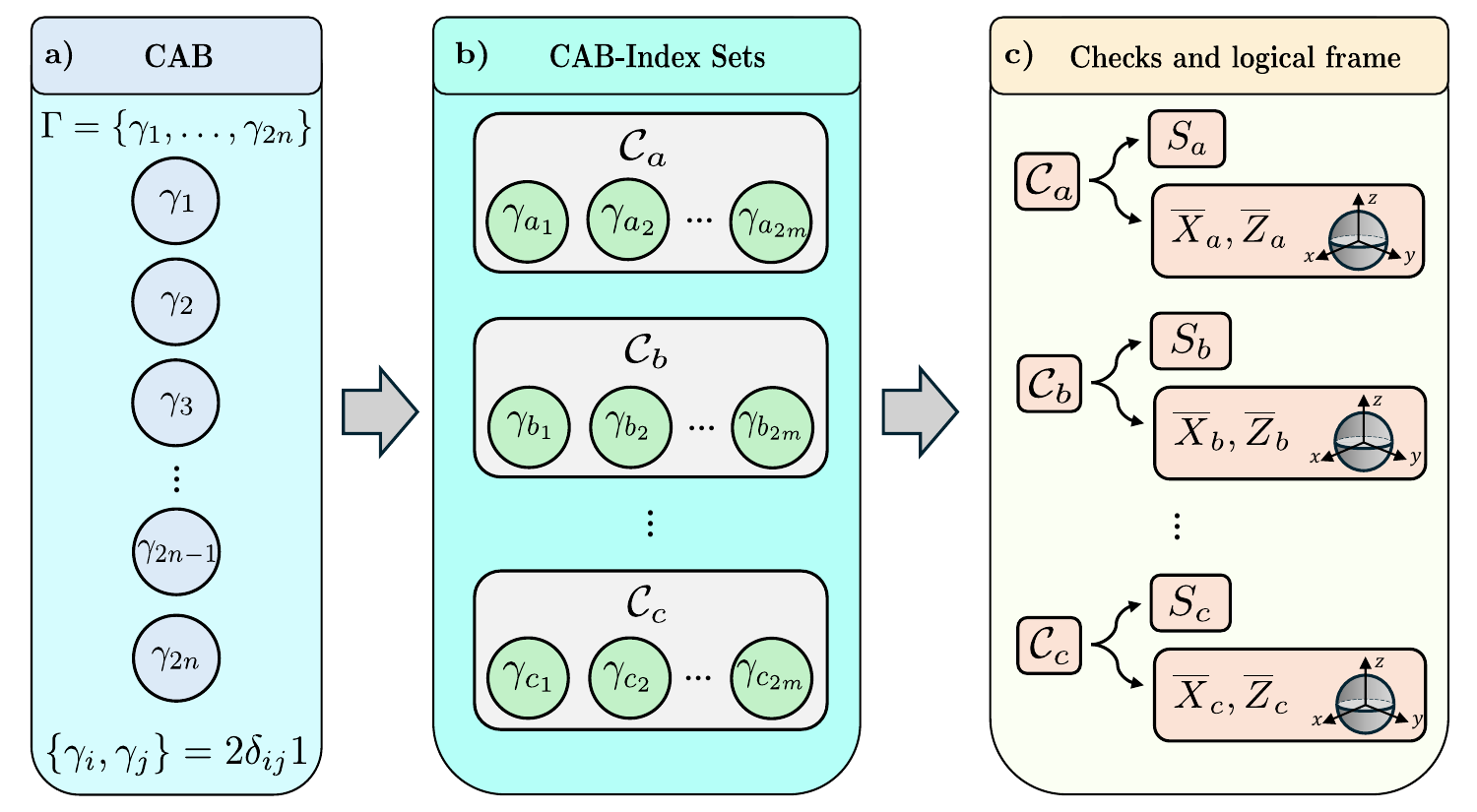}
\caption{Visual representation of a CAB presentation.
(a) A Clifford algebra basis
$\Gamma=\{\gamma_1,\ldots,\gamma_{2n}\}$ provides binary coordinates
for the Pauli group.
(b) CAB indices are organized into pairwise-disjoint sets
$\mathcal C_a$.
(c) The monomial on each set defines a stabilizer check, while
quadratic monomials within the same set define a logical Pauli
frame. Candidate syndrome-partition Pauli generators are expressed in the same
index language, so their stabilizer syndromes and products can be read
directly from CAB-index overlaps and symmetric differences.}
\label{fig:cab_presentation_workflow}
\end{figure*}

A CAB is a convenient generating set for the
Pauli group whose products can be labeled by sets of CAB indices.
Multiplication, commutation, and stabilizer syndrome can then be
computed from simple operations on these index sets. This allows
stabilizers, logical Pauli operators, and syndrome-carrying physical
operators to be characterized within the same coordinate system. We
first relate CAB coordinates to the usual binary symplectic Pauli representation
and then use them to construct Clifford and higher-level logical gates.
Figure~\ref{fig:cab_presentation_workflow} summarizes this organization.

% ------------------------------------------------------------
\subsection{Pauli coordinates and CAB coordinates}
\label{subsec:cab_definition}
% ------------------------------------------------------------

A Pauli string on $n$ qubits can be represented, modulo phase, by a
binary vector in $\mathbb F_2^{2n}$. The usual representation uses the
single-qubit $X_i$ and $Z_i$ operators as binary basis elements. This
description is especially useful when physical support is important,
since the entries of the binary vector directly identify the qubits on
which the Pauli acts.

For the constructions below, we use a different binary coordinate
system. A CAB on $n$ qubits is a set
\begin{align}
\Gamma
&=
\{\gamma_1,\ldots,\gamma_{2n}\}
\end{align}
of Pauli strings satisfying the anticommutation relations
\begin{align}
\{\gamma_i,\gamma_j\}
&=
2\delta_{ij}I .
\label{eq:cab_relations}
\end{align}
Thus each $\gamma_i$ squares to the identity, while every pair of
distinct CAB generators anticommutes. A standard example is the
Jordan--Wigner representation:
\begin{align}
\gamma_{2j-1}
&=
Z_1\cdots Z_{n-j}X_{n-j+1},
\nonumber\\
\gamma_{2j}
&=
Z_1\cdots Z_{n-j}Y_{n-j+1},
\qquad
j=1,\ldots,n .
\label{eq:jordan_wigner_cab}
\end{align}

We refer to a product of CAB generators as a \textit{CAB monomial}.
If $\mathcal A\subseteq\{1,\ldots,2n\}$ is a set of CAB indices, we
write
\begin{align}
\gamma_{\mathcal A}
&:=
\prod_{i\in\mathcal A}^{\rightarrow}\gamma_i ,
\label{eq:cab_monomial}
\end{align}
where the product is taken in increasing index order. The number of
indices $|\mathcal A|$ is the grade of the monomial. As
$\mathcal A$ ranges over all subsets of the $2n$ CAB indices, these
monomials range over the full $n$-qubit Pauli group modulo phase.
Thus a CAB provides a
different binary basis for the Pauli group. The proof and the
precise relation to standard symplectic Pauli coordinates are given in
Appendix~\ref{app:cab_algebra}. Whenever a CAB monomial is used as a
physical Pauli operator, its overall phase is chosen so that it is
Hermitian. For a grade-$g$ monomial we use
$i^{g(g-1)/2}\gamma_{\mathcal A}$. Any remaining $\pm1$ sign is written explicitly as
part of the stabilizer sign or the rotation direction; see
Appendix~\ref{app:cab_algebra}.

The usefulness of the CAB description comes from the way multiplication
and commutation translate into simple operations on index sets. Modulo
phase,
\begin{align}
\gamma_{\mathcal A}\gamma_{\mathcal B}
&\doteq
\gamma_{\mathcal A\triangle\mathcal B},
\label{eq:cab_product_symmetric_difference}
\end{align}
where $\mathcal A\triangle\mathcal B$ is the
\textit{symmetric difference}: the set of indices that occur in
exactly one of $\mathcal A$ or $\mathcal B$. For example, if
$\mathcal A=\{1,2,4\}$ and $\mathcal B=\{2,3\}$, then
$\mathcal A\triangle\mathcal B=\{1,3,4\}$. The common index $2$
disappears because $\gamma_2^2=I$. Symmetric difference is therefore
the set-theoretic version of addition modulo two.

Commutation is determined by a similarly simple parity rule. Moving
all generators in $\gamma_{\mathcal A}$ through those in
$\gamma_{\mathcal B}$ gives
\begin{align}
\gamma_{\mathcal A}\gamma_{\mathcal B}
&=
(-1)^{\chi_\Gamma(\mathcal A,\mathcal B)}
\gamma_{\mathcal B}\gamma_{\mathcal A},
\nonumber\\
\chi_\Gamma(\mathcal A,\mathcal B)
&=
|\mathcal A||\mathcal B|
-
|\mathcal A\cap\mathcal B|
\pmod 2 .
\label{eq:cab_commutation_rule}
\end{align}
Thus two CAB monomials commute when $\chi_\Gamma=0$ and anticommute
when $\chi_\Gamma=1$.

This immediately turns a stabilizer syndrome into an index-overlap
condition. If a stabilizer check is represented by
$S_a\doteq\gamma_{\mathcal C_a}$, then the $a$th component of the
syndrome of $\gamma_{\mathcal A}$ is simply
$\chi_\Gamma(\mathcal A,\mathcal C_a)$. The equal-syndrome condition of
Sec.~\ref{sec:gate_design_workflows} can therefore be tested directly
from CAB-index sets.

% ------------------------------------------------------------
\subsection{CAB presentations}
\label{subsec:disjoint_blocks}
% ------------------------------------------------------------

The object used in CAB-based gate constructions is a particular organization of stabilizer checks and
logical Pauli operators. We refer to this as a \textit{CAB
presentation}.

A particularly useful structured presentation chooses a collection of
pairwise-disjoint CAB-index sets
\begin{align}
\mathcal C_1,\ldots,\mathcal C_b
\subseteq
\{1,\ldots,2n\},
\end{align}
with each set having even cardinality,
$|\mathcal C_a|\in2\mathbb Z$. Each set defines a commuting
stabilizer check
$S_a=\eta_a\gamma_{\mathcal C_a}$, where the phase $\eta_a$ is chosen so that $S_a$ is a Hermitian Pauli
operator with the desired stabilizer sign. Because the sets are
disjoint and have even size, the CAB commutation rule guarantees that the resulting checks commute. 

The same CAB-index set also naturally organizes logical Pauli operators that commute
with those checks. Generally, a CAB monomial $\gamma_{\mathcal A}$ commutes with every associated stabilizer check whenever
$|\mathcal A\cap\mathcal C_a|$ is even for every CAB-index set
$\mathcal C_a$. Such monomials therefore lie in the Pauli normalizer.
For the constructions below we use the simplest nontrivial examples where quadratic monomials are supported within a single CAB-index set. For example, consider two CAB indices $i,j\in\mathcal C_a$.
The quadratic monomial $\gamma_i\gamma_j$ overlaps the set
$\mathcal C_a$ in two indices and therefore commutes with $S_a$. Since
the CAB-index sets are disjoint, it also commutes with every other
stabilizer check. Quadratic monomials formed from indices within one
CAB-index set therefore lie in the normalizer of the stabilizer group and can
be used to represent logical Pauli operators.

We now specialize to the case in which the CAB-index sets contain four elements. Let
\begin{align}
\mathcal C_a
&=
\{c_{a,1},c_{a,2},c_{a,3},c_{a,4}\}.
\label{eq:quartic_block_general}
\end{align}
This quartic choice is useful because it is the smallest even CAB-index set that can support a nontrivial logical degree of freedom. As we show below, a quartic set naturally includes exactly one logical qubit, giving a one-to-one correspondence between a CAB-index set, one stabilizer check, and one logical Pauli frame. It also provides the addressability of the logical gate constructions considered below. A quartic CAB presentation of this form is additional structure and is not assumed to exist for every stabilizer code. In particular, we do not assume that an arbitrary stabilizer code admits a single CAB in which all independent stabilizers are represented by pairwise-disjoint quartic sets. If these sets represent all $b=n-k$ independent checks, their
disjointness requires $4b\leq2n$, and hence $k\geq n/2$.
They are associated with $b$ logical qubits, and when they exhaust the $2n$ CAB
indices, $k=b=n/2$.

The corresponding quartic monomial defines one stabilizer check,
while the six possible quadratic monomials that can be formed from the set provide candidates for logical Pauli operators. For example, we may choose
\begin{align}
\overline X_a
&\doteq
\gamma_{c_{a,1}}\gamma_{c_{a,2}},
&
\overline Z_a
&\doteq
\gamma_{c_{a,2}}\gamma_{c_{a,3}},
\label{eq:quartic_block_logicals}
\end{align}
which commute with the stabilizer check and anticommute with one
another. The remaining quadratic pairs give stabilizer-equivalent
representatives of the same logical axes. Multiplication by the
quartic stabilizer replaces a pair of indices by its complementary
pair within the set. Thus, schematically,
\begin{align}
\gamma_{c_{a,1}}\gamma_{c_{a,2}}
&\sim
\gamma_{c_{a,3}}\gamma_{c_{a,4}},
\nonumber\\
\gamma_{c_{a,2}}\gamma_{c_{a,3}}
&\sim
\gamma_{c_{a,1}}\gamma_{c_{a,4}},
\nonumber\\
\gamma_{c_{a,1}}\gamma_{c_{a,3}}
&\sim
\gamma_{c_{a,2}}\gamma_{c_{a,4}},
\label{eq:quartic_pair_classes}
\end{align}
where $\sim$ denotes equality on the codespace up to phase and
stabilizer multiplication. The six quadratic pairs therefore collapse
into three logical Pauli operators, naturally identified with
$\overline X_a$, $\overline Z_a$, and $\overline Y_a$.

This is the key structural feature of the quartic CAB presentation. The same quartic CAB-index set $\mathcal C_a$ that defines the stabilizer check $S_a$ also determines the three quadratic equivalence classes identified with
$\overline X_a$, $\overline Y_a$, and $\overline Z_a$. Thus, the stabilizer check and the logical Pauli frame are tied to the same CAB-index set via the common label $a$. Several disjoint quartic CAB-index sets therefore provide, in one coordinate system, both a commuting set of stabilizer checks $S_1,\ldots,S_b$ and the associated logical Pauli operators
$(\overline X_1,\overline Z_1),\ldots,
(\overline X_b,\overline Z_b)$. 

We can also form stabilizer checks from larger disjoint CAB-index sets. Recall that a CAB for $q$ qubits consists of $2q$ CAB elements, so $2q$ such elements generate the Pauli algebra of $q$ qubits. If the product of all $2q$ elements is then imposed as one independent stabilizer check, the usual stabilizer counting gives $q-1$ remaining logical qubits. Thus a quartic CAB-index set contains one logical qubit, while sets of size six and eight contain two and three, respectively. Larger CAB-index sets can therefore associate several logical qubits with a single stabilizer check.

A simple example is the Iceberg family $[[2m,2m-2,2]]$ whose stabilizer group is generated by the two commuting checks $X^{\otimes 2m}$ and $Z^{\otimes 2m}$ \cite{SelfIceberg2024}. In an appropriate CAB presentation, these two checks can be represented by two disjoint CAB-index sets of size $2m$, so each set includes $m-1$ logical qubits, and together they give the expected $2m-2$ logical qubits. We focus instead on quartic sets because associating each CAB-index set with a single logical qubit makes logical addressability and the gate constructions below especially transparent.

A CAB presentation can also be iterated because the logical Pauli operators
of one encoded code themselves form a Pauli algebra. Begin with an
intermediate code containing $m$ logical qubits and choose a logical
Pauli frame
$(\overline X_1,\overline Z_1),\ldots,
(\overline X_m,\overline Z_m)$. We now treat these $m$ logical qubits
as the effective qubits for the next construction and choose a CAB
$\widetilde\Gamma$ for their encoded Pauli group. Selected indices of $\widetilde\Gamma$ are organized into disjoint
quartic CAB-index sets. The product of the four encoded CAB generators in one such set is imposed as a new stabilizer check on the intermediate code.
Quadratic products within that same set commute with the new check
and provide Pauli operators for the logical degrees of freedom that
remain after imposing the new check. Those surviving logical Pauli operators
then form the logical frame used to construct the encoded CAB at the
next encoding level. Repeating this step produces the recursive CAB structure used for the hypercube codes below; the general encoded-CAB construction is developed in Appendix~\ref{app:recursive_cab_scaffolds}. Therefore, quartic CAB-index sets can serve as building blocks for larger codes.

This shared description is what makes the CAB representation useful for gate construction. Relations among CAB-index sets simultaneously determine stabilizer syndromes and the logical Pauli content of the corresponding physical operators. We can therefore design syndrome partitions directly in the CAB coordinates while keeping their intended logical action explicit. In what follows, we exploit this connection to obtain analytic logical gate constructions.

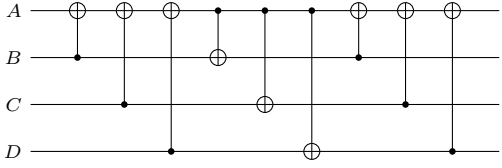
\begin{figure}[t]
\centering
\begingroup
\newcommand{\drawCNOT}[3]{%
    \def\targetR{0.17}
    \fill (#1,#2) circle[radius=0.07];
    \draw (#1,#2)--(#1,#3);
    \draw[fill=white] (#1,#3) circle[radius=\targetR];
    \begin{scope}[shift={(#1,#3)}]
        \draw (-\targetR,0)--(\targetR,0);
        \draw (0,-\targetR)--(0,\targetR);
    \end{scope}
}
\begin{tikzpicture}[x=0.62cm,y=0.62cm,every node/.style={font=\scriptsize}]
    \draw (0,0)--(10,0);
    \draw (0,-1)--(10,-1);
    \draw (0,-2)--(10,-2);
    \draw (0,-3)--(10,-3);

    \node[left] at (0,0) {$A$};
    \node[left] at (0,-1) {$B$};
    \node[left] at (0,-2) {$C$};
    \node[left] at (0,-3) {$D$};

    \drawCNOT{1}{-1}{0}
    \drawCNOT{2}{-2}{0}
    \drawCNOT{3}{-3}{0}

    \drawCNOT{4}{0}{-1}
    \drawCNOT{5}{0}{-2}
    \drawCNOT{6}{0}{-3}

    \drawCNOT{7}{-1}{0}
    \drawCNOT{8}{-2}{0}
    \drawCNOT{9}{-3}{0}
\end{tikzpicture}
\endgroup
\caption{Logical action of the four-qubit complement gate written as a sequence of CNOT operations.}
\label{fig:complement_cnot_circuit}
\end{figure}

% ------------------------------------------------------------
% ------------------------------------------------------------
\subsection{CAB presentations and syndrome-partition gates}
\label{subsec:cab_index_constructions}
% ------------------------------------------------------------

The CAB-index structure of a CAB presentation lets us build the syndrome carriers
and logical Pauli strings that appear in
Sec.~\ref{sec:gate_design_workflows} directly from CAB indices. Consider two quartic CAB-index sets
\begin{align}
\mathcal C_A
&=
\{a_1,a_2,a_3,a_4\},
&
\mathcal C_B
&=
\{b_1,b_2,b_3,b_4\}.
\end{align}
A quadratic monomial with one index in each set,
\begin{align}
E_{AB}^{jk}
&\doteq
\gamma_{a_j}\gamma_{b_k},
\end{align}
has odd CAB-index overlap with $\mathcal C_A$ and $\mathcal C_B$ and
even overlap with every other stabilizer CAB-index set. It therefore
anticommutes with $S_A$ and $S_B$ and commutes with the other checks. Its syndrome is therefore
\begin{align}
\boldsymbol\sigma(E_{AB}^{jk})
&=
\boldsymbol e_A+\boldsymbol e_B,
\label{eq:cab_bridge_syndrome}
\end{align}
where $\boldsymbol e_A\in\mathbb F_2^b$ is the unit syndrome vector
whose $A$th component is one and whose remaining components are zero. Every quadratic monomial joining the same two CAB-index sets has
this syndrome.

Multiplying two such errors removes the syndrome:
\begin{align}
E_{AB}^{jk}E_{AB}^{j'k'}
&\doteq
\left(
\gamma_{a_j}\gamma_{a_{j'}}
\right)
\left(
\gamma_{b_k}\gamma_{b_{k'}}
\right).
\label{eq:cab_bridge_pair_product}
\end{align}
The factors on the right are within-set quadratics and therefore
represent logical Pauli operators. Hence the two CAB-index sets determine the syndrome, while the particular indices chosen
at those endpoints determine the logical Pauli obtained when two
equal-syndrome errors are multiplied.

This gives a simple CAB design rule. Products that connect different sets determine the stabilizer syndrome, whereas quadratic products within a set determine the relative logical Pauli action. Syndrome-partition gates can then be obtained by arranging these two ingredients together.

% ............................................................
\subsubsection{Perfect matching}
\label{subsubsec:perfect_matching_templates}
% ............................................................

The simplest construction pairs corresponding indices of two quartic
CAB-index sets,
\begin{align}
\mathcal P_j
&=
\{a_j,b_j\},
\qquad
j=1,\ldots,4.
\label{eq:matching_position_supports}
\end{align}
The corresponding Pauli generators
$G_j\doteq\gamma_{\mathcal P_j}$ constitute an equal syndrome partition with syndrome $\boldsymbol e_A+\boldsymbol e_B$. Their CAB supports are pairwise
disjoint, so they commute, and every index in
$\mathcal C_A\cup\mathcal C_B$ occurs exactly once. Thus
\begin{align}
\mathcal P_1\triangle\mathcal P_2
\triangle\mathcal P_3\triangle\mathcal P_4
&=
\mathcal C_A\triangle\mathcal C_B,
\end{align}
and the four Pauli generators multiply, up to phase, to $S_AS_B$.

Relative to the first Pauli generator, the remaining three differ by the
logical products
$\overline X_A\overline X_B$,
$\overline Y_A\overline Y_B$, and
$\overline Z_A\overline Z_B$. The four Pauli generators therefore form the
logical square required by the equal-syndrome Clifford construction,
and the resulting logical action is
\begin{align}
\overline X_A
&\leftrightarrow
\overline X_B,
&
\overline Z_A
&\leftrightarrow
\overline Z_B.
\end{align}
Thus the perfect-matching template implements
$\overline{\mathrm{SWAP}}_{A,B}$.

The same matching extends to any even number of selected CAB-index sets by placing the same position from every selected set in one Pauli generator. Choosing the same CAB position in every set gives a convenient
canonical matching, but is not required by the equal-syndrome
condition. Other pairwise-disjoint matchings can permute the quadratic
CAB pairs that appear in products $G_jG_k$, thereby changing which
logical Pauli axes are related between the selected sets. As
discussed below, this freedom corresponds to local Clifford dressing of
the resulting logical gate. Every Pauli generator then has the same syndrome on the selected
checks, while the four Pauli generators together contain every CAB index of
those sets exactly once. For four CAB-index sets $A,B,C,D$, the resulting
logical Clifford acts, for example, as
\begin{align}
\overline X_A
&\mapsto
\overline X_B\overline X_C\overline X_D,
&
\overline Z_A
&\mapsto
\overline Z_B\overline Z_C\overline Z_D,
\end{align}
with the analogous transformation on the other sets. We call this
multi-set extension the \textit{complement} gate $\overline{\mathrm{COMP}}_{1,2,3,4}$. Its CNOT
decomposition is shown in
Fig.~\ref{fig:complement_cnot_circuit}, and the general CAB-index
construction is given in Appendix~\ref{app:cab_templates}.

\begin{figure}[t]
\centering
\begingroup
\newcommand{\drawCNOT}[3]{%
    \def\targetR{0.17}
    \fill (#1,#2) circle[radius=0.07];
    \draw (#1,#2)--(#1,#3);
    \draw[fill=white] (#1,#3) circle[radius=\targetR];
    \begin{scope}[shift={(#1,#3)}]
        \draw (-\targetR,0)--(\targetR,0);
        \draw (0,-\targetR)--(0,\targetR);
    \end{scope}
}
\begin{tikzpicture}[x=0.70cm,y=0.62cm,every node/.style={font=\scriptsize}]
    \draw (0,0)--(4,0);
    \draw (0,-1)--(4,-1);
    \draw (0,-2)--(4,-2);
    \draw (0,-3)--(4,-3);

    \node[left] at (0,0) {$A$};
    \node[left] at (0,-1) {$B$};
    \node[left] at (0,-2) {$C$};
    \node[left] at (0,-3) {$D$};

    \drawCNOT{1}{-1}{0}
    \drawCNOT{2}{-2}{0}
    \drawCNOT{3}{-3}{0}
\end{tikzpicture}
\endgroup
\caption{The multi-source CNOT with logical qubits $B$, $C$, and $D$ as controls and logical qubit $A$ as their common target.}
\label{fig:fanout_cnot_circuit}
\end{figure}
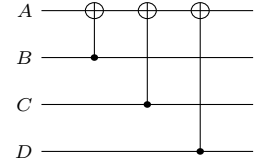

% ............................................................
\subsubsection{Nested index sets}
\label{subsubsec:nested_index_templates}
% ............................................................

A second CAB pattern produces a directed logical CNOT. Again consider
quartic CAB-index sets $\mathcal C_A$ and $\mathcal C_B$, and suppose we want
the check action
\begin{align}
S_A
&\mapsto
S_A,
&
S_B
&\mapsto
S_AS_B.
\label{eq:nested_target_check_action}
\end{align}
Every Pauli generator must therefore commute with $S_A$ and
anticommute with $S_B$, while the product of all four generators must
equal $S_A$ up to the phase fixed by the rotation signs. Choose
\begin{align}
\mathcal A
&=
\{a_2,a_3\},
&
\mathcal B_q
&=
\{b_3\}
\subset
\mathcal B_r
=
\{b_1,b_2,b_3\}.
\end{align}
The four Pauli-generator CAB-index sets are
\begin{align}
\mathcal P_1
&=
\mathcal B_q,
&
\mathcal P_2
&=
\mathcal A\cup\mathcal B_q,
\nonumber\\
\mathcal P_3
&=
\mathcal A\cup\mathcal B_r,
&
\mathcal P_4
&=
\mathcal C_A\cup\mathcal B_r.
\label{eq:nested_quartic_cnot_supports}
\end{align}
Each set contains an even number of indices from $\mathcal C_A$
and an odd number from $\mathcal C_B$, so all four Pauli generators form an equal syndrome partition. The nesting makes the corresponding CAB monomials
commute, and
\begin{align}
\mathcal P_1\triangle\mathcal P_2
\triangle\mathcal P_3\triangle\mathcal P_4
&=
\mathcal C_A,
\end{align}
so their product is proportional to $S_A$. In the logical frame of
Eq.~\eqref{eq:quartic_block_logicals}, the control pair
$\{a_2,a_3\}$ represents $\overline Z_A$, while
$\mathcal B_q\triangle\mathcal B_r=\{b_1,b_2\}$ represents
$\overline X_B$. The logical action is therefore
\begin{align}
\overline X_A
&\mapsto
\overline X_A\overline X_B,
&
\overline Z_B
&\mapsto
\overline Z_A\overline Z_B,
\nonumber\\
\overline Z_A
&\mapsto
\overline Z_A,
&
\overline X_B
&\mapsto
\overline X_B,
\end{align}
up to stabilizers and Pauli phases. Hence the nested-index template
implements
$\overline{\mathrm{CNOT}}_{A\rightarrow B}$.

The same pattern extends to several control logical qubits and one common target logical qubit. Let $\mathcal T$ denote the set of control logical qubits and $t$ the
common target. Each control $\overline X$ acquires the target
$\overline X$, while the target $\overline Z$ acquires the product of
the control $\overline Z$ operators. The resulting multi-source CNOT is
\begin{align}
\overline{\mathrm{CNOT}}_{\mathcal T\to t}
&=
\prod_{s\in\mathcal T}
\overline{\mathrm{CNOT}}_{s\to t},
\label{eq:nested_fanout_as_cnot_product}
\end{align}
as illustrated in Fig.~\ref{fig:fanout_cnot_circuit}, with the general
CAB-index formula given in Appendix~\ref{app:cab_templates}.

% ............................................................
\subsubsection{Index relabeling and logical Clifford generation}
\label{subsubsec}
% ............................................................

A quartic CAB-index set determines three logical Pauli operators, but it does not uniquely determine which of them is called $X$, $Y$, or $Z$. With the convention of Eq.~\eqref{eq:quartic_block_logicals},
\begin{align}
\overline X
&:\quad
\{1,2\}\sim\{3,4\},
\nonumber\\
\overline Z
&:\quad
\{2,3\}\sim\{1,4\},
\nonumber\\
\overline Y
&:\quad
\{1,3\}\sim\{2,4\}.
\end{align}
Permuting the four CAB indices leaves the quartic stabilizer check unchanged up to a phase, but permutes these three quadratic operators. Thus the same quartic CAB-index set can be presented with different assignments of its logical $X$, $Y$, and $Z$ axes.

This index permutation is a change of CAB presentation, not a physical gate. Its usefulness is that the analytic gate templates depend only on the index relations. After choosing a different within-set labeling, the same matching or nested-index template acts on different logical Pauli axes. The matching construction therefore gives SWAP-type gates in different local Clifford frames, while the nested construction yields the corresponding CNOT-, CZ-, or CY-type actions depending on the chosen control and target axes. Direct $\pi/4$ rotations about logical $X$ and $Z$ axes generate the
single-qubit Clifford group, while the nested-index construction supplies an
entangling CNOT-type gate. Together these operations generate the full logical
Clifford group. The explicit index permutations are given in Appendix~\ref{app:cab_relabelings}. In Appendix~\ref{app:cab_block_code_family}, we give a CAB-inspired distance-two error-detecting code whose perfect matching logical SWAP gates are individually addressable and are built from low-weight, support-disjoint physical Pauli generators.

% ............................................................
\subsubsection{Higher-level gates from CAB-index set products}
\label{subsubsec:cab_higher_level}
% ............................................................

The same quartic CAB-index structure also builds the non-Clifford
syndrome patterns of Sec.~\ref{sec:gate_design_workflows}. For instance, in  the common-syndrome construction, we can fix one detectable CAB monomial
and vary only its logical content. With four quartic CAB-index sets
$A,B,C,D$, choose $E\doteq
\gamma_{a_1}\gamma_{b_1}$,
and choose three commuting logical operators
$\overline Z_B,\overline Z_C,\overline Z_D$ represented by
within-set quadratics and commuting with $E$. Then
\begin{align}
G_{\boldsymbol u}
&\doteq
E\,
\overline Z_B^{u_1}
\overline Z_C^{u_2}
\overline Z_D^{u_3},
\qquad
\boldsymbol u\in\mathbb F_2^3,
\label{eq:cab_common_syndrome_cube}
\end{align}
all have the same syndrome as $E$, while their relative logical labels
are the eight products generated by
$\overline Z_B,\overline Z_C,\overline Z_D$. The CAB presentation has
therefore produced the rank-three common-syndrome logical cube of
Sec.~\ref{subsec:equal_angle_hierarchy}. With the corresponding signed
$\pi/8$ rotations, these Pauli generators implement a CCZ-type logical
gate. In this construction, one detectable error is fixed while the
within-set logical operators are varied.

The Pauli root construction that produces logical $\pi/4$ and $\pi/8$ rotations uses the opposite
arrangement in which the logical Pauli is fixed while the detectable errors are
varied to produce equal syndrome pairs. Choose three mutually commuting quadratic monomials
\begin{align}
E_1
&\doteq
\gamma_{a_1}\gamma_{b_1},
&
E_2
&\doteq
\gamma_{a_2}\gamma_{c_1},
&
E_3
&\doteq
\gamma_{a_3}\gamma_{d_1}.
\label{eq:cab_three_carriers}
\end{align}
Their CAB supports are disjoint, so they commute. Their syndromes are
\begin{align}
\boldsymbol s_1
&=
\boldsymbol e_A+\boldsymbol e_B,
&
\boldsymbol s_2
&=
\boldsymbol e_A+\boldsymbol e_C,
&
\boldsymbol s_3
&=
\boldsymbol e_A+\boldsymbol e_D,
\end{align}
which are independent. Let $L$ be a zero-syndrome logical Pauli that commutes with
$E_1,E_2,E_3$. The seven nonidentity products
\begin{align}
\{ E_1,
E_2,
E_3,
E_1E_2,
E_1E_3,
E_2E_3,
E_1E_2E_3 \}
\label{eq:cab_fano_carriers}
\end{align}
have the seven nonzero syndromes generated by
$\boldsymbol s_1,\boldsymbol s_2,\boldsymbol s_3$. Pairing every
product with the same logical Pauli, $\left(
E_{\boldsymbol u},
E_{\boldsymbol u}L
\right)$, gives two Pauli generators with equal syndrome whose product gives $L$.

Any two independent error directions give the three equal-syndrome pairs required for a logical $\pi/4$ rotation about $L$. Using all three directions gives the seven pairs required for the corresponding $\pi/8$ rotation. CAB multiplication therefore realizes the Fano-plane structure of Fig.~\ref{fig:Fano_plane} with the seven points representing the products in Eq.~\eqref{eq:cab_fano_carriers}, and each line, such as $\{E_1,E_2,E_1E_2\}$, representing an embedded Clifford gate.

Additionally, overlap patterns between different CAB-index sets can also be used to construct the syndrome relations required for unequal-syndrome lifting. For example, choose
three mutually commuting quadratic monomials
\begin{align}
F_{AB}
&\doteq
\gamma_{a_1}\gamma_{b_1},
&
F_{AC}
&\doteq
\gamma_{a_2}\gamma_{c_1},
&
F_{BC}
&\doteq
\gamma_{b_2}\gamma_{c_2}.
\end{align}
Their CAB supports can be chosen disjoint, and their syndromes satisfy
\begin{align}
\boldsymbol\sigma(F_{AB})
+
\boldsymbol\sigma(F_{AC})
+
\boldsymbol\sigma(F_{BC})
&=
\boldsymbol 0,
\label{eq:cab_triangle_syndrome_relation}
\end{align}
because every set label appears twice. If the same order-two logical gate is built on these three syndrome directions, every check will see either zero or two copies of its doubled-angle logical
action. The cancellation condition of
Theorem~\ref{thm:syndrome_module_lifting} is then satisfied directly by
the CAB-index set-overlap pattern. The quartic CAB presentation can therefore organize all of the
syndrome-partition mechanisms introduced in Sec.~\ref{sec:gate_design_workflows}.

% ============================================================
\subsection{Recursive CAB syndrome partitions on hypercube codes}
\label{sec:hypercube_design}
\label{sec:hypercube_results}
% ============================================================

We next apply the CAB construction to hypercube codes, a family built
recursively by concatenating small error-detecting codes
\cite{GotoManyHypercube2024}. CABs are particularly useful for these codes because their recursive
structure allows the same CAB organization to reappear at successive
encoded levels. Related many-hypercube and
subsystem constructions have also been developed to improve code
parameters and reduce syndrome-measurement overhead
\cite{GotoOptimizedManyHypercube2025,NakaiGotoSubsystemMHC2025}.
The small error-detecting codes underlying these constructions, as well as
the first error-correcting hypercube level, are also directly relevant
to recent experimental logical-qubit architectures
\cite{RinesLogicalArchitecture2025}.

Here we use the hierarchy generated from the $[[4,2,2]]$ seed code to
show how recursive CAB structure can be turned into logical gates. The recursion can be viewed as concatenation. At each level,
we take four copies of the previous hypercube code as the inner code
and, for each inner-code logical qubit, encode its four copies using an
outer $[[4,2,2]]$ code. The CAB construction is especially useful
because the same quartic CAB-index sets and gate templates can then be
reused at each concatenation level. Starting from simple equal-syndrome CNOT and SWAP constructions on the $[[4,2,2]]$ code,
we construct addressed Clifford gates at higher levels, low-weight
gates obtained by recursively promoting lower-level gates, and
analytic non-Clifford constructions for CCZ-type gates.

% ------------------------------------------------------------
\subsubsection{The $[[4,2,2]]$ seed: a physical logical square}
\label{subsec:hypercube_seed_cab}
% ------------------------------------------------------------

For the $[[4,2,2]]$ seed we use the CAB
\begin{align}
\gamma_1&=YYYX, &
\gamma_2&=YYYZ, &
\nonumber\\
\gamma_3&=YYXI, &
\gamma_4&=YYZI, &
\nonumber\\
\gamma_5&=YXII, &
\gamma_6&=YZII, &
\nonumber\\
\gamma_7&=XIII, &
\gamma_8&=ZIII.
\label{eq:hypercube_seed_gammas}
\end{align}
The two quartic CAB-index sets are
\begin{align}
\mathcal C_X
&=
\{1,4,5,8\},
&
\mathcal C_Z
&=
\{2,3,6,7\}.
\label{eq:hypercube_seed_blocks}
\end{align}
Taking the ordered products of the displayed CAB generators gives
$S_X=-XXXX$ and $S_Z=-ZZZZ$. Throughout the hypercube recursion, we take the code to be the simultaneous $+1$ eigenspace of these signed checks. A convenient logical Pauli frame is
\begin{align}
\gamma_1\gamma_4
&\doteq \overline X_1=IIXX,
&
\gamma_4\gamma_5
&\doteq \overline Z_1=IZZI,
\nonumber\\
\gamma_2\gamma_3
&\doteq \overline X_2=IIZZ,
&
\gamma_3\gamma_6
&\doteq \overline Z_2=IXXI.
\label{eq:hypercube_seed_logical_frame}
\end{align}

The CAB templates of
Sec.~\ref{subsec:cab_index_constructions} provide particularly simple
physical representatives on this seed. The perfect-matching template
between $\mathcal C_X$ and $\mathcal C_Z$ pairs the CAB indices as
$(1,2)$, $(4,3)$, $(5,6)$, and $(8,7)$. The corresponding quadratic
monomials reduce to the four single-qubit operators
$Y_1,Y_2,Y_3,Y_4$. All four have the same nonzero syndrome, and their
$\pi/4$ rotations implement $\overline{\mathrm{SWAP}}$.

For the recursive construction below it is useful to record the logical
square carried by these four physical $Y$ operators explicitly. Set
\begin{align}
E&=Y_1,
&
P&=\overline X_1\overline X_2,
&
Q&=\overline Z_1\overline Z_2.
\label{eq:hypercube_seed_square_axes}
\end{align}
Using the logical frame in Eq.~\eqref{eq:hypercube_seed_logical_frame},
the four Pauli generators may be ordered as
\begin{align}
\mathcal Q
&=
\{Y_1,Y_2,Y_4,Y_3\}
\equiv_{\mathcal C}
E\{I,-P,-Q,PQ\},
\label{eq:hypercube_seed_y_square}
\end{align}
where $A\equiv_{\mathcal C}B$ means
$A|\psi\rangle=B|\psi\rangle$ for every $|\psi\rangle\in\mathcal C$;
for an ordered collection, this equality is applied entry by entry.
Here $P=-Y_3Y_4$, $Q=-Y_2Y_3$, and $YYYY$ is a positive stabilizer. Thus $EP\equiv_{\mathcal C}-Y_2$ and $EQ\equiv_{\mathcal C}-Y_4$, and the four physical $Y$ Pauli generators form a four-vertex logical square.
The apparently unusual ordering $1,2,4,3$ is chosen so that the two
square directions are $P$ and $Q$; this ordering will make the
square-to-cube construction below transparent.

The nested-index template of Sec.~\ref{subsubsec:nested_index_templates} also has simple realizations on the seed code. Choosing $\mathcal C_A=\mathcal C_X$, $\mathcal C_B=\mathcal C_Z$,
$\mathcal A=\{5,8\}$, and
$\mathcal B_q=\{7\}\subset\mathcal B_r=\{3,6,7\}$ gives the four single-qubit Pauli generators $\{X_1,X_2,-X_3,X_4\}$, whose $\pi/4$ rotations implement
$\overline{\mathrm{CNOT}}_{2\to1}$. Interchanging the roles of the two CAB-index sets and choosing $\mathcal A=\{6,7\}$ and
$\mathcal B_q=\{8\}\subset\mathcal B_r=\{4,5,8\}$
instead gives
$\{Z_1,Z_2,-Z_3,Z_4\}$, whose $\pi/4$ rotations implement
$\overline{\mathrm{CNOT}}_{1\to2}$. These are again equal-syndrome logical squares of the form
$E\{I,B_1,B_2,B_1B_2\}$. 

\begin{figure}[t]
\centering
\includegraphics[width=0.95\columnwidth]{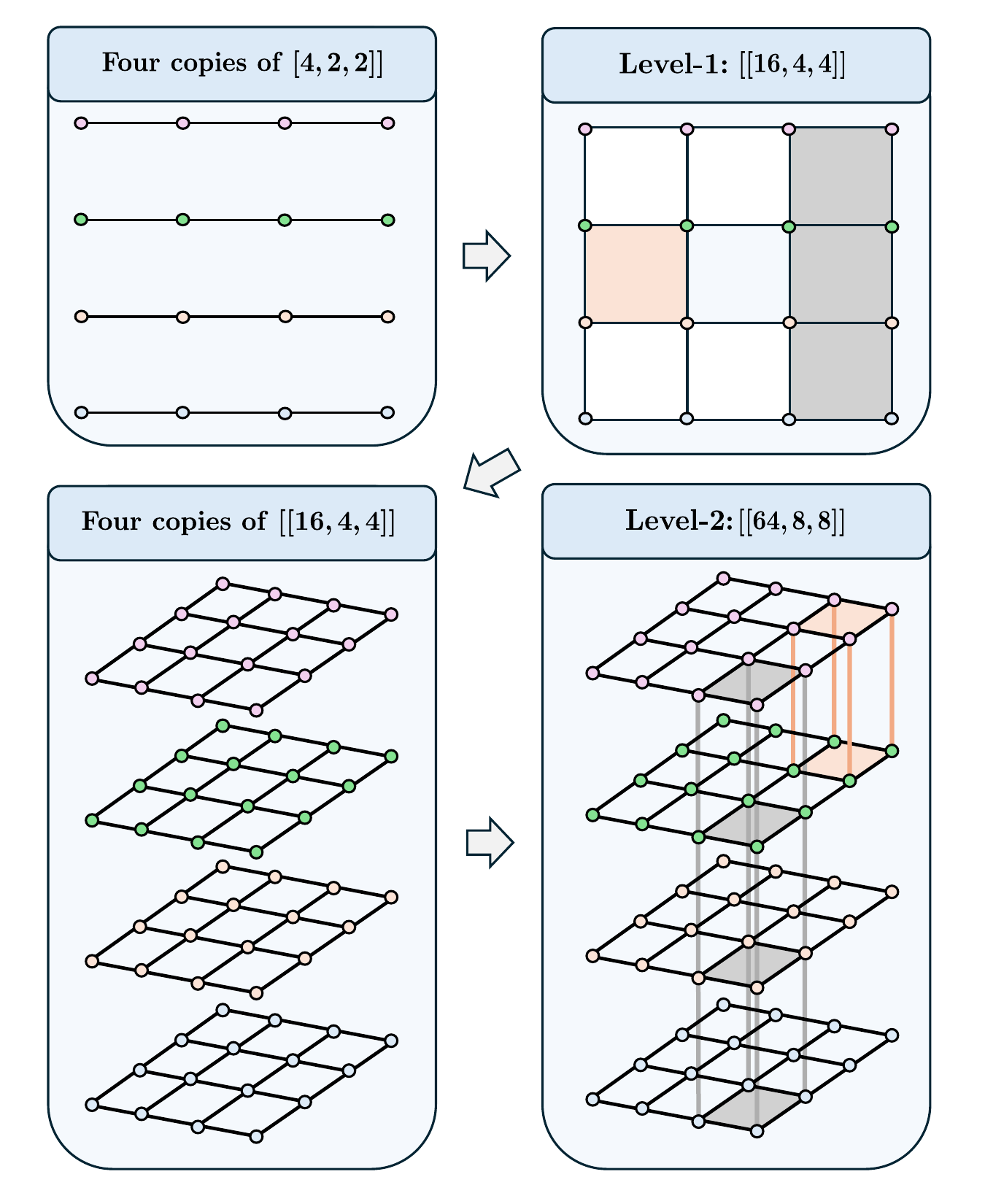}
\caption{Recursive concatenation of the first two hypercube levels.
Four copies of the $[[4,2,2]]$ inner code contain eight logical qubits
before the outer-code stabilizers are imposed. For each of the two
inner-code logical qubits, its four copies are encoded using an outer
$[[4,2,2]]$ code, producing the $[[16,4,4]]$ code. Repeating the same
construction with $[[16,4,4]]$ as the inner code produces the
$[[64,8,8]]$ code. At each level, the same quartic CAB organization is
applied to the encoded Pauli algebra of the inner code.}
\label{fig:hypercube_schematic}
\end{figure}

% ------------------------------------------------------------
\subsubsection{Recursive concatenation and encoded CABs}
\label{subsec:hypercube_recursive_construction}
% ------------------------------------------------------------

To construct the next level, we treat the previous hypercube code
$\mathcal H^{(\ell-1)}$ as the inner code and take four independent
copies, labeled $A,B,C,D$. If $q$ labels a logical qubit of the inner
code, then there are four corresponding logical qubits
$q_A,q_B,q_C,q_D$, with logical Pauli operators
$\overline X_q^{(r)}$ and $\overline Z_q^{(r)}$,
$r\in\{A,B,C,D\}$.

For each inner-code logical qubit $q$, these four copies are treated as
the four qubits of an outer $[[4,2,2]]$ code. We impose the two encoded
outer-code stabilizer checks
\begin{align}
T_q^X
&=
-\overline X_q^{(A)}
 \overline X_q^{(B)}
 \overline X_q^{(C)}
 \overline X_q^{(D)},
\nonumber\\
T_q^Z
&=
-\overline Z_q^{(A)}
 \overline Z_q^{(B)}
 \overline Z_q^{(C)}
 \overline Z_q^{(D)}.
\label{eq:hypercube_refinement_checks}
\end{align}
Because the outer $[[4,2,2]]$ code encodes two logical qubits into
four effective qubits, each inner-code logical-qubit label gives rise
to two logical qubits at the next concatenation level. For the first two levels,
\begin{align}
[[16,8,2]]
&\longrightarrow
[[16,4,4]],
&
[[64,16,4]]
&\longrightarrow
[[64,8,8]],
\label{eq:hypercube_first_refinements}
\end{align}
where the codes on the left are simply the four-copy inner-code systems
before the outer $[[4,2,2]]$ stabilizers are imposed. The recursive
construction is depicted in Fig.~\ref{fig:hypercube_schematic}.

The encoded CAB can then be obtained by a direct substitution rule. In the expressions of Eq.~\eqref{eq:hypercube_seed_gammas}, replace the
single-qubit Pauli acting on physical qubit $r$ by the corresponding
logical Pauli acting in inner-code copy $r$,
\begin{align}
X_r &\rightarrow \overline X_q^{(r)}, &
Y_r &\rightarrow \overline Y_q^{(r)}, &
Z_r &\rightarrow \overline Z_q^{(r)},
\label{eq:hypercube_encoded_substitution}
\end{align}
with $r=1,\ldots,4$. For example, the seed CAB generator
$\gamma_1=YYYX$ becomes
\begin{align}
\gamma_1^{(q)}
=
\overline Y_q^{(1)}
\overline Y_q^{(2)}
\overline Y_q^{(3)}
\overline X_q^{(4)}.
\end{align}
The remaining encoded CAB generators are obtained by the same
substitution. The two quartic CAB-index sets now multiply to the two
outer-code checks $T_q^X$ and $T_q^Z$, while their quadratic products
provide the two logical qubits encoded by that outer $[[4,2,2]]$ code.
In other words, we have rebuilt the same $[[4,2,2]]$ CAB in the
logical Pauli algebra of the inner code.

The CAB-index sets themselves remain
$\mathcal C_X=\{1,4,5,8\}$ and
$\mathcal C_Z=\{2,3,6,7\}$. Therefore the perfect-matching and
nested-index templates use exactly the same CAB-index relations as on
the physical $[[4,2,2]]$ seed; only the Pauli operators represented by
the $\gamma_j$ are now encoded operators of the inner code. This is
why the same equal-syndrome logical squares and gate templates recur at
successive concatenation levels.

For example, the perfect-matching template again uses the index pairs $(1,2), (4,3), (5,6),(8,7)$.
For the outer code associated with lower-level logical qubit $q$, the
four Pauli generators are therefore
\begin{align}
\{\gamma_1^{(q)}\gamma_2^{(q)},
\gamma_4^{(q)}\gamma_3^{(q)},
\gamma_5^{(q)}\gamma_6^{(q)},
\gamma_8^{(q)}\gamma_7^{(q)}\}.
\label{eq:hypercube_encoded_matching_generators}
\end{align}
Reducing these quadratic products using the chosen physical
representatives of the lower-level logical Paulis gives the physical
Pauli generators of the encoded SWAP LGA. The nested-index CNOT
generators are obtained from the same encoded CAB in exactly the same
way using the index pairings of
Sec.~\ref{subsubsec:nested_index_templates}.

At the first concatenation level, the inner $[[4,2,2]]$ code has two
logical qubits. The four copies of inner logical qubit $1$ are encoded
by one outer $[[4,2,2]]$ code and produce the new logical qubits
$(1,2)$. Independently, the four copies of inner logical qubit $2$
produce the new logical qubits $(3,4)$.

Applying the nested-index template to either outer code gives CNOTs in
both directions within the corresponding logical pair. With the seed
logical frame of Eq.~\eqref{eq:hypercube_seed_logical_frame}, the
required inner-code logical $X$ and $Z$ representatives have physical
weight two. Since the four inner-code copies have disjoint physical
support, each addressed CNOT is realized by four mutually disjoint
weight-two Pauli generators and has Pauli-rotation depth one. The
perfect-matching template similarly gives SWAPs on the same two
logical pairs using four mutually disjoint weight-three Pauli
generators.

The same concatenation is repeated with $[[16,4,4]]$ as the inner
code. Since the inner code has four logical qubits, there are four
independent outer $[[4,2,2]]$ encodings, one for each inner logical
qubit. Each outer code produces two logical qubits, giving the eight
logical qubits of the $[[64,8,8]]$ code naturally grouped as
$(1,2)$, $(3,4)$, $(5,6)$, and $(7,8)$.

The nested-index template gives both CNOT directions on each of these
four pairs using four mutually disjoint weight-four Pauli generators. The weight follows from the encoded-CAB substitution: The physical
single-qubit Pauli operators appearing in the seed template are
replaced by logical Pauli representatives of the $[[16,4,4]]$ inner
code, chosen here with physical weight four. Because the four
inner-code copies have disjoint physical support, the resulting four
Pauli generators are also mutually disjoint.
Since the code distance is eight, these constructions saturate the
equal-syndrome bound $w_{\max}\ge d/2$. Perfect matching gives an
addressed SWAP on each of the same four pairs using four mutually
disjoint weight-seven Pauli generators. These analytic templates act naturally within the four logical pairs
defined by the outer $[[4,2,2]]$ encodings. 

The logical pairs identified above are distinguished by the recursive
concatenation because the CAB templates give especially simple
low-weight, mutually disjoint physical realizations on those pairs.
The CAB-index templates themselves are not restricted to these natural
pairs and can also specify CNOTs between logical qubits belonging to
different pairs. For such choices, however, the most direct physical
representatives need not remain mutually disjoint or comparably low
weight. The CAB-guided searches discussed below are used to find
improved physical realizations for these additional pairs.

The same templates are not restricted to two logical CAB-index sets. The
pairwise constructions above use two CAB-index sets and give SWAP or CNOT.
Applying the same CAB-index pattern to several logical qubits at once
produces the corresponding multi-qubit gates. In particular, perfect
matching across several CAB-index sets gives the complement family, while the
multi-source version of the nested-index construction gives the multi-source CNOT. On $[[16,4,4]]$, matching across all four CAB-index sets gives a four-logical-qubit complement gate with four disjoint
weight-two Pauli generators. The same construction gives larger complement gates on
$[[64,8,8]]$; their physical weights are summarized in
Table~\ref{tab:hypercube_main_results}. The corresponding complement
and multi-source CNOT families, together with the bounded physical searches and
resulting logical-gate catalogs, are described in
Appendix~\ref{app:hypercube_catalogs}.

% ------------------------------------------------------------
\subsubsection{Promoting lower-level gates}
\label{subsec:hypercube_promoted_gates}
% ------------------------------------------------------------

There is a second way to carry gates through the recursive
concatenation. Instead of constructing a new gate from the outer CAB,
we may apply the same inner-code gate independently in each of the four
inner-code copies. Because the four copies have disjoint physical
support, this preserves the Pauli-generator weight, although the total
number of generators increases by a factor of four. For the promoted
gates considered below, the parallel inner-code operation preserves
the outer $[[4,2,2]]$ stabilizers and therefore induces a logical
operation on the concatenated code.

For example, applying the seed SWAP independently in the four
$[[4,2,2]]$ inner-code copies uses sixteen disjoint weight-one
rotations. After the outer $[[4,2,2]]$ stabilizers are imposed, the
resulting logical operation on $[[16,4,4]]$ is
\begin{align}
\overline{\mathrm{SWAP}}_{1,3}
\overline{\mathrm{SWAP}}_{2,4}.
\end{align}
The two seed CNOT constructions similarly become products of two CNOTs acting
between the corresponding logical pairs. Promotion retains the seed's weight-one physical Pauli generators, while the
resulting logical operation acts simultaneously on two logical pairs.

The same construction can be repeated at the next level. Promoting the seed SWAP LGA through both levels gives sixty-four disjoint weight-one
rotations on $[[64,8,8]]$, with the SWAP LGA acting as
\begin{align}
\overline{\mathrm{SWAP}}_{1,5}
\overline{\mathrm{SWAP}}_{2,6}
\overline{\mathrm{SWAP}}_{3,7}
\overline{\mathrm{SWAP}}_{4,8}.
\end{align}
Alternatively, a gate already constructed on $[[16,4,4]]$ may be
promoted independently in the four $[[16,4,4]]$ inner-code copies. In
that case the Pauli generator weight is inherited from the $[[16,4,4]]$
gate; for example, its four-qubit complement becomes
\begin{align}
\overline{\mathrm{COMP}}_{1,3,5,7}
\overline{\mathrm{COMP}}_{2,4,6,8}
\end{align}
using sixteen weight-two Pauli generators.

\begin{table*}[t]
\centering
\footnotesize
\renewcommand{\arraystretch}{1.12}
\begin{tabular*}{\textwidth}{@{\extracolsep{\fill}}llccc@{}}
\toprule
Code & analytic construction & $N_{\rm gen}$ & $w_{\max}$ & $D$ \\
\midrule
$[[4,2,2]]$
& directed CNOT / SWAP
& $4$ & $1$ & $1$ \\
\midrule
$[[16,4,4]]$
& addressed CNOT
& $4$ & $2$ & $1$ \\
$[[16,4,4]]$
& addressed SWAP
& $4$ & $3$ & $1$ \\
$[[16,4,4]]$
& four-qubit complement
& $4$ & $2$ & $1$ \\
$[[16,4,4]]$
& recursive CCZ-type gate
& $16$ & $2$ & $2$ \\
\midrule
$[[64,8,8]]$
& addressed CNOT
& $4$ & $4$ & $1$ \\
$[[64,8,8]]$
& addressed SWAP
& $4$ & $7$ & $1$ \\
$[[64,8,8]]$
& disjoint complement
& $4$ & $4$, $6$, or $9$ & $1$ \\
$[[64,8,8]]$
& recursive CCZ-type gate
& $16$ & $4$ & $2$ \\
\bottomrule
\end{tabular*}
\caption{Analytic CAB syndrome-partition gates on the
$[[4,2,2]]$ seed and the first two error-correcting hypercube codes.
Here $N_{\rm gen}$ is the number of Pauli generators in the LGA,
$w_{\max}$ is the maximum Pauli-generator weight, and $D$ is the
Pauli-rotation depth.}
\label{tab:hypercube_main_results}
\end{table*}
% ------------------------------------------------------------
\subsubsection{Recursive CCZ-type gates}
\label{subsec:hypercube_recursive_ccz}
% ------------------------------------------------------------

We can also analytically construct a recursive CCZ-type gate on these codes using CAB. This is easiest to see first on the
code formed from four $[[4,2,2]]$ inner-code copies, with parameters $[[16,8,2]]$ before the outer $[[4,2,2]]$ stabilizers in
Eq.~\eqref{eq:hypercube_refinement_checks} are imposed. The four inner-code copies are denoted $A,B,C,D$. In this example, each inner-code copy will use a copy of the perfect matching SWAP gate generated by weight-one $Y$ strings, but the nested index CNOT constructions work equally well. Writing $P_R=\overline X_{R,1}\overline X_{R,2}$, and $Q_R=\overline Z_{R,1}\overline Z_{R,2}$
with $R\in\{A,B,C,D\}$, we may write a logical square as
\begin{align}
\mathcal Q_R
&=
\{Y_{R,1},Y_{R,2},Y_{R,4},Y_{R,3}\} \nonumber\\
&\equiv_{\mathcal K}
E_R\{I,-P_R,-Q_R,P_RQ_R\},
\label{eq:hypercube_local_y_squares}
\end{align}
where $E_R=Y_{R,1}$ is the syndrome carrying Pauli generator, and $\mathcal K$ is the codespace of those four inner-code copies. Every Pauli generator in $\mathcal Q_R$ has weight one and
has the same local syndrome given by
\begin{align}
\boldsymbol\eta_R
&=
\boldsymbol e_{S_X^{(R)}}+
\boldsymbol e_{S_Z^{(R)}}.
\label{eq:hypercube_local_y_syndrome}
\end{align}
Thus the four inner-code copies supply four identical logical squares.

We next combine the squares in $A$ and $B$. There are two useful
matchings of their vertices. Pairing corresponding vertices gives the
\emph{straight} square 
\begin{align}
\mathcal P_{AB}^{\rm s}
=\{&Y_{A,1}Y_{B,1},
Y_{A,2}Y_{B,2},
Y_{A,4}Y_{B,4},
Y_{A,3}Y_{B,3}\},
\label{eq:hypercube_ab_straight_plane}
\end{align}
while exchanging the two columns of one square gives the
\emph{crossed} square
\begin{align}
\mathcal P_{AB}^{\rm c}
=\{&Y_{A,1}Y_{B,2},
Y_{A,2}Y_{B,1},
Y_{A,4}Y_{B,3},
Y_{A,3}Y_{B,4}\}.
\label{eq:hypercube_ab_crossed_plane}
\end{align}
All eight operators have weight two, commute, and have the same syndrome
$\boldsymbol\eta_A+\boldsymbol\eta_B$ on the four-copy code. Taking $E_{AB}=Y_{A,1}Y_{B,1}$ as a reference, multiplying any of the
eight Pauli generators by $E_{AB}$ gives a zero-syndrome operator and
therefore a logical Pauli operator. Up to stabilizers and Pauli signs,
these relative logical Paulis are the eight products generated by
$P_A$, $P_B$, and $Q_AQ_B$. These three operators are independent and
commute. Consequently the
union
$\mathcal P_{AB}^{\rm s}\cup\mathcal P_{AB}^{\rm c}$ is a rank-three logical cube. At angle $\pi/8$, these eight signed Pauli generators can then provide a CCZ-type gate on the four-copy code. Repeating the
same construction on $C$ and $D$ gives a second cube with disjoint
physical support with respect to the cube on $A$ and $B$.

The important step is what happens when the outer $[[4,2,2]]$
stabilizers are imposed to form the $[[16,4,4]]$ code. The new checks do not destroy the two cubes. Instead, each cube splits
into one straight and one crossed four-generator square, giving four
squares in total:
$\mathcal P_{AB}^{\rm s}$,
$\mathcal P_{AB}^{\rm c}$,
$\mathcal P_{CD}^{\rm s}$, and
$\mathcal P_{CD}^{\rm c}$. For the
particular $Y$ square above, every straight Pauli generator commutes with both
new outer $Z$-type checks, whereas every crossed Pauli generator
anticommutes with both. Let $\boldsymbol e_{T_1^Z}$ and $\boldsymbol e_{T_2^Z}$ denote the unit syndrome vectors for the two new outer $Z$-type checks
\begin{align}
\boldsymbol\tau
&=
\boldsymbol e_{T_1^Z}+
\boldsymbol e_{T_2^Z}.
\label{eq:hypercube_refinement_tau}
\end{align}
The four equal-syndrome squares in the $[[16,4,4]]$ code then have syndromes
\begin{align}
\boldsymbol s_{AB}^{\rm s}
&=\boldsymbol\eta_A+\boldsymbol\eta_B,
&
\boldsymbol s_{AB}^{\rm c}
&=\boldsymbol\eta_A+\boldsymbol\eta_B+\boldsymbol\tau,
\nonumber\\
\boldsymbol s_{CD}^{\rm s}
&=\boldsymbol\eta_C+\boldsymbol\eta_D,
&
\boldsymbol s_{CD}^{\rm c}
&=\boldsymbol\eta_C+\boldsymbol\eta_D+\boldsymbol\tau.
\label{eq:hypercube_four_plane_syndromes}
\end{align}
Thus imposing the outer checks reorganizes the two logical cubes of the four-copy code into four equal-syndrome logical squares. These four squares provide Clifford gates that can be used in the
unequal-syndrome lift of Sec.~\ref{subsec:syndrome_module_lifting},
as we now explain.

Doubling the angle on any one of these four squares from $\pi/8$ to
$\pi/4$ gives an equal-syndrome Clifford gate. With the displayed generators and positive rotation directions, the
straight and crossed $AB$ squares can be written as
\begin{align}
\mathcal P_{AB}^{\nu}
&\equiv_{\mathcal K}
E_{AB}^{\nu}
\{I,P_AP_B,Q_AQ_B,P_AP_BQ_AQ_B\},\nonumber\\
&\;\;\;\;\nu\in\{\mathrm{s},\mathrm{c}\},
\end{align}
where $E_{AB}^{\mathrm{s}}=Y_{A,1}Y_{B,1}$ and
$E_{AB}^{\mathrm{c}}=Y_{A,1}Y_{B,2}$. Their logical action is therefore the same and may be written as $\overline R_{AB}=
\exp(i\pi\Pi_{AB})$, where
\begin{align}
\Pi_{AB}
=
\frac14(I+P_AP_B)(I+Q_AQ_B)
\end{align}
is a projector. The reference Paulis commute with $\Pi_{AB}$, so
$\exp(i\pi E_{AB}^{\nu}\Pi_{AB})=\exp(i\pi\Pi_{AB})$. Hence $\overline R_{AB}^{\,2}=I$.

The same holds for the straight and crossed $CD$ squares. Moreover,
the outer stabilizers give
\begin{align}
P_AP_BP_CP_D
&=T_1^XT_2^X,
&
Q_AQ_BQ_CQ_D
&=T_1^ZT_2^Z,
\label{eq:hypercube_ab_cd_stabilizer_equivalence}
\end{align}
so $P_AP_B\sim P_CP_D$ and $Q_AQ_B\sim Q_CQ_D$ on the
concatenated codespace. All four squares therefore implement the same
order-two logical Clifford action,
\begin{align}
\overline R_{AB}^{\rm s}
=
\overline R_{AB}^{\rm c}
=
\overline R_{CD}^{\rm s}
=
\overline R_{CD}^{\rm c}
&\equiv \overline R,
&
\overline R^2&=I.
\label{eq:hypercube_common_clifford_face_action}
\end{align}

The four syndrome vectors satisfy the lifting condition,
\begin{align}
\boldsymbol s_{AB}^{\rm s}
+
\boldsymbol s_{AB}^{\rm c}
+
\boldsymbol s_{CD}^{\rm s}
+
\boldsymbol s_{CD}^{\rm c}
&=
\boldsymbol 0.
\label{eq:hypercube_four_plane_zero_sum}
\end{align}
Therefore, every stabilizer check sees the same order-two Clifford
$\overline R$ an even number of times. The check-conditioned product in
Theorem~\ref{thm:syndrome_module_lifting} is the identity for every
check, so the four logical squares may be run at half angle. Their union
is exactly the sixteen-generator $\pi/8$ LGA obtained from the two
logical cubes of the four-copy code. Choose all sixteen physical rotation directions to be positive. In the
fixed logical frame and signed-check convention above, the encoded
operator is
\begin{align}
\overline U
&=
I-\frac14
\left(I+\overline X_1\overline X_3\right)
\left(I+\overline X_2\overline X_4\right)
\left(I-\overline Z_2\overline Z_4\right).
\label{eq:hypercube_recursive_ccz_projector}
\end{align}
The three logical Paulis
$\overline X_1\overline X_3$,
$\overline X_2\overline X_4$, and
$\overline Z_2\overline Z_4$
commute and are independent. The operator applies phase $-1$ when
their eigenvalues are $+1,+1,-1$, respectively, and phase $+1$
otherwise. Choosing these three operators as logical $Z$ axes
identifies the action as a CCZ gate up to logical Pauli conjugation,
with the remaining logical qubit unaffected. We therefore refer to
it as CCZ-type.

The same construction fixes the gate's physical properties. The seed CNOT and SWAP logical gates used here have weight-one Pauli generators, so pairing vertices across $A$ and $B$ or
across $C$ and $D$ gives weight-two Pauli generators. The two straight squares
have disjoint support and can be executed together, and the two crossed
squares can likewise be executed together. The resulting
$[[16,4,4]]$ construction therefore has sixteen total rotations, all with Pauli generator weight two, implemented at a Pauli-rotation depth of two. 

The argument depends only on the equal-syndrome rank-two square structure, not on the seed vertices having physical weight one. It therefore recurses at the encoded level. Before the second outer-code encoding, the four copies of the
$[[16,4,4]]$ inner code form a $[[64,16,4]]$ four-copy code. Choose a
weight-two encoded square on each of the four $[[16,4,4]]$ copies. Pair the corresponding vertices across $AB$ and $CD$ exactly as
in Eqs.~\eqref{eq:hypercube_ab_straight_plane} and
\eqref{eq:hypercube_ab_crossed_plane}. The resulting sixteen Pauli generators have weight four and again form two rank-three logical cubes of the four-copy code. The new outer-code checks split those cubes into four
equal-syndrome square faces whose syndromes sum to zero. Choosing the
stabilizer-equivalent rotation signs so that the four doubled-angle
faces represent the same Clifford action gives the same
lifting argument at the encoded level. Consequently, the concatenated $[[64,8,8]]$ code has a recursive CCZ-type construction with sixteen total rotations, all with Pauli generator weight four, implemented at a Pauli-rotation depth of two.

% ------------------------------------------------------------
\subsubsection{Analytic summary and additional addressability}
\label{subsec:hypercube_analytic_summary}
% ------------------------------------------------------------

The recursive analytic constructions are summarized in
Table~\ref{tab:hypercube_main_results}. Here $D$ denotes the Pauli-rotation depth of the displayed construction,
i.e., the minimum number of Pauli-rotation layers required for its fixed
set of Pauli generators.

The CAB-index templates specify logical Clifford gates algebraically
between chosen logical qubits. For the natural logical pairs identified
by the recursive construction, these templates already give low-weight,
Pauli-rotation depth one physical realizations. Applying the same index relations to
other logical pairs still specifies the desired logical gate, but the
most direct physical representatives need not remain pairwise disjoint
or have comparably low weight.

We therefore use the CAB presentation to guide the same equal-syndrome
stabilizer-factorization search introduced in
Sec.~\ref{subsec:qldpc_clifford_search}. The difference is that the CAB
construction now supplies the stabilizer relation to be factorized and
the common syndrome required of its Pauli generators. We search for alternative
low-weight, mutually commuting, pairwise-disjoint physical
factorizations of that CAB-specified stabilizer. If the CAB-specified stabilizer does not admit a suitable factorization directly, we multiply it by a bounded product of additional stabilizer generators and search again. This changes the physical stabilizer representative being factorized without changing its action on the codespace, and can expose lower-weight or more nearly disjoint Pauli generator decompositions. The dressing rules and search
bounds are given in Appendix~\ref{app:hypercube_search_domains}. The CAB construction provides the target logical gate, along with a stabilizer relation and syndrome pattern that guide the physical search. These constraints strongly favor the intended operation, and many of the resulting low-weight, disjoint factorizations realize exactly the addressed Clifford predicted by the CAB template. The stabilizer product and syndrome alone, however, do not uniquely determine the logical action, because different factorizations can have different relative logical Pauli products. We therefore determine the logical Clifford action of every valid LGA independently, which also reveals additional gates within the same bounded search.

On the $[[16,4,4]]$ code, the analytic CAB construction gives both CNOT
orientations and a SWAP on the natural logical pairs $(1,2)$ and
$(3,4)$. The CAB-guided search finds the same addressed gates on the
additional pairs $(1,4)$ and $(2,3)$. These additional realizations
connect the addressed logical-gate graph across all four logical qubits.

At the next level, the analytic CAB construction gives both CNOT
orientations and a SWAP on the four disjoint pairs of logical qubits $(1,2), (3,4), (5,6),(7,8)$.
The supplementary search finds both CNOT orientations on twelve
additional logical pairs,
\begin{align}
&(1,3),\ (1,6),\ (1,8),\ (2,4),\ (2,5),\ (2,7),
\nonumber\\
&(3,6),\ (3,8),\ (4,5),\ (4,7),\ (5,7),\ (6,8).
\end{align}
Together with the four analytic pairs, this gives individually
addressed CNOTs on sixteen of the twenty-eight logical pairs and makes
the CNOT addressability graph connected.

In addition to the addressed CNOT and SWAP
realizations targeted by the CAB templates, the same bounded searches
also produce multi-source CNOT, complement, and other Clifford transformations.
The retained catalogs contain $168$ physical realizations with Pauli-rotation depth one
representing $66$ distinct logical Clifford gates on
$[[16,4,4]]$, and $246$ verified realizations representing $118$ such
operations on $[[64,8,8]]$.

The analytic and numerical parts can therefore play complementary roles.
The CAB presentation specifies the logical gate, its syndrome pattern,
and the stabilizer relation that must be factorized. The bounded search
then uses that information to find low-weight, support-disjoint physical
representatives and, in doing so, can reveal additional Clifford
operations within the same finite search domain. Further details of the
searches and retained catalogs are given in
Appendix~\ref{app:hypercube_catalogs}.

% ------------------------------------------------------------
\subsection{An analytic hierarchy in middle Reed--Muller codes}
\label{subsec:middle_rm_nonclifford}
% ------------------------------------------------------------

The recursive hypercube construction above realizes CCZ-type gates with
$w_{\max}=d/2$ by combining several Clifford LGAs through the
unequal-syndrome lifting condition. The middle Reed--Muller family gives
a complementary construction in which a single equal-syndrome logical
cube is represented by mutually disjoint physical Pauli generators. This
produces a Pauli-rotation depth one hierarchy of controlled-$Z$ gates while
again attaining $w_{\max}=d/2$.

The middle Reed--Muller codes are self-dual CSS codes with a rich
logical Clifford structure
\cite{gong2024computation,tansuwannont2026construction,
albert2026beyond}. Diagonal non-Clifford transversal gates in quantum Reed--Muller
codes have also been studied extensively
\cite{rengaswamy2020optimality,barg2026geometric}. For the middle
family, the conventional single-qubit diagonal constructions remain at
the Clifford level. Here we instead use mutually disjoint multiqubit
Pauli rotations.

The construction is easiest to see for the $[[16,6,4]]$ tesseract
code. Label the physical qubits by
$x_1x_2x_3x_4\in\mathbb F_2^4$ and, for each
$\boldsymbol u=(u_1,u_2,u_3)\in\mathbb F_2^3$, define
\begin{align}
G_{\boldsymbol u}
&=
Z_{u_1u_2u_3 0}
Z_{u_1u_2u_3 1}.
\label{eq:tesseract_pair_Pauli generators}
\end{align}
These are the eight edges of the tesseract parallel to the
$x_4$ direction. They have weight two and pairwise-disjoint supports
whose union is all sixteen physical qubits.

The eight Pauli generators also share one nonzero stabilizer syndrome.
Choosing $G_{\boldsymbol 0}$ as a reference, the seven relative
products $G_{\boldsymbol 0}G_{\boldsymbol u}$ have zero syndrome and,
modulo stabilizers, run through the seven nonidentity products of three independent
commuting logical $Z$ operators, as shown in Appendix~\ref{app:middle_rm_hierarchy}. The physical Pauli generators therefore form
a rank-three logical cube. Applying the Walsh-sign construction of
Sec.~\ref{subsec:equal_angle_hierarchy} at angle $\pi/8$ gives
\begin{align}
U
&=
\prod_{\boldsymbol u\in\mathbb F_2^3}
\exp\left[
 i\frac{\pi}{8}
 (-1)^{|\boldsymbol u|}
 G_{\boldsymbol u}
\right],
\label{eq:tesseract_ccz_layer}
\end{align}
which realizes a logical CCZ-type gate. The syndrome and logical-cube
identities underlying this construction are proved in
Appendix~\ref{app:middle_rm_hierarchy}.

The same construction extends uniformly to the middle Reed--Muller
family
\begin{align}
QRM(2r)
&=
\left[
\left[
2^{2r},
\binom{2r}{r},
2^r
\right]
\right].
\label{eq:middle_rm_parameters_nonclifford}
\end{align}
Divide the $2r$ Boolean coordinates into $r+1$ fixed coordinates $C$
and $r-1$ free coordinates $W$. For each
$\boldsymbol u\in\mathbb F_2^{r+1}$, fix the coordinates in $C$ to
$\boldsymbol u$ and allow the coordinates in $W$ to vary. The resulting
$Z$-type Pauli generator is
\begin{align}
G_{\boldsymbol u}
&=
\prod_{\boldsymbol x:\,\boldsymbol x_C=\boldsymbol u}
Z_{\boldsymbol x}.
\label{eq:middle_rm_slice_Pauli generator_main}
\end{align}
There are $2^{r+1}$ such Pauli generators. Their supports are pairwise disjoint and together cover all physical qubits, and each generator has weight $2^{r-1}=\frac d2$.

These Pauli generators share one nonzero syndrome, and their relative products
realize all products of $r+1$ independent commuting logical $Z$
operators. They therefore form a rank-$(r+1)$ logical cube. The full
polynomial-overlap argument and the identification of the logical axes
are given in Appendix~\ref{app:middle_rm_hierarchy}. Applying the same Walsh-sign construction at angle $\pi/2^{r+1}$ gives
\begin{align}
U_r
&=
\prod_{\boldsymbol u\in\mathbb F_2^{r+1}}
\exp\left[
 i(-1)^{|\boldsymbol u|}
 \frac{\pi}{2^{r+1}}
 G_{\boldsymbol u}
\right],
\label{eq:middle_rm_hierarchy_gate_main}
\end{align}
which realizes a logical $C^rZ$-type phase gate on the $r+1$ logical axes.
The first few members are summarized in
Table~\ref{tab:middle_rm_nonclifford_family}.

\begin{table}[t]
\centering
\begin{tabular*}{\columnwidth}{@{\extracolsep{\fill}}cccccc@{}}
\hline\hline
$r$ & Code & Pauli generators & $w_{\max}$ & Angle & Logical gate \\
\hline
$1$ & $[[4,2,2]]$     & $4$  & $1$ & $\pi/4$  & $CZ$ \\
$2$ & $[[16,6,4]]$    & $8$  & $2$ & $\pi/8$  & $CCZ$ \\
$3$ & $[[64,20,8]]$   & $16$ & $4$ & $\pi/16$ & $C^3Z$ \\
$4$ & $[[256,70,16]]$ & $32$ & $8$ & $\pi/32$ & $C^4Z$ \\
\hline\hline
\end{tabular*}
\caption{Equal-syndrome gates with Pauli-rotation depth one in the middle quantum
Reed--Muller family. Each construction uses mutually disjoint Pauli generators
that partition the physical qubits.}
\label{tab:middle_rm_nonclifford_family}
\end{table}

For every $r$, the code distance is $d=2^r$ and every Pauli generator has
weight $d/2$. The construction therefore saturates the equal-syndrome
weight bound of Eq.~\eqref{eq:equal_syndrome_weight_bound} while
remaining Pauli-rotation depth one. This hierarchy lies outside the
single-qubit diagonal transversal constructions studied in
Refs.~\cite{rengaswamy2020optimality,barg2026geometric}. For the
self-dual middle family $m=2r$, those constructions reach only
Clifford-level diagonal gates, whereas the disjoint multiqubit Pauli
rotations used here realize controlled-$Z$ gates at arbitrarily high
levels of the Clifford hierarchy.

% ============================================================
\section{Fault-propagation structure and fault tolerance}
\label{sec:fault_diagnostics}
% ============================================================

\subsection{Depth-one fault propagation and Pauli-rotation faults}

Many constructions above have Pauli-rotation depth one, so all of their
Pauli rotations have pairwise-disjoint physical support. Although these
LGAs are not transversal in the usual sense, their disjoint supports
restrict fault propagation because a single-qubit fault can overlap at
most one active Pauli generator. For instance, a weight-$w$ Pauli error can overlap at
most $w$ active Pauli generators. We therefore track how physical
Pauli errors propagate through each depth-one LGA, treating every
multiqubit Pauli rotation as an ideal operation. 

For an input Pauli $P$, if $P$
anticommutes with the signed generator $\epsilon G$, then
\begin{align}
e^{i\epsilon\theta G}P e^{-i\epsilon\theta G}
&=
\cos(2\theta)P
+
i\epsilon\sin(2\theta)GP.
\label{eq:single_fault_propagation}
\end{align}
If $P$ commutes with $G$, it is unchanged. Thus a fault may split into
two Pauli branches, but every branch remains confined to the support of
the single generator that the fault encounters.

We exhaustively propagate all weight-one and weight-two Pauli errors through every depth-one logical gate found in our searches, and all weight-three Pauli errors for every distinct logical action found. For every code considered, errors within the correctable range $w \leq \left \lfloor \frac{d-1}{2} \right \rfloor$ remain correctable after propagation through the ideal logical gate. This remains true even when the weight of the error grows substantially. For example, in the $[[64,8,8]]$ code, some logical gates are generated by weight-four Pauli strings. A weight-two error can share support on two such generators and propagate to a Pauli of weight eight while the
resulting error remains correctable. More generally, this behavior is expected as a logical gate preserves the correctability of an error set according to the Knill--Laflamme conditions, regardless of the physical weights of the propagated errors.

For Clifford LGAs, the propagated errors are again Pauli operators, so this correctability is directly visible from the stabilizer syndromes. For non-Clifford LGAs, a propagated Pauli error can instead become a coherent sum of Pauli components. The full conjugated error remains correctable, but at weights two and three some individual Pauli components of the non-Clifford transformed Pauli errors share a stabilizer syndrome while belonging to different stabilizer cosets. Therefore, ordinary stabilizer syndrome extraction will not always work for these non-Clifford propagated errors.

More generally, if a multiqubit Pauli rotation is implemented
as a single physical operation, a failure of that operation may
produce an arbitrary error over its entire support. For a depth-one
layer, let
\begin{align}
\mathcal B=\{B_1,\ldots,B_m\}
\end{align}
denote the disjoint physical gate supports, together with a one-qubit
part for every idle qubit. Following Ref.~\cite{dua2026qubitcss}, the \emph{partition distance} of $\mathcal B$ is
\begin{align}
d_{\mathrm{part}}(\mathcal B)
=
\min_{\overline P\in\mathcal N(\mathcal S)\setminus\mathcal S}
\left|
\left\{
B\in\mathcal B:
B\cap\operatorname{supp}(\overline P)\neq\varnothing
\right\}
\right|,
\label{eq:partition_distance}
\end{align}
namely, the minimum number of Pauli-rotation supports whose union can support a
nontrivial logical Pauli operator. A nontrivial logical Pauli changes the encoded information without changing the stabilizer syndrome. The partition distance counts the
fewest rotation supports or idle-qubit locations that can together
support such an undetectable, logically nontrivial Pauli error. Correcting faults at $s$ unknown
locations requires distinguishing any two allowed errors, which can
together involve up to $2s$ locations. Thus arbitrary faults at up to
$s$ locations are correctable when $d_{\mathrm{part}}\geq2s+1$.
In particular, partition distance two guarantees detection, but not
correction, of every single-location fault, whereas partition distance
at least three permits correction of one such fault.

A useful feature of the syndrome-partition framework is that partition
distance is directly built into the algebraic structure of an LGA. For any subset
$J$ of Pauli generators,
\begin{align}
\boldsymbol{\sigma}_{\boldsymbol S}
\left(
\prod_{j\in J}G_j
\right)
&=
\bigoplus_{j\in J}
\boldsymbol{\sigma}_{\boldsymbol S}(G_j).
\label{eq:syndrome_xor_normalizer}
\end{align}
Thus a collection of generators can multiply to a logical Pauli only
if their syndrome vectors sum to zero.

Suppose the construction contains a set $J$ of $m$ generators whose
syndromes sum to zero and whose product represents a nontrivial logical
Pauli. Their supports then contain a logical operator, so $d_{\mathrm{part}}\le m$. Furthermore, if every support in the partition $\mathcal B$ has size at most $w_{\max}$ and $(m-1)w_{\max}<d$, then any union of fewer than $m$ parts has physical weight below the
code distance and therefore cannot contain a nontrivial logical Pauli and $d_{\mathrm{part}}=m$.
The order of the first logical zero-syndrome relation can therefore be
used as an algebraic design constraint for the partition distance.

The equal-syndrome construction used extensively above provides the
simplest example. Any two generators with the same syndrome have a
zero-syndrome product, and our nontrivial constructions require at
least one such pair to differ by a nontrivial logical Pauli. For
$w_{\max}<d$, no individual generator support can contain a logical
operator, so these support-disjoint layers necessarily have $d_{\mathrm{part}}=2$. More generally, one could design searches in which the first allowed
logical zero-syndrome relation involves three or more generators,
thereby targeting larger partition distance directly through the
syndrome constraints. 

This result is useful both as a diagnostic and as a design principle.
A layer with $d_{\mathrm{part}}=2$ can detect an arbitrary fault on one
unknown Pauli-rotation, but cannot in general determine which of two
possible gate supports suffered the fault. Additional location information can remove this ambiguity. Such
information may be provided directly by the hardware, for example
through erasure conversion \cite{wu2022erasure}, or by a protected
implementation when a multiqubit rotation is decomposed into smaller operations. Flagged or verified ancilla constructions provide
possible routes in the latter setting
\cite{chamberland2018flag,chao2020flag,prabhu2023fault}.

\subsection{Intermediate correction for higher-depth gates}

Further considerations are needed for gates with Pauli-rotation depth greater than one. Rotations within one Pauli-rotation
layer have disjoint support, but a data qubit may participate in
another rotation in a later layer. An error that remains detectable
after one layer can therefore spread further when subsequent
overlapping rotations are applied.

A natural approach is pieceable fault tolerance, in which a
nontransversal logical gate is divided into pieces separated by
intermediate error correction \cite{yoder2016universal}. The
support-disjoint Pauli-rotation layers found here provide natural
candidate pieces. This does not replace the requirement that each
physical Pauli rotation itself be implemented fault-tolerantly; rather,
the intermediate correction prevents faults from continuing to spread
through later overlapping layers. However, care is needed because these intermediate layers need not preserve the codespace and therefore may modify the stabilizers available for intermediate error correction.

Conveniently, the syndrome data we use to guide the search for logical gates also determines which stabilizer checks remain
available throughout such an implementation. Let
\begin{align}
\Sigma
&=
\begin{bmatrix}
\boldsymbol{\sigma}_{\boldsymbol S}(G_1)
&
\cdots
&
\boldsymbol{\sigma}_{\boldsymbol S}(G_q)
\end{bmatrix},
&
\rho_{\mathrm{syn}}
&=
\operatorname{rank}_{\mathbb F_2}\Sigma .
\label{eq:syndrome_span_rank}
\end{align}
Let $\boldsymbol c\in\mathbb F_2^{n-k}$ specify a stabilizer product $S(\boldsymbol c)
=
\prod_{a=1}^{n-k} S_a^{c_a}$.
This stabilizer product commutes with every Pauli generator in the LGA
exactly when $\boldsymbol c^{T}\Sigma=0$.
Thus the stabilizer combinations that remain unchanged throughout the
LGA are labeled by vectors in the left null space of $\Sigma$. The rank--nullity theorem therefore gives
\begin{align}
\operatorname{rank}\mathcal S_{\mathrm{const}}
=
n-k-\rho_{\mathrm{syn}},
\label{eq:constant_check_rank}
\end{align}
where $\mathcal S_{\mathrm{const}}$ is the subgroup of stabilizers that
commute with every Pauli generator. These checks remain unchanged throughout
the gate and can therefore remain available for intermediate syndrome
extraction.

This number is particularly simple for the LGA constructions we considered above. Here $\rho_{\mathrm{syn}}$ is simply the number of
independent syndrome vectors carried by the Pauli generators, which by
definition is the rank of $\Sigma$. A
nontrivial equal-syndrome LGA has $\rho_{\mathrm{syn}}=1$; the
six-generator Clifford-root construction has syndrome-span rank two;
and the seven-pair Fano construction has rank three. Thus only one,
two, or three independent stabilizer directions, respectively, can
change while these gates are applied, while all remaining
$n-k-\rho_{\mathrm{syn}}$ independent stabilizer combinations remain
unchanged. Whether these unchanged checks suffice to distinguish the relevant
faults, and how to measure them fault tolerantly, must still be
established for each logical gate.

% ============================================================
\section{Discussion and Conclusions}
\label{sec:discussion}
% ============================================================

The syndrome-partition framework is formulated at the level of
commuting multiqubit Pauli rotations. This level of description is
useful because it exposes the stabilizer structure of the logical gate
without committing to a particular hardware decomposition, but it also
sets the scope of the present results. In particular, disjoint physical
supports do not by themselves imply a shallow circuit once each
multiqubit rotation is compiled into elementary operations. The
resulting circuit depth, connectivity requirements, and opportunities
for parallel execution will depend on the native interactions of the
underlying platform. Developing architecture-aware decompositions of
the constructions reported here is therefore an important next step.

The same distinction applies to the fault analysis. Treating each
multiqubit Pauli rotation as an elementary operation, we find that the
single-fault errors considered remain distinguishable by their
stabilizer syndromes after propagation through the LGA and are
therefore correctable within this model. This indicates that the separate support structure of many constructions can offer advantageous fault-propagation properties. It does not, however, constitute a full
circuit-level fault-tolerance proof. Decomposing a multiqubit rotation
into elementary gates can introduce additional fault locations and
correlated propagation that are not captured here. Establishing fault-tolerant implementations on specific
architectures will require analyzing these decompositions together with
syndrome extraction and recovery.

The computational results should likewise be interpreted within their
stated search domains. The qLDPC searches are exhaustive over the
specified Pauli generator weights, Pauli sectors, syndrome relations, and
stabilizer-equivalent representatives, but they do not establish global
optimality over all possible physical realizations. Broader Pauli ans\"atze, different syndrome embeddings, and alternative
stabilizer dressings may produce still simpler realizations. More
generally, the framework separates the algebraic question of whether a
code-preserving logical gate exists from the subsequent optimization of
its physical implementation.

One conceptual feature of the framework is that several apparently
different routes to logical gates at higher levels of the Clifford hierarchy can be formulated using
the same syndrome data. Equal-syndrome factorizations directly expose
logical structure, equal-syndrome pairs provide systematic rotations
about chosen logical Pauli axes, and relations among lower-level gates
can be used to construct operations higher in the Clifford hierarchy.
These mechanisms are algebraically distinct, but the same
codespace-preservation condition governs them all. This is particularly
useful beyond the Clifford group, where ordinary symplectic stabilizer
methods no longer provide a complete description of the conjugated
checks.

Fault-tolerance considerations also suggest a natural extension of
the syndrome-partition approach. Many of the constructions studied here
rely on pairwise syndrome relations, but more general relations involving
larger collections of Pauli rotations could be used to design gates with
greater tolerance to faults in the physical rotations. This suggests
incorporating fault-tolerance criteria directly into future
syndrome-partition searches, alongside the generator weight and
Pauli-rotation depth already considered here. The many physically
distinct realizations of the same logical operation found by our searches
provide additional freedom for such optimization, including the choice
of how errors can spread and which stabilizer checks remain available
between successive layers.

Several extensions are therefore natural. The search methods can be
applied to larger qLDPC families and to higher levels of the Clifford
hierarchy, while a broader theoretical question is to identify which
features of a stabilizer code predict the existence of low-weight,
low-overlap syndrome partitions. Extending the construction principles
beyond LGAs would further enlarge the class
of accessible physical ans\"atze. On the analytic side, Clifford
algebra coordinates may reveal additional recursive code--gate
constructions, while the independent Reed--Muller results suggest that
useful syndrome-partition structure may also be identifiable directly in other algebraic code families. These directions point toward a
more systematic form of code--gate co-design in which error-correction
properties and logical operations are optimized together.

Machine learning and AI-assisted search methods could help explore logical gate ansätze beyond the equal-syndrome constructions considered in this work. Once an LGA has been specified, we can verify the codespace-preservation conditions exactly. However, the more challenging task involves identifying effective collections of Pauli generators and syndrome relations. AI-based approaches could help propose more general syndrome partitions and optimize the constructions for physical weight, Pauli-rotation depth, and other fault-tolerance criteria.

The search space can also expand by relaxing the commuting-generator assumption in our LGA constructions and integrating syndrome-based searches with code automorphisms. This would enable the joint exploration of Pauli-rotation layers, permutations, and logical actions. An alternative direction is to invert the design process considered here by starting with the logical gates and using them to impose stabilizer relations. These gates could then be completed into codes with the desired distance and other relevant properties. The syndrome-partition framework can also be integrated with modular code-construction approaches, such as Quantum LEGO~\cite{CaoPRXQuantum2022} and its Power-up framework~\cite{cao2026quantum}. Such investigations might include useful syndrome-partition structures on small code blocks and determining the compatibility or symmetry conditions that ensure the persistence of corresponding logical operations when these blocks are assembled into larger codes. We leave such explorations and AI-assisted generalizations of and improvements to our framework for future research.

In summary, we have introduced syndrome partitions as a general design
principle for direct logical gates built from commuting Pauli
rotations. The central result is an exact codespace-preservation
condition, valid at arbitrary rotation angles, that depends directly on
the stabilizer syndromes of the Pauli generators. The condition
applies uniformly to CSS and non-CSS stabilizer codes and at both
Clifford and non-Clifford angles. Its equal-syndrome specialization
turns Clifford gate design into a stabilizer-factorization problem,
bounds the minimum generator weight in terms of the code distance, and
organizes systematic constructions of logical Pauli rotations and
gates higher in the Clifford hierarchy.

We demonstrated the framework in complementary computational and
analytic settings. Searches on CSS bivariate-bicycle and non-CSS Mirror
qLDPC codes uncovered large families of Clifford operations together with
direct controlled-phase and logical $T$-type rotations. Clifford algebra
bases provide an analytic route to syndrome-partition solutions,
producing addressed CNOT, SWAP, controlled, and CCZ-type operations on
recursive hypercube codes. Independently, middle quantum Reed--Muller
codes admit disjoint controlled-$Z$ constructions whose physical
weights saturate the general equal-syndrome bound.

The same syndrome information also exposes structure relevant to fault
tolerance. Relations among the logical gate generator syndromes constrain how small
a collection of Pauli-rotation supports can carry a logical operator,
while the independent syndrome directions determine which stabilizer
combinations remain unchanged throughout a gate and may therefore be
available for intermediate error correction.

Together, these results show that the stabilizer description of a
quantum code can serve as a common framework for error correction,
logical-gate construction, and fault-tolerance-aware design. Syndrome
partitions therefore provide a route toward co-designing the
error-correcting, computational, and fault-tolerance properties of
encoded quantum systems.

\begin{acknowledgments}
This material is based upon work supported by the U.S. Department of Energy, Office of Science, Office of Advanced Scientific Computing Research, as part of the Genesis Mission National Science and Technology Challenges, under Award Number DE-SC0026692.
\end{acknowledgments}

\newpage
\bibliographystyle{apsrev4-2}
\bibliography{refs}

\appendix
\newpage

% ============================================================
\section{Syndrome-partition proofs}
\label{app:desired_syndrome_proof}
% ============================================================

This appendix proves the statements used in
Sec.~\ref{sec:gate_design_workflows}. We first prove the
arbitrary-angle codespace-preservation condition, and the equal syndrome weight bound. We then derive the signed $\pi/4$ preservation condition used in the Clifford searches and the
sign construction for higher controlled phases. We conclude with the
unequal-syndrome lifting theorem and the logical-derivative rule.

% ------------------------------------------------------------
\subsection{Arbitrary-angle preservation condition}
\label{app:arbitrary_angle_proof}
% ------------------------------------------------------------

Consider the LGA
\begin{align}
U
&=
\prod_{j=1}^{q}
\exp\left(
i\epsilon_j\theta_jG_j
\right),
\qquad
\epsilon_j\in\{\pm1\},
\label{eq:app_general_rotation_layer}
\end{align}
with pairwise-commuting Hermitian Pauli generators $G_j$. For a
stabilizer generator $S_a$, let $\mathcal J_a$ contain precisely those
indices $j$ for which $G_j$ anticommutes with $S_a$.

A rotation whose generator commutes with $S_a$ leaves the check
unchanged. If $G_j$ anticommutes with $S_a$, moving the check through
that rotation produces a doubled-angle factor
$e^{2i\epsilon_j\theta_jG_j}$. Since the $G_j$ commute with one another,
these factors can be collected to give
\begin{align}
U^\dagger S_aU
&=
S_aV_a,
&
V_a
&=
\prod_{j\in\mathcal J_a}
e^{2i\epsilon_j\theta_jG_j}.
\label{eq:app_check_conditioned_unitary}
\end{align}

We can now prove
Theorem~\ref{thm:arbitrary_angle_syndrome_partition}. Suppose first
that $U$ preserves the codespace $\mathcal{C}$. Then
$U^\dagger S_aU|\psi\rangle=|\psi\rangle$ for every
$|\psi\rangle\in\mathcal C$. Using
Eq.~\eqref{eq:app_check_conditioned_unitary} gives
$S_aV_a|\psi\rangle=|\psi\rangle$. Multiplying by $S_a$ and using
$S_a|\psi\rangle=|\psi\rangle$ yields
\begin{align}
V_a|\psi\rangle
&=
|\psi\rangle .
\end{align}
Conversely, if every $V_a$ acts trivially on the codespace, then
\begin{align}
U^\dagger S_aU|\psi\rangle
&=
S_a|\psi\rangle
=
|\psi\rangle
\end{align}
for every stabilizer generator. Hence $U|\psi\rangle$ remains in the
simultaneous $+1$ eigenspace of the stabilizer group and therefore in
the codespace.

% ------------------------------------------------------------
\subsection{Equal-syndrome consequences}
\label{app:equal_syndrome_reduction}
% ------------------------------------------------------------

Let Eq.~\eqref{eq:app_check_conditioned_unitary} be a nontrivial code-preserving LGA whose commuting Pauli generators
all carry the same nonzero syndrome. Choose one generator,
$E=G_1$, as a reference and define $L_j=EG_j$.
Because $E$ and $G_j$ have the same syndrome, each $L_j$ has zero
syndrome and therefore lies in the Pauli normalizer.

Suppose, for contradiction, that every pair $G_iG_j$ represents the
trivial logical Pauli class. Then in particular every $L_j=G_1G_j$ acts as a scalar on the codespace. The equal-syndrome decomposition of
Sec.~\ref{subsec:equal_syndrome_general} then gives the same phase to
every logical basis state. The LGA can therefore induce only an
overall phase and cannot implement a nontrivial logical operation. Hence, any nontrivial equal-syndrome LGA must contain at least
one pair $G_iG_j$ whose class in the Pauli normalizer modulo
stabilizers is nontrivial. By definition of the code distance, $d \leq \operatorname{wt}(G_iG_j)$. If $w_{\max}
=
\max_j \operatorname{wt}(G_j)$ then $\operatorname{wt}(G_iG_j)
\leq
\operatorname{wt}(G_i)+\operatorname{wt}(G_j)$.
Therefore $d\leq2w_{\max}$ and thus
\begin{align}
w_{\max}
&\geq
\left\lceil\frac d2\right\rceil .
\end{align}

We next prove Corollary~\ref{thm:desired_syndrome_partition}.
For an equal-syndrome LGA at physical angle $\pi/4$, every affected
stabilizer check selects all $q$ rotations. Its check-conditioned
operation is therefore
\begin{align}
U^2
&=
\prod_{j=1}^{q}
e^{i\epsilon_j(\pi/2)G_j}
=
i^q
\left(\prod_{j=1}^{q}\epsilon_j\right)
\left(\prod_{j=1}^{q}G_j\right),
\end{align}
where we used the mutual commutativity of the $G_j$. The arbitrary-angle preservation theorem then implies that the LGA
preserves the specified $+1$ codespace if and only if
\begin{align}
i^q
\left(\prod_{j=1}^{q}\epsilon_j\right)
\left(\prod_{j=1}^{q}G_j\right)
&\in
\mathcal S.
\end{align}
Thus the product of the Pauli generators must represent a
stabilizer up to its Hermitian sign, and the rotation directions must
supply precisely the sign required for the resulting operator to lie
in the $+1$ stabilizer group.

% ------------------------------------------------------------
\subsection{Higher-level equal-syndrome phases}
\label{app:logical_sign_construction}
% ------------------------------------------------------------

We prove the two algebraic statements used in
Sec.~\ref{subsec:equal_angle_hierarchy}: the existence of a suitable
syndrome carrier and the logical sign construction for
C${}^{m-1}$Z.

First consider the syndrome carrier. Let
$S_1,\ldots,S_r$, with $r=n-k$, be independent stabilizer generators.
Complete them to a canonical symplectic Pauli frame containing
destabilizers $D_1,\ldots,D_r$ and physical representatives of a
logical Pauli basis. The frame may be chosen so that $D_a$
anticommutes only with $S_a$ and commutes with every logical-basis
representative. Hence, for any syndrome
$\boldsymbol v\in\mathbb F_2^r$,
\begin{align}
E_{\boldsymbol v}
&=
\prod_{a=1}^{r}D_a^{v_a}
\end{align}
has syndrome $\boldsymbol v$ and commutes with all logical operators
expressed in this canonical frame. Thus, for any chosen commuting
logical Pauli classes, one may choose physical representatives and a
nonzero syndrome carrier $E$ that commute.

Now let $B_1,\ldots,B_m$ be independent commuting logical Paulis and
write their $2^m$ products as
\begin{align}
L_{\boldsymbol u}
&=
\prod_{a=1}^{m}B_a^{u_a},
\qquad
\boldsymbol u\in\mathbb F_2^m.
\label{eq:app_logical_sign_family}
\end{align}
For a simultaneous eigenvalue assignment
$\boldsymbol z\in\mathbb F_2^m$, where $B_a$ has eigenvalue
$(-1)^{z_a}$, the operator $L_{\boldsymbol u}$ has eigenvalue
$(-1)^{\boldsymbol u\cdot\boldsymbol z}$. At physical angle
$\theta=\pi/2^m$, define
\begin{align}
A(\boldsymbol z)
&=
\sum_{\boldsymbol u\in\mathbb F_2^m}
\epsilon_{\boldsymbol u}
(-1)^{\boldsymbol u\cdot\boldsymbol z}.
\label{eq:app_signed_logical_sum}
\end{align}
Choose $\epsilon_{\boldsymbol u}=
(-1)^{\boldsymbol u\cdot\boldsymbol z^\ast}$
for a selected eigenvalue assignment $\boldsymbol z^\ast$. Orthogonality of
the binary characters gives
\begin{align}
A(\boldsymbol z)
&=
2^m\delta_{\boldsymbol z,\boldsymbol z^\ast}.
\label{eq:app_sign_orthogonality}
\end{align}
The effective angle $\theta A(\boldsymbol z)$ is therefore $\pi$ for
the selected eigenvalue assignment and zero for every other assignment. The resulting
logical operation contributes a minus sign only to
$\boldsymbol z^\ast$.

Taking
$B_a=\overline Z_a$ and
$\boldsymbol z^\ast=(1,\ldots,1)$ therefore gives
$\overline{\mathrm{C}{}^{m-1}\mathrm{Z}}$. In this case
\begin{align}
\epsilon_{\boldsymbol u}
&=
(-1)^{|\boldsymbol u|}.
\end{align}

Finally, the same argument proves the Pauli generator-count bound quoted in
Sec.~\ref{subsec:equal_angle_hierarchy}. For a uniform-angle
equal-syndrome LGA at $\theta=\pi/2^m$,
$A(\boldsymbol z)$ is a sum of $q$ numbers equal to $\pm1$, so
\begin{align}
|A(\boldsymbol z)|
&\leq q.
\end{align}
Code preservation requires
$A(\boldsymbol z)$ to be divisible by $2^m$ for every eigenvalue
assignment, while a nontrivial logical phase requires
$A(\boldsymbol z)/2^m$ to be odd for at least one assignment. Hence
\begin{align}
|A(\boldsymbol z)|
&\geq 2^m
\end{align}
for some $\boldsymbol z$, and therefore $q
\geq 2^m$. The construction above uses exactly $2^m$ generators and saturates
this bound within the commuting equal-syndrome, uniform-angle setting.

% ------------------------------------------------------------
\subsection{Proof of the syndrome-partition lifting theorem}
\label{app:module_lifting_proof}
% ------------------------------------------------------------

We prove Theorem~\ref{thm:syndrome_module_lifting}.
For stabilizer check $S_a$, every Pauli generator $G_{\mu j}$ of
$W_\mu$ has the same syndrome bit $(\boldsymbol s_\mu)_a$.
Hence $S_a$ either selects the entire doubled-angle operation
\begin{align}
R_\mu
&=
W_\mu^2
\end{align}
or does not select it. The check-conditioned unitary of the combined
LGA is therefore
\begin{align}
V_a
&=
\prod_\mu
R_\mu^{\,(\boldsymbol s_\mu)_a}.
\label{eq:app_module_conditioned}
\end{align}

By assumption, each $R_\mu$ preserves the codespace, with exact
logical action $\overline R_\mu$. Restricting the equation above to
the codespace gives
\begin{align}
V_a|_{\mathcal C}
&=
\prod_\mu
\overline R_\mu^{\,(\boldsymbol s_\mu)_a}.
\end{align}
The arbitrary-angle syndrome-partition theorem then implies that the
combined LGA preserves the codespace if and only if
\begin{align}
\prod_\mu
\overline R_\mu^{\,(\boldsymbol s_\mu)_a}
&=
I
\qquad
\forall a,
\end{align}
which is Eq.~\eqref{eq:module_logical_cancellation}. 

% ============================================================
\section{Clifford syndrome-partition search}
\label{app:clifford_search_methods}
% ============================================================

This appendix describes the general computational procedure used to
identify code-preserving Clifford syndrome partitions. The same machinery underlies the direct qLDPC searches and, with the CAB-guided restrictions introduced later, the hypercube searches. The combinatorial stage works in the binary Pauli group modulo scalar phase. Each accepted support pattern is then lifted to Hermitian Pauli representatives, and a linear system over $\mathbb F_2$ determines rotation signs that preserve the specified $+1$ stabilizer sector. We first formulate the binary stabilizer
factorization problem, then describe the recursive exact-cover
enumeration and the rotation-sign determination.

% ------------------------------------------------------------
\subsection{Stabilizer-factorization search}
\label{app:stabilizer_factorization_search}
% ------------------------------------------------------------

A convenient way to organize an equal-syndrome Clifford search is to
begin with a stabilizer multiplier $R$ represented in the binary Pauli
group modulo scalar phase and a desired number $q$ of Pauli generators. We seek
a partition of its nonidentity support,
\begin{align}
\operatorname{supp}(R)
&=
Q_1\sqcup Q_2\sqcup\cdots\sqcup Q_q,
\label{eq:app_physical_support_partition}
\end{align}
and define the binary Pauli generator $G_j=R|_{Q_j}$ by restricting the Pauli
letter pattern of $R$ to the qubits in $Q_j$. Because the supports are
disjoint, the Pauli generators commute automatically and their phase-free
product is $R$. The remaining combinatorial condition is that the
Pauli generators satisfy the chosen weight restrictions and carry the same
nonzero stabilizer syndrome,
\begin{align}
\boldsymbol{\sigma}_{\boldsymbol S}(G_1)
&=
\cdots
=
\boldsymbol{\sigma}_{\boldsymbol S}(G_q)
\neq
\boldsymbol 0.
\label{eq:app_compilation_equal_syndrome}
\end{align}
At the Clifford angle, such a factorization is a binary candidate for a
code-preserving LGA. The Hermitian phases of the Pauli generators and the
rotation directions are fixed in the later signed codespace verification
step so that every selected stabilizer check returns specifically to
the $+1$ sector.

This multiplier-first organization is particularly effective when the
stabilizer element $R$ is already known or when a manageable set of
candidate multipliers can be enumerated. The complementary
Pauli generator-first organization used in some of the direct qLDPC searches
begins instead with physical Pauli generators, groups them by syndrome, and
tests whether suitable products lie in the stabilizer group. The two
procedures impose the same algebraic preservation condition but explore
the physical search space in different ways.

% ------------------------------------------------------------
\subsection{Admissible Pauli generators and exact-cover enumeration}
\label{app:exact_cover_search}
% ------------------------------------------------------------

For a fixed multiplier $R$, we first enumerate the admissible
Pauli generators. A candidate Pauli generator is specified by a subset
$Q\subseteq\operatorname{supp}(R)$ and inherits from $R$ the physical
Pauli acting on every qubit in $Q$. Candidates are discarded if they
violate the specified weight or any additional support
restriction. Their full stabilizer syndromes are then computed and the candidates
are grouped by nonzero syndrome. The exact-cover recursion can be
applied independently within each syndrome class; when an analytic
construction supplies a target syndrome, the search may instead be
restricted to that class from the outset.

The remaining problem is an exact cover: choose exactly $q$
admissible Pauli generators whose supports are mutually disjoint and whose
union is $\operatorname{supp}(R)$. We solve this problem by recursive
backtracking. At each stage, the algorithm chooses an uncovered
physical qubit and branches over the surviving admissible Pauli generators
containing that qubit. Once one Pauli generator is selected, all candidates
overlapping it are removed from that branch.

The implementation uses a deterministic recursive cover. At each stage
it chooses the lowest-index uncovered qubit and branches over admissible
Pauli generators that contain that qubit and are contained in the remaining
uncovered support. A branch is terminated when the remaining support
contains fewer qubits than the number of Pauli generators still required. In
the final slot, the entire uncovered mask must itself be one of the
admissible Pauli generators. A branch is accepted only when every nonidentity
position of $R$ has therefore been covered exactly once using the
prescribed number of Pauli generators. This recursion enumerates the unordered
covers in the stated finite Pauli generator domain; the current implementation
does not rely on a minimum-remaining-values heuristic.

% ------------------------------------------------------------
\subsection{Choosing the \texorpdfstring{$\pi/4$}{pi/4} rotation signs}
\label{app:rotation_sign_selection}
% ------------------------------------------------------------

A physical factorization determines the Pauli generators of the LGA,
but not yet the signs of the individual rotations. Let $M_{aj}=1$ when
Pauli generator $G_j$ anticommutes with stabilizer generator $S_a$. A check $a$ selects the product
\begin{align}
P_a
&=
\prod_{j:M_{aj}=1}G_j.
\end{align}
For a valid Clifford LGA, this product must reduce, including its
phase, to a stabilizer after the rotation signs are included. For the equal-syndrome candidates considered here, this is the
check-by-check form of the preservation condition in
Eq.~\eqref{eq:general_syndrome_partition_condition}.

For every affected check, the binary product $P_a$ is first required to
represent a stabilizer class. After lifting the Pauli generators to canonical
Hermitian representatives, define $\lambda_a=\pm1$ through
\begin{align}
i^{m_a}P_a
&=
\lambda_a R_a,
\qquad
R_a\in\mathcal S,
\label{eq:app_check_product_phase}
\end{align}
where $m_a=\sum_jM_{aj}$ and $R_a$ carries the phase for which it acts
as $+1$ on the codespace. Writing the unknown rotation directions as
$\epsilon_j=(-1)^{x_j}$ and
$\lambda_a=(-1)^{b_a}$ reduces the preservation conditions to
\begin{align}
M\boldsymbol x
&=
\boldsymbol b
\pmod 2.
\label{eq:app_rotation_sign_linear_system}
\end{align}

We solve this system by Gaussian elimination over $\mathbb F_2$. An
inconsistent system rejects the candidate. If solutions exist, any one
of them gives a valid choice of rotation directions, while the null
space describes the remaining sign freedom. The implementation then verifies the signed LGA directly: every
stabilizer must map back into the stabilizer group with the correct
$+1$ phase. Thus the binary exact-cover enumeration and the signed
codespace verification are separate steps, and the reported physical
support catalogs do not depend on an implicit Pauli-phase convention.

% ============================================================
\section{Direct Clifford syndrome-partition searches on qLDPC codes}
\label{app:qldpc_clifford_search_details}
% ============================================================

This appendix gives the code definitions and finite search domains used
for the direct Clifford results of
Sec.~\ref{subsec:qldpc_clifford_search}. The search does not assume a
target logical gate. We first enumerate Pauli generators within a stated domain and organize them by stabilizer syndrome. We then test
equal-syndrome collections of Pauli generators and verify directly that the
resulting signed $\pi/4$ LGA preserves the $+1$ codespace.
For each code-preserving LGA, we compute its action on the fixed
logical Pauli frame used in the main text.

% ------------------------------------------------------------
\subsection{Code instances}
\label{app:qldpc_code_instances}
% ------------------------------------------------------------

We use the self-dual bivariate-bicycle $[[16,4,4]]$ and
$[[64,8,8]]$ codes of Ref.~\cite{liang2026self}. In both codes,
integer coordinates $u=(u_x,u_y)\in\mathbb Z^2$ label pairs of physical
qubits $v_1$ and $v_2$, denoted $(v_1,u)$ and $(v_2,u)$. The infinite coordinate grid is
made finite by declaring coordinates related by specified translations
to label the same pair of qubits.

For the $[[16,4,4]]$ code,
\begin{align}
u
&\equiv
u+m(0,4)+n(2,2),
\qquad m,n\in\mathbb Z .
\label{eq:app_bb16_lattice_identification}
\end{align}
Thus shifting by $(0,4)$ or by $(2,2)$ returns to the same physical
qubits. Every coordinate can be reduced to one of the eight labels
$(x,y)$ with $x\in\{0,1\}$ and $y\in\{0,1,2,3\}$. Since each label
contains two physical qubits, the code has sixteen qubits. These coordinate relations can be denoted
$\mathbb Z^2/\langle(0,4),(2,2)\rangle$.

For the $[[64,8,8]]$ code,
\begin{align}
u
&\equiv
u+m(0,8)+n(4,4),
\qquad m,n\in\mathbb Z .
\label{eq:app_bb64_lattice_identification}
\end{align}
Every coordinate can then be reduced to one of the thirty-two labels
$(x,y)$ with $x\in\{0,1,2,3\}$ and
$y\in\{0,1,\ldots,7\}$. With two physical qubits at each label, this
gives sixty-four qubits. Equivalently, the coordinate set is written
$\mathbb Z^2/\langle(0,8),(4,4)\rangle$.

Both codes use
\begin{align}
f(x,y)
&=
1+x+y+y^{-1},
\label{eq:app_bb_seed_polynomial}
\end{align}
together with
$\overline f(x,y)=1+x^{-1}+y^{-1}+y$.
Each monomial specifies a displacement from a coordinate $u$: the
four terms of $f$ correspond to
$(0,0)$, $(1,0)$, $(0,1)$, and $(0,-1)$, while the four terms of
$\overline f$ correspond to
$(0,0)$, $(-1,0)$, $(0,-1)$, and $(0,1)$.

The $Z$-type stabilizer associated with coordinate $u$ therefore acts
on four $v_1$ qubits and four $v_2$ qubits,
\begin{align}
S_Z(u)
={}&
Z_{v_1,u}
Z_{v_1,u+(1,0)}
Z_{v_1,u+(0,1)}
Z_{v_1,u+(0,-1)}
\nonumber\\
&
Z_{v_2,u}
Z_{v_2,u+(-1,0)}
Z_{v_2,u+(0,-1)}
Z_{v_2,u+(0,1)} .
\label{eq:app_bb_z_stabilizer}
\end{align}
The corresponding $X$-type stabilizer acts on the same eight physical
qubits with $Z$ replaced by $X$.

Whenever a shifted coordinate falls outside the representative labels
listed above, Eqs.~\eqref{eq:app_bb16_lattice_identification} or
\eqref{eq:app_bb64_lattice_identification} determine which physical
qubit it denotes. Evaluating $S_X(u)$ and $S_Z(u)$ at every distinct
coordinate label then generates the stabilizer checks.

For the non-CSS Mirror $[[36,6,6]]$ code of
Ref.~\cite{khesin2026mirror}, the physical qubits lie on the periodic lattice
$\mathbb Z_6\times\mathbb Z_6$. The lattice patterns used to build the stabilizers are
\begin{align}
A
&=
\{(1,2),(4,3),(4,4)\},
\nonumber\\
B
&=
\{(2,4),(3,1),(4,1)\}.
\label{eq:app_mirror_seed_patterns}
\end{align}
For each lattice site $u$, the corresponding stabilizer is
\begin{align}
S(u)&=
(-i)^{m(u)}Z(A+u)X(B-u),\nonumber\\
&\;\;m(u) = \left | (A+u)\cap(B-u) \right|,
\label{eq:app_mirror_stabilizer}
\end{align}
where the phase $(-i)^{m(u)}$ ensures that each translated stabilizer is Hermitian, and all coordinates taken modulo six. Translating $S(u)$ over the
lattice generates the stabilizer group.

% ------------------------------------------------------------
\subsection{Search domain and signed codespace verification}
\label{app:qldpc_clifford_search_algorithm}
% ------------------------------------------------------------

The direct Clifford searches use four pairwise support-disjoint Pauli
generators with one common nonzero syndrome,
\begin{align}
\boldsymbol{\sigma}_{\boldsymbol S}(G_1)
&=
\cdots
=
\boldsymbol{\sigma}_{\boldsymbol S}(G_4)
\neq
\boldsymbol 0,
\nonumber\\
G_1G_2G_3G_4
&\sim
S,
\qquad
S\in\mathcal S.
\label{eq:app_qldpc_clifford_conditions}
\end{align}
The physical candidates are generated either from stabilizer
multipliers or directly from bounded-weight Pauli generators, depending
on the code. Every surviving partition is then assigned rotation
directions using the signed codespace verification procedure of
Appendix~\ref{app:rotation_sign_selection}, and its logical action is
determined only after exact preservation of the chosen $+1$ codespace
has been verified.

\begin{table*}[t]
\centering
\footnotesize
\renewcommand{\arraystretch}{1.10}

\begin{tabular*}{\textwidth}
{@{\extracolsep{\fill}}l l l@{}}
\toprule
Code & Pauli generators $G_j$ & logical Pauli images \\
\midrule

self-dual BB $[[16,4,4]]$
&
\parbox[t]{0.34\textwidth}{\raggedright
$+X_1X_3$,
$+X_6X_{10}$,
$+X_9X_{11}$,
$+X_2X_{16}$}
&
\parbox[t]{0.38\textwidth}{\raggedright
$Z_1\mapsto -X_3X_4Z_1$,
$Z_2\mapsto -X_3X_4Z_2$,
$Z_3\mapsto -X_1X_2Z_3$,
$Z_4\mapsto -X_1X_2Z_4$}
\\

\midrule

self-dual BB $[[64,8,8]]$
&
\parbox[t]{0.34\textwidth}{\raggedright
$+X_1X_{33}$,
$+X_8X_{18}X_{28}X_{40}X_{41}X_{51}$,
$+X_{19}X_{21}X_{42}X_{43}X_{44}X_{53}$,
$+X_2X_{10}X_{11}X_{20}X_{52}X_{61}$}
&
\parbox[t]{0.38\textwidth}{\raggedright
$Z_1\mapsto -X_2X_4X_8Z_1$,
$Z_2\mapsto -X_1X_5X_7Z_2$,
$Z_4\mapsto -X_1X_5X_7Z_4$,
$Z_5\mapsto -X_2X_4X_8Z_5$,
$Z_7\mapsto -X_2X_4X_8Z_7$,
$Z_8\mapsto -X_1X_5X_7Z_8$}
\\

\midrule

Mirror $[[36,6,6]]$
&
\parbox[t]{0.34\textwidth}{\raggedright
$+X_{15}X_{18}X_{32}X_{35}$,
$+X_{14}X_{17}X_{33}X_{36}$,
$+Z_{19}Z_{22}Z_{25}Z_{28}$,
$+Z_1Z_4Z_7Z_{10}$}
&
\parbox[t]{0.38\textwidth}{\raggedright
$X_1\mapsto X_1X_2Z_3Z_6$,
$X_3\mapsto -Y_1X_2X_3Z_6$,
$X_6\mapsto Y_1X_6Z_3$,
$Z_1\mapsto X_2Z_1Z_3Z_6$,
$Z_2\mapsto Y_1Z_2Z_3$}
\\

\bottomrule
\end{tabular*}

\caption{Representative Pauli-rotation depth one equal-syndrome Clifford
LGAs from the direct qLDPC searches. Each row gives four signed
physical Pauli generators and the exact changed images in a fixed logical Pauli frame, including signs. Omitted logical
generators are unchanged modulo stabilizers.}
\label{tab:external_signed_representatives}
\end{table*}

% ------------------------------------------------------------
\subsection{The \texorpdfstring{$[[16,4,4]]$}{[[16,4,4]]} BB search}
\label{app:bb16_clifford_search}
% ------------------------------------------------------------

For the $[[16,4,4]]$ self-dual BB code, the stabilizer-first search
uses the complete stabilizer group of $2^{12}$ elements. For each stabilizer, every factorization into four pairwise support-disjoint
Pauli generators of weight at most four is enumerated, and only factorizations
with one common nonzero syndrome are retained. This exhausts the stated
finite domain. The search yields $52{,}056$ nontrivial physical
partitions realizing $291$ distinct fixed-frame logical Clifford gates. Solutions occur already at $w_{\max}=2=d/2$.

% ------------------------------------------------------------
\subsection{The \texorpdfstring{$[[64,8,8]]$}{[[64,8,8]]} BB search}
\label{app:bb64_clifford_search}
% ------------------------------------------------------------

For the $[[64,8,8]]$ code, enumerating the full stabilizer group is
impractical. We therefore begin with the stabilizer generators and form
all stabilizers that are products of one, two, or three generators.
Each generator has weight eight, so every stabilizer considered in this
search has weight at most $24$. Products can have smaller weight when
two generators act with the same Pauli on the same qubit, since those
factors cancel.

We next use the translation symmetry of the code to avoid repeating the
same factorization search at different physical locations. As described
in Appendix~\ref{app:qldpc_code_instances}, the $[[64,8,8]]$ code has
thirty-two distinct coordinate labels
\begin{align}
(x,y),
\qquad
x\in\{0,1,2,3\},
\qquad
y\in\{0,\ldots,7\},
\end{align}
with two physical qubits associated with each coordinate.

A translation is specified by
\begin{align}
(\Delta x,\Delta y),
\qquad
\Delta x\in\{0,1,2,3\},
\qquad
\Delta y\in\{0,\ldots,7\}.
\end{align}
The same coordinate shift is applied to both qubits associated with each
coordinate. To translate $(x,y)$, we first add
$(\Delta x,\Delta y)$. If the resulting $x$ coordinate is less than
four, no further change is needed. If it is four or larger, we subtract
four from $x$ and four from $y$, because coordinates differing by
$(4,4)$ label the same physical qubits. The $y$ coordinate is then
reduced modulo eight. For example, translation by $(1,0)$ sends
$(3,y)$ to $(0,y-4)$, with the second coordinate understood modulo
eight.

Applying the same translation to every qubit in a stabilizer gives
another stabilizer with the same weight and the same factorization
problem shifted to a different location in the code. We therefore group
stabilizers that are related by one of these thirty-two translations
and search only one stabilizer from each translation orbit.

For each stabilizer $S$ selected in this way, we enumerate
factorizations
\begin{align}
S
&=
\pm G_1G_2G_3G_4
\end{align}
in which the four Pauli generators are pairwise support-disjoint, each
Pauli generator has weight at most six, and all four Pauli generators have the same
nonzero stabilizer syndrome.

Whenever such a factorization is found, we apply all thirty-two
translations to the four Pauli generators to recover the corresponding
physical realizations at every translated location in the code. If two
translations produce the same unordered set of four physical Pauli generators, that physical realization is counted only once. For every distinct realization,
we then verify directly, including the rotation signs, that the $\pi/4$ LGA preserves the $+1$ codespace, and we compute its logical
Clifford action in the fixed logical Pauli frame.

Within this bounded search domain, we find $3{,}072$ distinct physical
syndrome partitions realizing $112$ fixed-frame logical Clifford
gates. The successful sorted generator weight profiles are
$(2,6,6,6)$ and $(6,6,6,6)$, up to permutation. No partition with
$w_{\max}\leq5$ is found within the same search domain. Thus six is the
smallest value of $w_{\max}$ found under the stated restrictions.

% ------------------------------------------------------------
\subsection{The \texorpdfstring{$[[36,6,6]]$}{[[36,6,6]]} Mirror search}
\label{app:mirror_clifford_search}
% ------------------------------------------------------------

For the non-CSS Mirror code, we use two complementary finite searches.
The first searches directly over all physical Pauli generators of exact
weight four. There are
\begin{align}
\binom{36}{4}3^4
&=
4{,}771{,}305
\end{align}
such Pauli strings, obtained by choosing four physical qubits and
placing an $X$, $Y$, or $Z$ operator on each. We compute the full
stabilizer syndrome of every Pauli generator and group Pauli generators with the same
nonzero syndrome. Each syndrome class is then searched exhaustively for
four pairwise support-disjoint Pauli generators whose product is a stabilizer.

The Mirror code also has four lattice translations that preserve the
stabilizer group, $(0,0),(0,3),(3,0),(3,3)$, with coordinates
understood modulo six. We identify complete four-generator realizations related
by these translations and count each translated set only once. The
exact-weight-four search yields $17$ distinct translation classes of
physical realizations. These realize $17$ distinct logical
Clifford gates, all with generator weight profile $(4,4,4,4)$. Thus
this search exhausts the entire exact-weight-four physical Pauli
domain.

The second search allows the four Pauli generators to have different weights.
We begin with the $36$ stabilizer generators of the Mirror code and
form all stabilizers that can be written as products of one, two, or
three generators. The generators have weight five or six, so every
nonidentity stabilizer in this bounded search has weight at most $18$.

As in the first search, stabilizers related by one of the four lattice
translations above lead to the same factorization problem in shifted
coordinates. We therefore search only one stabilizer from each set of
translation-related stabilizers. This leaves $1{,}965$ distinct
translation classes of stabilizers to be searched. Each stabilizer $S$
is searched for factorizations $S=G_1G_2G_3G_4$ into four pairwise
support-disjoint Pauli generators of weight at most six that share one
common nonzero stabilizer syndrome. Whenever a valid factorization is
found, the complete four-generator realization is translated using the four
lattice symmetries above to generate the corresponding physical
solutions elsewhere in the code. Translation-related copies of the same physical realization are counted only once.

This mixed-weight search yields $30$ distinct translation classes of
physical realizations. Eighteen have generator weight profile $(2,4,6,6)$,
and twelve have profile $(4,4,4,6)$. These $30$ physical classes
realize eight distinct fixed-frame logical Clifford gates, and each
of the eight gates has at least one physical realization with weights $(2,4,6,6)$.

% ============================================================
\section{Unequal-syndrome lifts from Clifford partitions}
\label{app:unequal_lifting_examples}
% ============================================================

The lifting search starts from the $\pi/4$ Clifford partitions of Appendix~\ref{app:qldpc_clifford_search_details}. For each physical partition we record its common nonzero syndrome and its exact logical Clifford action, including phase. Because each partition is an equal-syndrome code-preserving LGA, its logical action squares to the identity on the codespace. We therefore group the partitions according to their exact logical action and collect the distinct syndromes on which each action occurs. For each logical action, we then search for triples or quartets of syndromes whose binary sum vanishes.

\begin{table*}[t]
\centering
\footnotesize
\renewcommand{\arraystretch}{1.10}

\begin{tabular*}{\textwidth}
{@{\extracolsep{\fill}}lrrll@{}}
\hline\hline
Code
& 3-LGA relations
& 4-LGA relations
& Lift type
& Physical realization \\
\hline

$[[16,4,4]]$ BB
& $5{,}112$
& $339{,}513$
& CCZ-type
& $16$ rotations; $w_{\max}=2$, $D=2$ ($8+8$) \\

$[[64,8,8]]$ BB
& $0$
& $7{,}104$
& CCZ-type
& $16$ rotations; $w_{\max}=6$, $D=2$ ($8+8$) \\

$[[36,6,6]]$ Mirror
& $8$
& $6$
& CS-type
& $12$ rotations; $w_{\max}=6$, $D=3$ ($4+4+4$) \\

\hline\hline
\end{tabular*}

\caption{Results of the unequal-syndrome lifting search. For each exact
logical Clifford action found in the direct searches, we collect the
syndromes on which that same action occurs. The 3-LGA and 4-LGA
columns count triples and quartets of distinct syndromes whose binary
sum is zero. The final two columns summarize one verified physical lift
obtained from each code.}
\label{tab:app_unequal_relation_search}
\end{table*}

For BB16, one selected four-LGA construction is
\begin{align}
\mathcal B_A
&=
\{
X_3X_4,\,
X_2X_{13},\,
X_1X_{14},\,
X_{15}X_{16}
\},
\nonumber\\
\mathcal B_B
&=
\{
X_7X_8,\,
X_9X_{10},\,
X_6X_{11},\,
X_5X_{12}
\},
\nonumber\\
\mathcal B_C
&=
\{
X_1X_4,\,
X_3X_{14},\,
X_2X_{15},\,
X_{13}X_{16}
\},
\nonumber\\
\mathcal B_D
&=
\{
X_5X_8,\,
X_6X_9,\,
X_{10}X_{11},\,
X_7X_{12}
\}.
\label{eq:app_bb16_lifting_partitions}
\end{align}
Each set contains four mutually support-disjoint weight-two Pauli
generators. Their common syndromes are distinct and satisfy
\begin{align}
\boldsymbol s_A
\oplus
\boldsymbol s_B
\oplus
\boldsymbol s_C
\oplus
\boldsymbol s_D
&=
\boldsymbol 0.
\label{eq:app_bb16_syndrome_relation}
\end{align}
All four sets implement the same exact
order-two logical Clifford operation. Running the same Pauli generators at angle $\pi/8$ therefore satisfies
Theorem~\ref{thm:syndrome_module_lifting}, and implements a CCZ-type logical operation.  The sixteen physical rotations divide into two Pauli-rotation layers,
\begin{align}
\mathcal L_1
&=
\mathcal B_A\cup\mathcal B_B,
&
\mathcal L_2
&=
\mathcal B_C\cup\mathcal B_D,
\end{align}
so the construction has $w_{\max}=2$ and Pauli-rotation depth $D=2$.

For BB64, a four-LGA relation gives sixteen physical $\pi/8$
rotations with $w_{\max}=6$, arranged into two Pauli-rotation layers of
eight rotations each. The construction therefore has Pauli-rotation
depth two and again yields a CCZ-type operation. For the non-CSS Mirror
code, a three-LGA relation gives twelve physical $\pi/8$ rotations
with $w_{\max}=6$, arranged into three Pauli-rotation layers of four
rotations each. The construction therefore has Pauli-rotation depth
three and implements a CS-type controlled-phase operation.

% ============================================================
\section{Direct non-Clifford searches on qLDPC codes}
\label{app:nonclifford_search_details}
% ============================================================

This appendix specifies the finite search domains underlying the direct
$\pi/8$ qLDPC constructions of
Sec.~\ref{subsec:bb_nonclifford}. We search equal-syndrome
eight-generator families whose relative logical Pauli labels are the
products of three independent commuting logical Paulis.

% ------------------------------------------------------------
\subsection{Search criterion}
\label{app:nonclifford_search}
% ------------------------------------------------------------

For each code below, we first define a finite set of physical Pauli
generators to be searched. The choice of this set differs from code to
code and is stated explicitly in the corresponding subsection.

Within the chosen set, we compute the stabilizer syndrome of every
physical Pauli generator and group together those with the same
nonzero syndrome. For one such syndrome class, choose a generator
$G_0$. If $G$ is any other generator in the same class, then
$G_0G$ has zero syndrome and therefore lies in the Pauli normalizer.
Reducing $G_0G$ modulo stabilizers determines the logical Pauli by
which $G$ differs from $G_0$.

We search for eight generators in one syndrome class whose relative
logical Paulis are
\begin{align}
\{
I,\,
B_1,\,
B_2,\,
B_3,\,
B_1B_2,\,
B_1B_3,\,
B_2B_3,\,
B_1B_2B_3
\},
\label{eq:app_rank_three_logical_family}
\end{align}
for three independent commuting logical Paulis
$B_1,B_2,B_3$. Such a family has the logical structure required by the
rank-three equal-syndrome $\pi/8$ construction of
Sec.~\ref{subsec:bb_nonclifford}. 

Finally, we examine the physical supports of the eight generators.
We first test whether all eight are mutually support-disjoint. If not,
we test whether they can be partitioned into two Pauli-rotation layers

% ------------------------------------------------------------
\subsection{The \texorpdfstring{$[[16,4,4]]$}{[[16,4,4]]} BB code}
\label{app:bb16_nonclifford_search}
% ------------------------------------------------------------

For the $[[16,4,4]]$ self-dual BB code, the search exhausts all
nonidentity physical Pauli strings of weights one through three,
including mixed $X$, $Y$, and $Z$ strings. Every rank-three logical
family contained in a nonzero syndrome class is considered, and every
candidate is tested first for fully disjoint support and then for a
partition into two Pauli-rotation layers.

No Pauli-rotation depth one realization occurs in this search. A
Pauli-rotation depth two solution is found using eight weight-three $X$-type
generators, with supports
\begin{align}
&\{1,2,5\},\quad
\{4,5,15\},\quad
\{11,15,16\},\quad
\{1,11,14\},
\nonumber\\
&\{3,9,14\},\quad
\{9,13,16\},\quad
\{4,7,13\},\quad
\{2,3,7\}.
\label{eq:app_bb16_pi8_supports}
\end{align}
With this ordering, the rotation directions are
\[
(+,-,-,+,-,+,+,-).
\]
The first, third, fifth, and seventh generators are mutually
support-disjoint, as are the remaining four. The eight generators have
one common nonzero syndrome and the relative logical labels of
Eq.~\eqref{eq:app_rank_three_logical_family}. The LGA implements a CCZ-type logical gate. Hence the minimum Pauli-rotation depth
within the complete all-Pauli $w_{\max}\leq3$ domain is two.

% ------------------------------------------------------------
\subsection{The \texorpdfstring{$[[64,8,8]]$}{[[64,8,8]]} BB code}
\label{app:bb64_nonclifford_search}
% ------------------------------------------------------------

For the $[[64,8,8]]$ code, we perform a cumulative pure-$Z$ search at
$w_{\max}=4,5,6$. At each cutoff, every pure-$Z$ Pauli generator with weight
at most $w_{\max}$ is included. The Pauli generators are grouped by nonzero
syndrome, all rank-three logical families in those syndrome classes
are enumerated, and each family is tested first for fully disjoint
support and then for two Pauli-rotation layers.

No rank-three family occurs through $w_{\max}=5$. The first solutions
occur at $w_{\max}=6$. None is fully support-disjoint, but
Pauli-rotation depth two solutions exist. One example solution has supports
\begin{align}
&\{6,7,8,11,27,43\},\quad
\{29,36,37,38,45,61\},
\nonumber\\
&\{25,33,34,40,41,57\},\quad
\{2,3,4,15,31,47\},
\nonumber\\
&\{4,5,6,9,25,41\},\quad
\{27,34,35,36,43,59\},
\nonumber\\
&\{31,38,39,40,47,63\},\quad
\{1,2,8,13,29,45\}.
\label{eq:app_bb64_w6_supports}
\end{align}
With this ordering, the rotation directions are
\[
(+,-,-,+,-,+,+,-).
\]
The first four generators are mutually support-disjoint, as are the
final four. The resulting LGA yields a CCZ-type
logical gate. 

% ------------------------------------------------------------
\subsection{Translation-restricted searches on larger BB codes}
\label{app:large_bb_nonclifford_search}
% ------------------------------------------------------------

For the $[[96,12,8]]$ and $[[128,16,8]]$ self-dual BB codes, searching
over all possible sets of eight physical Pauli generators is too large.
We therefore look for a more structured type of solution built from
pure-$Z$ generators related by lattice translations.

The $[[96,12,8]]$ code is defined on $\mathbb Z^2/\langle(0,12),(4,4)\rangle$ with
\begin{align}
f(x,y)=1+x+x^{-1}y^{-1}+x^{-1}y,
\end{align}
while the $[[128,16,8]]$ code is defined on $\mathbb Z^2/\langle(0,8),(8,4)\rangle$ with
\begin{align}
f(x,y)=1+x+x^2y+x^{-1}y.
\end{align}
In both cases the second polynomial entering the self-dual BB
construction is the spatial inverse of $f$.

We use translations that return to the starting point after four
applications. A translation $t=(\Delta x,\Delta y)$ acts in the same
way on both BB qubit species. Because of the twisted periodic boundary
conditions, if the $x$ period is $L_x$ and
$x+\Delta x=\nu L_x+x'$ with $0\leq x'<L_x$, then
\[
(x,y)
\longmapsto
\left(
x',
\,y+\Delta y-4\nu
\right),
\]
with the second coordinate reduced modulo the corresponding $y$
period.

For BB96, the twelve translations with this four-step property are
\[
\begin{split}
&(0,3),(0,9),(1,1),(1,4),(1,7),(1,10),\\
&(2,5),(2,11),(3,0),(3,3),(3,6),(3,9),
\end{split}
\]
while for BB128 they are
\[
\begin{split}
&(0,2),(0,6),(2,1),(2,3),(2,5),(2,7),\\
&(4,0),(4,4),(6,1),(6,3),(6,5),(6,7).
\end{split}
\]
We search all twelve translations for each code.

For a chosen translation $t$, we start from a pure-$Z$ Pauli string
$F$ and generate the four translated strings
\begin{align}
F,\qquad
tF,\qquad
t^2F,\qquad
t^3F,
\qquad
t^4F=F.
\label{eq:app_order_four_Pauli generator_orbit}
\end{align}
We keep this set only if all four strings carry the same nonzero
stabilizer syndrome. We then compare them logically by multiplying
each translated string by $F$. The four resulting logical classes must
have the form
\[
\{I,B_1,B_2,B_1B_2\}
\]
for two independent commuting logical Paulis $B_1$ and $B_2$.

We next choose a second pure-$Z$ string and repeat the same construction
using the same translation $t$. Its four translated copies must carry
the same syndrome as the first four and must supply the remaining
logical classes
\[
\{B_3,B_1B_3,B_2B_3,B_1B_2B_3\}.
\]
Together, the two sets of four strings then have the eight relative
logical labels of Eq.~\eqref{eq:app_rank_three_logical_family}. The two
starting strings are allowed to have different weights. Every completed
eight-generator family is tested for mutually disjoint physical
support and its signed $\pi/8$ logical action is verified exactly.

For each of the twelve translations, we repeat this search while
increasing the allowed seed weight through
$w_{\max}=4,5,6,7,8$. No solution is found through
$w_{\max}=7$ for either code. The first solutions appear when
weight-eight strings are included, and the resulting eight generators
are mutually support-disjoint.

For the $[[96,12,8]]$ code, one successful choice is
$t=(1,10)$. Two sets of four translated weight-eight $Z$ generators
give eight mutually disjoint Pauli generators with one common
nonzero syndrome. Their product is a positive stabilizer of weight
$64$, and direct verification of the signed $\pi/8$ LGA gives a CCZ-type logical action.

For the $[[128,16,8]]$ code, one successful choice is
$t=(0,2)$. Again, two sets of four translated weight-eight $Z$
generators give eight mutually disjoint Pauli generators with one
common nonzero syndrome. Their product is a positive stabilizer of
weight $64$, and the signed $\pi/8$ LGA is verified as CCZ-type.

The absence of solutions through $w_{\max}=7$ applies only to the
pure-$Z$ translation-based search defined above. It does not rule out
other eight-generator constructions on these codes.

% ------------------------------------------------------------
\subsection{The non-CSS \texorpdfstring{$[[36,6,6]]$}{[[36,6,6]]}
Mirror code}
\label{app:mirror_nonclifford_search}
% ------------------------------------------------------------

For the non-CSS Mirror $[[36,6,6]]$ code, enumerating every mixed-Pauli
string through weight five gives too large a search space. We therefore
use the code's four verified lattice translations, $(0,0), (0,3), (3,0), (3,3)$
with coordinates understood modulo six, to reduce the set of Pauli
strings that enter the search.

For each physical Pauli string, we compute its stabilizer syndrome and
then translate the string by each of the four shifts above. We keep the
string only if all four translated copies have the same stabilizer
syndrome as the original. This is only a restriction used to make the
finite search manageable; the translated copies are not required to
appear together in the final eight-generator gate.

We apply this test separately at each Pauli weight. No strings survive
at weights one, two, or three. At weight four, $243$ strings survive,
and at weight five, $32$ survive. Searching the weight-four set finds
no eight-generator family with the logical structure of
Eq.~\eqref{eq:app_rank_three_logical_family}. When the weight-five
strings are included, the cumulative search contains $275$ physical
Pauli strings, of which $267$ have nonzero syndrome. Two valid
eight-generator families are found. Neither is fully support-disjoint,
but both can be partitioned into two Pauli-rotation layers of four
rotations each.

One verified family is
\begin{align}
G_1&=Z_2Y_{17}Y_{25}X_{30}Y_{31},
&
G_2&=X_9Y_{10}Y_{16}Z_{23}Y_{32},
\nonumber\\
G_3&=Y_2X_{19}Y_{29}Y_{30}Z_{33},
&
G_4&=X_4Y_8Y_9Z_{18}Y_{23},
\nonumber\\
G_5&=Y_7X_{12}Y_{13}Z_{20}Y_{35},
&
G_6&=Z_5Y_{14}X_{27}Y_{28}Y_{34},
\nonumber\\
G_7&=X_1Y_{11}Y_{12}Z_{15}Y_{20},
&
G_8&=Y_5X_{22}Y_{26}Y_{27}Z_{36}.
\label{eq:app_mirror_pi8_Pauli generators}
\end{align}
With this ordering, the rotation directions are $(+,-,-,+,-,+,+,-)$. The eight generators commute and carry the same nonzero stabilizer
syndrome. Taking $G_1$ as the reference, their relative logical
Paulis have the rank-three structure of
Eq.~\eqref{eq:app_rank_three_logical_family}. Their total product is a positive stabilizer of weight $32$, and direct
verification of the signed $\pi/8$ LGA gives a CCZ-type logical action. Their supports divide into
\begin{align}
\{G_1,G_2,G_5,G_6\},
\qquad
\{G_3,G_4,G_7,G_8\},
\label{eq:app_mirror_pi8_support_groups}
\end{align}
so the gate has Pauli-rotation depth two.

\section{CAB algebra and encoded CAB presentations}
\label{app:cab_algebra}
% ============================================================

This appendix gives the algebraic details behind the CAB coordinates
introduced in Sec.~\ref{subsec:cab_definition}. We first show that CAB
monomials provide a complete set of binary coordinates for the Pauli
group modulo phase and relate those coordinates explicitly to the
standard binary symplectic representation. We then collect the
multiplication, commutation, syndrome, and Hermiticity rules used
throughout the analytic constructions. Finally, we show that every
stabilizer code is Clifford-equivalent to a canonical recursive
encoded-CAB.

% ------------------------------------------------------------
\subsection{CAB monomials as Pauli coordinates}
\label{app:cab_coordinate_proof}
% ------------------------------------------------------------

Let
\begin{align}
\Gamma
&=
\{\gamma_1,\ldots,\gamma_{2n}\}
\end{align}
be a CAB on $n$ qubits, so that every $\gamma_i$ is a Hermitian Pauli
operator satisfying
\begin{align}
\{\gamma_i,\gamma_j\}
&=
2\delta_{ij}I.
\end{align}
For a subset
$\mathcal A\subseteq\{1,\ldots,2n\}$, define the ordered monomial
$\gamma_{\mathcal A}$ as in
Eq.~\eqref{eq:cab_monomial}.

We first show that two different CAB supports cannot represent the same
Pauli operator modulo phase. It is enough to prove that no nonempty CAB
monomial is proportional to the identity.

Let $g=|\mathcal A|$. If $j\notin\mathcal A$, moving $\gamma_j$
through the $g$ factors in $\gamma_{\mathcal A}$ produces the sign
$(-1)^g$. If instead $j\in\mathcal A$, it anticommutes with the other
$g-1$ factors and produces $(-1)^{g-1}$. For a nonempty proper subset
$\mathcal A$, there are CAB indices both inside and outside the set.
If $\gamma_{\mathcal A}$ were proportional to the identity, it would
have to commute with both types, forcing both $g$ and $g-1$ to be even,
which is impossible. If $\mathcal A$ contains all $2n$ indices, then
$g=2n$ and the monomial anticommutes with every $\gamma_j$, so it is
again nontrivial.

Thus only the empty CAB product is proportional to the identity.
Consequently, if
$\gamma_{\mathcal A}\doteq\gamma_{\mathcal B}$, then
\begin{align}
\gamma_{\mathcal A\triangle\mathcal B}
&\doteq
I,
\end{align}
where $\triangle$ denotes symmetric difference. The preceding argument
forces
$\mathcal A\triangle\mathcal B=\emptyset$, and therefore
$\mathcal A=\mathcal B$.

There are $2^{2n}=4^n$ possible CAB supports and exactly $4^n$
$n$-qubit Pauli operators modulo phase. Hence
\begin{align}
\mathcal A
&\longmapsto
\gamma_{\mathcal A}
\end{align}
is a bijection between binary CAB supports and the Pauli group modulo
phase. A CAB is therefore a genuine binary coordinate system for the
full Pauli group modulo phase.

% ------------------------------------------------------------
\subsection{Relation to standard symplectic Pauli coordinates}
\label{app:cab_symplectic_coordinates}
% ------------------------------------------------------------

The usual binary representation assigns to an $n$-qubit Pauli operator
a vector
\begin{align}
\boldsymbol p
&=
(\boldsymbol x|\boldsymbol z)
\in
\mathbb F_2^{2n},
\end{align}
with multiplication modulo phase represented by binary vector
addition. Two Paulis with vectors $\boldsymbol p$ and
$\boldsymbol q$ commute or anticommute according to their standard
symplectic product
\begin{align}
\langle\boldsymbol p,\boldsymbol q\rangle
&=
\boldsymbol p^T
\Lambda
\boldsymbol q,
&
\Lambda
&=
\begin{pmatrix}
0&I\\
I&0
\end{pmatrix}.
\label{eq:app_standard_symplectic_form}
\end{align}
Let $\boldsymbol t_i$ denote the standard symplectic coordinate vector
of the CAB generator $\gamma_i$, and collect these vectors as the
columns of
\begin{align}
T_\Gamma
&=
\begin{pmatrix}
\boldsymbol t_1&
\boldsymbol t_2&
\cdots&
\boldsymbol t_{2n}
\end{pmatrix}.
\label{eq:app_cab_coordinate_matrix}
\end{align}
If $\boldsymbol a\in\mathbb F_2^{2n}$ is the binary support vector of
$\mathcal A$, then the corresponding ordinary Pauli coordinate is
simply
\begin{align}
\boldsymbol p
&=
T_\Gamma\boldsymbol a.
\label{eq:app_cab_to_symplectic_coordinates}
\end{align}
The bijection proved above implies that $T_\Gamma$ is invertible.
Thus standard Pauli coordinates and CAB coordinates contain exactly
the same information and are related by an invertible binary change of
basis.

The CAB anticommutation relations determine the symplectic form in
these new coordinates. Define the $2n\times2n$ binary matrix
$\Omega_\Gamma$ by
\begin{align}
(\Omega_\Gamma)_{ij}
&=
\begin{cases}
0,&i=j,\\
1,&i\neq j.
\end{cases}
\label{eq:app_cab_commutation_matrix}
\end{align}
Pairwise anticommutation of distinct CAB generators is equivalent to
\begin{align}
T_\Gamma^T\Lambda T_\Gamma
&=
\Omega_\Gamma.
\label{eq:app_cab_pullback_symplectic_form}
\end{align}
Hence, if $\boldsymbol a$ and $\boldsymbol b$ are the binary support
vectors of two CAB monomials, their Pauli commutation parity is
\begin{align}
\boldsymbol a^T\Omega_\Gamma\boldsymbol b
&=
|\mathcal A||\mathcal B|
-
|\mathcal A\cap\mathcal B|
\pmod 2,
\label{eq:app_cab_commutation_bilinear}
\end{align}
which is precisely the CAB commutation rule used in the main text. CAB
coordinates therefore replace the usual $X/Z$ basis by a binary basis
whose symplectic form has the particularly simple all-to-all
anticommutation structure of
Eq.~\eqref{eq:app_cab_commutation_matrix}.

% ------------------------------------------------------------
\subsection{Multiplication, syndrome, and Hermiticity of CAB monomials}
\label{app:cab_operational_rules}
% ------------------------------------------------------------

Because every CAB generator squares to the identity, multiplication
modulo phase corresponds to addition of CAB support vectors, or
equivalently to symmetric difference of their supports:
\begin{align}
\gamma_{\mathcal A}\gamma_{\mathcal B}
&\doteq
\gamma_{\mathcal A\triangle\mathcal B}.
\label{eq:app_cab_symmetric_difference}
\end{align}

The stabilizer syndrome follows from the same commutation rule. If
\begin{align}
S_a
&\doteq
\gamma_{\mathcal C_a},
\end{align}
then the corresponding syndrome component of
$\gamma_{\mathcal A}$ is
\begin{align}
\sigma_a(\gamma_{\mathcal A})
&=
|\mathcal A||\mathcal C_a|
-
|\mathcal A\cap\mathcal C_a|
\pmod2.
\label{eq:app_cab_syndrome_rule}
\end{align}
For the even CAB-index sets used throughout the CAB constructions,
$|\mathcal C_a|$ is even, so the first term vanishes modulo two. The
syndrome is then determined simply by whether the CAB support overlaps
each CAB-index set in an odd or even number of indices.

The ordered CAB monomial $\gamma_{\mathcal A}$ is not necessarily
Hermitian as written. If $g=|\mathcal A|$, reversing the order of the
$g$ mutually anticommuting Hermitian factors gives
\begin{align}
\gamma_{\mathcal A}^\dagger
&=
(-1)^{g(g-1)/2}
\gamma_{\mathcal A}.
\end{align}
A convenient Hermitian involution proportional to the same Pauli
operator is therefore
\begin{align}
\widehat\gamma_{\mathcal A}
&=
i^{g(g-1)/2}
\gamma_{\mathcal A},
&
\widehat\gamma_{\mathcal A}^2
&=I,
\label{eq:app_cab_hermitian_representative}
\end{align}
up to an additional sign $\pm1$. Throughout the paper, CAB calculations
of multiplication, commutation, and syndrome are performed modulo
phase. When a CAB monomial is used as a physical stabilizer check or
Pauli generator, its Hermitian phase is fixed explicitly; any
remaining $\pm1$ rotation orientation is treated separately.

% ------------------------------------------------------------
\subsection{Encoded CAB presentations and recursive encodings}
\label{app:recursive_cab_scaffolds}
% ------------------------------------------------------------

The CAB construction can be repeated inside an encoded Pauli group.
Suppose an intermediate stabilizer code contains $m$ surviving logical
qubits with a chosen logical Pauli frame. Their encoded Pauli group
modulo phase is isomorphic to the Pauli group on $m$ ordinary qubits
and therefore admits an encoded CAB
\begin{align}
\widetilde\Gamma
&=
\{\widetilde\gamma_1,\ldots,\widetilde\gamma_{2m}\}.
\end{align}
Encoded stabilizer constraints and the remaining logical degrees of
freedom can consequently be organized using the same CAB-index set
language as at the physical level.

The recursive hypercube construction of the main text uses this fact
with additional physical structure. The following proposition records
the more general algebraic existence statement. It does not assert
that the required Clifford change of frame is local, low weight, or
geometrically natural.

\begin{proposition}[Recursive CAB encoding up to Clifford equivalence]
\label{prop:recursive_cab_scaffold_arbitrary_rank}
Every $[[n,k]]$ stabilizer code is Clifford-equivalent to a canonical
recursive encoding in which stabilizer constraints are introduced by
pairwise parity checks on the surviving encoded qubits. At every
nonterminal encoding step, the checks introduced in that round can
be represented as pairwise-disjoint quartic CAB-index sets of a CAB
for the current encoded Pauli group.
\end{proposition}

To construct the canonical code, begin with $n$ unconstrained
qubit degrees of freedom. Choose any collection of disjoint pairs. For
a pair $(a,b)$, impose the parity check $Z_aZ_b$.
This removes one degree of freedom while leaving one encoded qubit
associated with the pair. For example, $Z_a$ may be retained as a
logical $Z$, while $X_aX_b$ serves as a logical $X$. Unpaired qubits
are carried forward unchanged.

The surviving objects again form a Pauli group on fewer qubits, so the
same construction can be repeated at the encoded level. Each pairwise
parity check reduces the number of surviving encoded qubits by one and
adds one independent stabilizer. Repeating the procedure reduces the
number of degrees of freedom from $n$ to any prescribed $k>0$ while
introducing exactly $n-k$ independent checks. For $k=0$, one may
continue until a single encoded qubit remains and fix it with one final
encoded Pauli check.

At an encoding step with $m$ surviving encoded qubits, choose an
encoded CAB so that each encoded $Z_\alpha$ is represented, modulo
phase, by a quadratic CAB monomial,
\begin{align}
\overline Z_\alpha
&\doteq
i\widetilde\gamma_{2\alpha-1}
\widetilde\gamma_{2\alpha}.
\end{align}
Such a choice is always possible, for example by applying a
Jordan--Wigner CAB construction to the chosen encoded Pauli frame.
The pair check
$\overline Z_\alpha\overline Z_\beta$ is then represented by a
quartic CAB monomial. Checks introduced from disjoint encoded pairs in
the same encoding step therefore occupy disjoint quartic CAB-index
sets.

The resulting code is an $[[n,k]]$ stabilizer code. Conversely,
standard stabilizer normal form implies that the stabilizer group of
any $[[n,k]]$ code is related to this canonical stabilizer group by a
Clifford transformation
\cite{gottesman1997stabilizer,calderbank1998quantum}. If $U$ is such a
Clifford, conjugating the canonical checks, logical frame, and CAB
generators by $U$ transfers the recursive encoded-CAB presentation to
the desired code.

This establishes the existence of recursive encoded-CAB coordinates for
an arbitrary stabilizer code. Conjugation by a Clifford encoding circuit preserves the CAB
anticommutation relations, the quartic index sets, and all algebraic
relations among the corresponding checks and logical operators. However, it does
not preserve their physical simplicity. The conjugated CAB generators
may become high-weight and nonlocal, and monomials supported on
disjoint CAB-index sets may overlap on the same physical qubits.
Therefore this existence result is algebraic. The useful additional
property of the codes studied in the main text is that their relevant
CAB monomials also admit low-weight, support-structured physical Pauli strings.

\section{CAB-index templates and logical Clifford structure}
\label{app:cab_templates}
% ============================================================

This appendix gives the general CAB-index constructions underlying
Sec.~\ref{subsec:cab_index_constructions}. We first extend the
perfect-matching construction from two quartic CAB-index sets to an arbitrary
number of even sets of equal size and derive the resulting complement action. We then
give the corresponding multi-control extension of the nested-index
construction, which produces multi-source CNOTs. Finally, we describe the
action of CAB-index relabelings on the three logical Pauli axes of a
quartic CAB-index set and explain how these ingredients organize the logical
Clifford group algebraically. Throughout this appendix, let
\begin{align}
\mathcal C_a
&=
\{c_{a,1},c_{a,2},c_{a,3},c_{a,4}\}
\end{align}
denote pairwise-disjoint quartic CAB-index sets. Each set defines a
stabilizer check $S_a\doteq\gamma_{\mathcal C_a}$ and one logical
qubit. We use the same logical-frame convention as in the main text,
so that the six quadratic pairs within a set fall into the three
classes associated with $\overline X_a$, $\overline Y_a$, and
$\overline Z_a$.

The CAB-index constructions below determine the syndrome pattern and
the logical Clifford action modulo stabilizers and Pauli phases.  The
rotation directions $\epsilon_j=\pm1$ are fixed separately by the
signed equal-syndrome $\pi/4$ preservation condition of
Eq.~\eqref{eq:general_syndrome_partition_condition}.  This distinction
is important: the CAB-index pattern determines the algebraic gate
structure, while the sign choice ensures preservation of the specified
$+1$ stabilizer eigenspace.

% ------------------------------------------------------------
\subsection{Perfect matching and complement gates}
\label{app:cab_matching_complement}
% ------------------------------------------------------------

Consider first an even set of selected logical qubits
$\mathcal R$. For each CAB position $j=1,\ldots,4$, take the
corresponding index from every selected set,
\begin{align}
\mathcal P_j
&=
\{c_{a,j}:a\in\mathcal R\}.
\label{eq:app_general_matching_sets}
\end{align}
The associated Pauli generators are
$G_j\doteq\gamma_{\mathcal P_j}$. Each Pauli generator contains exactly one
CAB index from every selected CAB-index set and none from the
remaining sets. Since the CAB-index sets have even size, the CAB
commutation rule shows that every $G_j$ anticommutes with precisely the
checks in $\mathcal R$. Thus all four Pauli generators have the same nonzero
syndrome.

The four CAB-index sets $\mathcal P_j$ are pairwise disjoint.  This is
an index-space statement and should not be confused with disjoint
physical-qubit support after the CAB monomials are mapped to Pauli
strings.  Their CAB monomials commute because $|\mathcal R|$ is even:
for two different positions $j$ and $k$, the index sets are disjoint
and both have even cardinality. Finally, every CAB index of every
selected set occurs once among the four Pauli generators, so
\begin{align}
\mathcal P_1\triangle\mathcal P_2
\triangle\mathcal P_3\triangle\mathcal P_4
&=
\bigtriangleup_{a\in\mathcal R}\mathcal C_a.
\label{eq:app_matching_product}
\end{align}
The product of the four Hermitian Pauli representatives is therefore
proportional to the stabilizer product
$\prod_{a\in\mathcal R}S_a$.  The four Pauli generators consequently define
an equal-syndrome Clifford template; choosing their rotation directions
to satisfy Eq.~\eqref{eq:general_syndrome_partition_condition} gives an
exact codespace-preserving $\pi/4$ LGA.

The projective logical action follows directly from the same index
structure. Consider a quadratic representative of one logical Pauli
axis on a selected set $a$. If its CAB pair occupies positions $j$
and $k$, it anticommutes with precisely the two matching Pauli generators
$G_j$ and $G_k$. Conjugation by those two rotations multiplies it,
modulo stabilizers and Pauli phase, by $G_jG_k$. The two CAB indices
belonging to set $a$ cancel, leaving the same quadratic pair on every
other selected set. Hence, for any logical Pauli axis
$P\in\{X,Y,Z\}$,
\begin{align}
\overline P_a
&\longmapsto
\prod_{\substack{b\in\mathcal R\\b\neq a}}
\overline P_b,
\qquad
a\in\mathcal R,
\label{eq:app_general_complement_action}
\end{align}
while logical operators on sets outside $\mathcal R$ are fixed. For
$|\mathcal R|=2$, Eq.~\eqref{eq:app_general_complement_action}
exchanges the two logical qubits and gives the SWAP construction of the
main text. For larger even sets it gives the complement transformation.
The exact Pauli signs in the logical images are fixed by the chosen
rotation directions.

The requirement that $|\mathcal R|$ be even is essential for this
four-generator construction. If an odd number of sets were selected,
the sets $\mathcal P_j$ would have odd cardinality and disjoint CAB
monomials from different positions would anticommute.

% ------------------------------------------------------------
\subsection{Nested index sets and multi-source CNOTs}
\label{app:cab_nested_fanout}
% ------------------------------------------------------------

The nested construction admits a similarly direct multi-set
extension. Let $\mathcal T$ be a set of control logical qubits and let
$t$ denote a target qubit. Within every control set $s\in\mathcal T$,
choose a two-index set $\mathcal A_s$ representing the logical
$\overline Z_s$ axis. Define
\begin{align}
\mathcal A
&=
\bigcup_{s\in\mathcal T}\mathcal A_s.
\end{align}
Inside the target set choose two nested odd subsets
\begin{align}
\mathcal B_q
&\subset
\mathcal B_r
\subset
\mathcal C_t,
\end{align}
with $|\mathcal B_q|=1$ and $|\mathcal B_r|=3$, such that
$\mathcal B_q\triangle\mathcal B_r$ represents
$\overline X_t$. The four Pauli generator index sets are
\begin{align}
\mathcal P_1
&=
\mathcal B_q,
&
\mathcal P_2
&=
\mathcal A\cup\mathcal B_q,
\nonumber\\
\mathcal P_3
&=
\mathcal A\cup\mathcal B_r,
&
\mathcal P_4
&=
\left(
\bigcup_{s\in\mathcal T}\mathcal C_s
\right)
\cup
\mathcal B_r.
\label{eq:app_general_fanout_sets}
\end{align}

Each Pauli generator has even CAB-index overlap with every source
set and odd overlap with the target set. Thus all four commute with
the control checks, anticommute with the target check, and share the same
nonzero syndrome. The nesting also guarantees pairwise commutation of
the Pauli generators. Every $\mathcal P_j$ has odd cardinality, while the
intersection of every pair has odd cardinality, so the CAB commutation
parity
$|\mathcal P_i||\mathcal P_j|-|\mathcal P_i\cap\mathcal P_j|$
vanishes modulo two.

Under symmetric difference, the repeated control pairs and both target
subsets cancel, leaving
\begin{align}
\mathcal P_1\triangle\mathcal P_2
\triangle\mathcal P_3\triangle\mathcal P_4
&=
\bigtriangleup_{s\in\mathcal T}\mathcal C_s.
\label{eq:app_fanout_product}
\end{align}
The product of the four Hermitian representatives is therefore
proportional to the stabilizer product
$\prod_{s\in\mathcal T}S_s$.  As above, the CAB-index sets define the
Clifford template and the rotation directions are then chosen to
satisfy the exact $+1$-codespace phase condition.

With the stated choice of control $\overline Z$ axes and target
$\overline X$ axis, the projective logical action is
\begin{align}
\overline X_s
&\longmapsto
\overline X_s\overline X_t,
&
\overline Z_s
&\longmapsto
\overline Z_s,
\qquad s\in\mathcal T,
\nonumber\\
\overline X_t
&\longmapsto
\overline X_t,
&
\overline Z_t
&\longmapsto
\left(
\prod_{s\in\mathcal T}\overline Z_s
\right)
\overline Z_t.
\label{eq:app_general_fanout_action}
\end{align}
Thus, up to the Pauli phases fixed by the rotation signs,
\begin{align}
\overline{\mathrm{CNOT}}_{\mathcal T\rightarrow t}
&=
\prod_{s\in\mathcal T}
\overline{\mathrm{CNOT}}_{s\rightarrow t}.
\end{align}
For a single control set, the construction reduces to the directed
CNOT template of the main text. The choices of $\overline Z_s$ and
$\overline X_t$ are a frame convention. Other choices of quadratic
pairs within the control and target sets give locally
Clifford-related versions of the same nested construction. We make this
freedom explicit in the next subsection.

% ------------------------------------------------------------
\subsection{Index relabeling and logical Pauli axes}
\label{app:cab_relabelings}
% ------------------------------------------------------------

A quartic CAB-index set contains six quadratic pairs, but multiplication by
the quartic stabilizer identifies complementary pairs. The three
resulting classes may be written
\begin{align}
X &: \quad
\{1,2\}\sim\{3,4\},
\nonumber\\
Y &: \quad
\{1,3\}\sim\{2,4\},
\nonumber\\
Z &: \quad
\{2,3\}\sim\{1,4\}.
\label{eq:app_quartic_axis_classes}
\end{align}
Permuting the four CAB indices preserves the quartic CAB-index set and permutes
these three quadratic classes. The full permutation group $S_4$
therefore acts on the three logical Pauli axes. The permutations that
act trivially on all three classes form the Klein four group
\begin{align}
V_4
&=
\{
e,\,
(12)(34),\,
(13)(24),\,
(14)(23)
\}.
\end{align}
Consequently, the nontrivial action on logical axes is
\begin{align}
S_4/V_4
&\simeq
S_3.
\label{eq:app_s4_v4_axis_action}
\end{align}
Thus every permutation of the logical $X$, $Y$, and $Z$ axes can be
represented by a relabeling of the four CAB indices. For reference, one
convenient set of representatives is shown in
Table~\ref{tab:app_cab_axis_relabelings}.

\begin{table}[t]
\centering
\caption{Representative quartic CAB-index permutations and their
induced permutations of the three logical Pauli-axis classes. Other
representatives differ by elements of the kernel $V_4$.}
\label{tab:app_cab_axis_relabelings}

\begin{tabular*}{\columnwidth}
{@{\extracolsep{\fill}}cc@{}}
\hline\hline
CAB-index permutation & Logical-axis action \\
\hline
$e$     & $(X,Y,Z)\mapsto(X,Y,Z)$ \\
$(12)$  & $(X,Y,Z)\mapsto(X,Z,Y)$ \\
$(23)$  & $(X,Y,Z)\mapsto(Y,X,Z)$ \\
$(13)$  & $(X,Y,Z)\mapsto(Z,Y,X)$ \\
$(123)$ & $(X,Y,Z)\mapsto(Z,X,Y)$ \\
$(132)$ & $(X,Y,Z)\mapsto(Y,Z,X)$ \\
\hline\hline
\end{tabular*}

\end{table}

This statement should be interpreted carefully. An index relabeling is
first of all a change in how the CAB presentation identifies quadratic
pairs with logical Pauli axes. It is not, by itself, a claim that a
physical single-qubit Clifford gate has been applied to the encoded
state. Its usefulness for gate construction comes from applying a CAB
template after the relabeling. A relabeled matching pairs different
logical Pauli axes between two sets and therefore produces a locally
Clifford-dressed SWAP. Similarly, relabeling a nested template allows any quadratic logical axis of a control set to replace the $\overline Z$ axis in Eq.~\eqref{eq:app_general_fanout_sets}, and any target-set axis to replace $\overline X$. Applying the template after relabeling selects a different set of physical CAB monomials and yields the corresponding locally Clifford-dressed transformation.

% ============================================================
\section{CAB-derived \texorpdfstring{$[[4m,2m,2]]$}{[[4m,2m,2]]}
code family}
\label{app:cab_block_code_family}
% ============================================================

This appendix gives the explicit CAB-derived code family referenced in
Sec.~\ref{subsec:cab_index_constructions}. The construction provides a simple
example in which the CAB-index structure can be translated directly
into useful physical support properties. For every positive integer
$m$, it produces a $[[4m,2m,2]]$ stabilizer code with fully addressable
logical SWAPs and complement gates on arbitrary even sets of logical
qubits. Each such gate is realized by four mutually disjoint Pauli
rotations, with maximum Pauli generator weight at most $m$.

% ------------------------------------------------------------
\subsection{Code construction and parameters}
\label{app:cab_block_code_construction}
% ------------------------------------------------------------

Set $N=4m$, $r=2m$
and use the reversed Jordan--Wigner CAB
\begin{align}
\gamma_{2j-1}
&=
Z^{N-j}XI^{j-1} \nonumber\\
\gamma_{2j}
&=
Z^{N-j}YI^{j-1}
\qquad
j=1,\ldots,N.
\label{eq:app_reverse_jw_cab}
\end{align}
The $2N=8m$ CAB indices are divided into $r=2m$ disjoint quartic
CAB-index sets
\begin{align}
\mathcal C_t
&=
\{t,t+r,t+2r,t+3r\},
\qquad
t=1,\ldots,r.
\label{eq:app_4m_cab_blocks}
\end{align}
Each set defines a Hermitian stabilizer check
$S_t=\eta_t\gamma_{\mathcal C_t}$, with the phase $\eta_t$ chosen so
that the desired codespace is the $+1$ eigenspace.

The sets are disjoint and have even cardinality, so the corresponding
CAB monomials commute. They are also independent since a product of any
nonempty collection of the checks contains the union of the
corresponding CAB-index sets and therefore cannot be proportional to the
identity. The code consequently has $r=2m$ independent stabilizers on
$N=4m$ physical qubits and encodes $k=N-r=2m$
logical qubits.

A convenient logical Pauli frame is obtained directly within each
quartic CAB-index set. For example, one may choose
\begin{align}
\overline X_t
&\doteq
i\gamma_t\gamma_{t+r},
&
\overline Z_t
&\doteq
i\gamma_{t+r}\gamma_{t+2r}.
\label{eq:app_4m_logical_frame}
\end{align}
Both operators overlap $\mathcal C_t$ evenly and hence commute with
$S_t$, while they have no CAB-index overlap with any other check. Their
CAB supports intersect in one index, so they anticommute with one
another. Logical operators belonging to different sets commute.
Thus each quartic CAB-index set organizes one of the $2m$ encoded logical
qubits in the manner described in
Appendix~\ref{app:cab_templates}.

% ------------------------------------------------------------
\subsection{Distance}
\label{app:cab_block_code_distance}
% ------------------------------------------------------------

The family has distance two. We first rule out any weight-one operator
in the Pauli normalizer.

In the reversed Jordan--Wigner coordinates, a single-qubit $X$ or $Y$
operator has an odd number of CAB generators in its monomial
representation. The quartic sets $\mathcal C_t$ partition the entire
CAB, so at least one set must have odd overlap with that monomial.
Since the CAB-index sets have even size, odd CAB overlap is exactly
the condition for anticommutation. Every physical weight-one $X$ or
$Y$ therefore has nonzero stabilizer syndrome. A physical $Z_j$ is, up to phase,
$i\gamma_{2j-1}\gamma_{2j}$. The two CAB indices belong to distinct
quartic CAB-index sets in Eq.~\eqref{eq:app_4m_cab_blocks}, so $Z_j$ has odd
overlap with each of those two checks and likewise has nonzero
syndrome. Thus no nonidentity weight-one Pauli lies in the normalizer.

A weight-two normalizer element is obtained by pairing corresponding
physical qubits in the two halves of the code. Direct multiplication
of the four Jordan--Wigner generators in
Eq.~\eqref{eq:app_4m_cab_blocks} shows that every stabilizer has two
identical Pauli patterns separated by $2m$ qubits. Consequently, $X_jX_{j+2m}$ with $1\leq j\leq2m$, commutes with every stabilizer check: any local anticommutation with
one copy of the stabilizer pattern is repeated in the other copy.

These operators are not stabilizers. In CAB coordinates, stabilizer
elements are sums of complete quartic CAB-index set-incidence vectors, whereas
the incidence vector of
$X_jX_{j+2m}$ is not a union of complete sets. It therefore
represents a nontrivial logical Pauli. Hence $d\leq2$, while the
weight-one argument gives $d\geq2$, proving $\mathcal Q_m=[[4m,2m,2]]$

% ------------------------------------------------------------
\subsection{Row-cell structure of the Jordan--Wigner CAB}
\label{app:4m_row_cell_structure}
% ------------------------------------------------------------

The useful physical weight bounds follow from a simple geometric
property of the chosen Jordan--Wigner ordering. Arrange the CAB
indices in four rows of length $r=2m$,
\begin{align}
\gamma_{t+(j-1)r},
\qquad
t=1,\ldots,r,
\qquad
j=1,\ldots,4.
\end{align}
Each row corresponds to a disjoint set of $m$ physical qubits. Within a
row, the two CAB indices $2c-1$ and $2c$ occupy the same physical
\emph{row cell}. It is therefore convenient to define
\begin{align}
c(t)
&=
\left\lceil\frac{t}{2}\right\rceil,
\qquad
c(t)\in\{1,\ldots,m\}.
\label{eq:app_row_cell_coordinate}
\end{align}

Consider two CAB indices $t$ and $u$ in the same row. In the product of
their Jordan--Wigner strings, the common $Z$ tails cancel everywhere
outside the interval between their row cells. The resulting Pauli
therefore has support precisely on that interval, giving
\begin{align}
\operatorname{wt}
\left(
i\gamma_{t+(j-1)r}
\gamma_{u+(j-1)r}
\right)
&=
|c(t)-c(u)|+1.
\label{eq:app_row_cell_pair_weight}
\end{align}
When the two CAB indices occupy the same row cell, the product reduces
to a single physical $Z$ and has weight one. Equation~\eqref{eq:app_row_cell_pair_weight} is the physical support
law behind the gate constructions below. Since a row contains only
$m$ physical cells, any such row-supported Pauli generator has weight at most $m=\frac{N}{4}$.

% ------------------------------------------------------------
\subsection{Fully addressable logical SWAPs}
\label{app:4m_swap_family}
% ------------------------------------------------------------

Choose any two sets $\mathcal C_t$ and $\mathcal C_u$. Their
row-wise perfect matching gives the four Pauli generators
\begin{align}
P_j^{(t,u)}
&=
i\gamma_{t+(j-1)r}
\gamma_{u+(j-1)r},
\qquad
j=1,\ldots,4.
\label{eq:app_4m_swap_Pauli generators}
\end{align}
Each Pauli generator contains one CAB index from $\mathcal C_t$ and one from
$\mathcal C_u$, so all four carry the same syndrome, supported on the
two checks $S_t$ and $S_u$. Their CAB-index sets are pairwise disjoint,
and because every Pauli generator has even CAB grade, the four Pauli operators
commute.

Their product contains every index of the two quartic CAB-index sets exactly
once and therefore satisfies
\begin{align}
\prod_{j=1}^{4}P_j^{(t,u)}
&\doteq
S_tS_u.
\end{align}
This is precisely the perfect-matching construction of
Appendix~\ref{app:cab_matching_complement}, so the corresponding signed
$\pi/4$ LGA implements the logical SWAP between encoded qubits $t$
and $u$.

The physical support is particularly simple. Each $P_j^{(t,u)}$ lives
in a different Jordan--Wigner row, so the four rotations are mutually
disjoint in physical-qubit support. Moreover,
Eq.~\eqref{eq:app_row_cell_pair_weight} gives
\begin{align}
\operatorname{wt}\left(P_j^{(t,u)}\right)
&=
|c(t)-c(u)|+1
\leq
m
=
\frac N4.
\label{eq:app_4m_swap_weight}
\end{align}
Thus every pair of the $2m$ encoded qubits admits an addressable
SWAP LGA with Pauli-rotation depth one with four Pauli generators.

The weight distribution can also be read off directly from the row
cells. There are $m$ pairs occupying the same cell and hence
requiring weight-one Pauli generators. For each
$2\leq w\leq m$, there are $4(m-w+1)$ pairs whose four SWAP Pauli generators have weight $w$. These
counts sum to $\binom{2m}{2}$, confirming that every logical pair is
represented.

\begin{table*}[t]
\centering
\footnotesize
\renewcommand{\arraystretch}{1.12}

\begin{tabular*}{\textwidth}
{@{\extracolsep{\fill}}l l c l@{}}
\toprule
Logical operation
& Physical Pauli generators
& $w_{\max}$
& Stabilizer-check action \\
\midrule

$\overline{\mathrm{SWAP}}_{\mathrm{AB}}$
&
$Z_2,\ Z_4,\ Z_6,\ Z_8$
& 1
&
$S_{\mathrm A}\leftrightarrow S_{\mathrm B}$
\\

$\overline{\mathrm{SWAP}}_{\mathrm{AC}}$
&
$Y_1X_2,\ Y_3X_4,\ Y_5X_6,\ Y_7X_8$
& 2
&
$S_{\mathrm A}\leftrightarrow S_{\mathrm C}$
\\

$\overline{\mathrm{SWAP}}_{\mathrm{AD}}$
&
$X_1X_2,\ X_3X_4,\ X_5X_6,\ X_7X_8$
& 2
&
$S_{\mathrm A}\leftrightarrow S_{\mathrm D}$
\\

$\overline{\mathrm{SWAP}}_{\mathrm{BC}}$
&
$Y_1Y_2,\ Y_3Y_4,\ Y_5Y_6,\ Y_7Y_8$
& 2
&
$S_{\mathrm B}\leftrightarrow S_{\mathrm C}$
\\

$\overline{\mathrm{SWAP}}_{\mathrm{BD}}$
&
$X_1Y_2,\ X_3Y_4,\ X_5Y_6,\ X_7Y_8$
& 2
&
$S_{\mathrm B}\leftrightarrow S_{\mathrm D}$
\\

$\overline{\mathrm{SWAP}}_{\mathrm{CD}}$
&
$Z_1,\ Z_3,\ Z_5,\ Z_7$
& 1
&
$S_{\mathrm C}\leftrightarrow S_{\mathrm D}$
\\

\midrule

$\overline{\mathrm{COMP}}_{\mathrm{ABCD}}$
&
$Z_1Z_2,\ Z_3Z_4,\ Z_5Z_6,\ Z_7Z_8$
& 2
&
$\displaystyle
S_a\mapsto
\prod_{b\in\{\mathrm A,\mathrm B,\mathrm C,\mathrm D\}\setminus\{a\}}
S_b$
\\
\bottomrule
\end{tabular*}

\caption{Representative four-generator Clifford gates for the
$[[8,4,2]]$ member of the CAB-derived family. The six SWAP rows
demonstrate full pairwise addressability of the four encoded qubits.
The final row realizes the even-set complement on all four logical qubits.
Within every row of the table, the four physical Pauli generators have
pairwise-disjoint support.}
\label{tab:842_representative_layers}
\end{table*}

% ------------------------------------------------------------
\subsection{Even-set complement gates}
\label{app:4m_complement_family}
% ------------------------------------------------------------

The same organization realizes the complement template on multiple CAB-index sets. Let $\mathcal R\subseteq\{1,\ldots,2m\}$ contain an even
number of logical qubits. In each CAB row define
\begin{align}
Q_j^{(\mathcal R)}
&\doteq
\prod_{t\in\mathcal R}
\gamma_{t+(j-1)r},
\qquad
j=1,\ldots,4,
\label{eq:app_4m_complement_Pauli generators}
\end{align}
with the overall phase chosen to make each Pauli generator Hermitian.

Every $Q_j^{(\mathcal R)}$ contains one CAB index from each selected logical qubit and therefore carries the same syndrome, with one syndrome bit
on every check in $\mathcal R$. Since $|\mathcal R|$ is even, the four
row Pauli generators have even CAB grade. Their CAB supports are disjoint, so
they commute. Their product is
\begin{align}
\prod_{j=1}^{4}Q_j^{(\mathcal R)}
&\doteq
\prod_{t\in\mathcal R}S_t.
\label{eq:app_4m_complement_product}
\end{align}
They therefore realize the even-set complement action derived in
Appendix~\ref{app:cab_matching_complement}. For any logical Pauli axis
$P\in\{X,Y,Z\}$ on a selected logical qubit,
\begin{align}
\overline P_t
&\longmapsto
\prod_{\substack{u\in\mathcal R\\u\neq t}}
\overline P_u.
\label{eq:app_4m_complement_action}
\end{align}

The Jordan--Wigner support remains favorable even when many logical
qubits participate. Order the selected CAB positions in one row as
$t_1<\cdots<t_{2s}$. Pairing consecutive Jordan--Wigner factors shows
that the physical support of $Q_j^{(\mathcal R)}$ is the symmetric
difference of the row-cell intervals joining
$c(t_1)$ to $c(t_2)$, $c(t_3)$ to $c(t_4)$, and so forth. Repeated
endpoints cancel automatically. Most importantly, this entire support
remains inside a single row of $m$ physical qubits. Hence
\begin{align}
\operatorname{wt}
\left(
Q_j^{(\mathcal R)}
\right)
&\leq
m
=
\frac N4.
\label{eq:app_4m_complement_weight_bound}
\end{align}
The four Pauli generators occupy four different physical rows and are
therefore mutually disjoint. Thus every even-set complement gate in
this family has Pauli-rotation depth one and obeys the same $N/4$
maximum-generator weight bound as the pairwise SWAP construction.

% ------------------------------------------------------------
\subsection{The \texorpdfstring{$[[8,4,2]]$}{[[8,4,2]]} example}
\label{app:842_cab_family_example}
% ------------------------------------------------------------

The first four-logical-qubit example has $m=2$, so $N=8$ and $r=4$. Label the
four CAB-index sets
\begin{align}
\mathcal C_{\mathrm A}
&=
\{1,5,9,13\},
&
\mathcal C_{\mathrm B}
&=
\{2,6,10,14\},
\nonumber\\
\mathcal C_{\mathrm C}
&=
\{3,7,11,15\},
&
\mathcal C_{\mathrm D}
&=
\{4,8,12,16\}.
\label{eq:app_842_cab_blocks}
\end{align}
With the reversed Jordan--Wigner CAB, one convenient choice of
stabilizer generators is
\begin{align}
S_{\mathrm A}
&=
IYZXIYZX,
&
S_{\mathrm B}
&=
IXZYIXZY,
\nonumber\\
S_{\mathrm C}
&=
YZXIYZXI,
&
S_{\mathrm D}
&=
XZYIXZYI.
\label{eq:app_842_stabilizers}
\end{align}

Here each Jordan--Wigner row contains only two physical qubits, so the
general bound gives $w_{\max}\leq2$. Table
\ref{tab:842_representative_layers} lists the six pairwise logical
SWAPs and the four-logical-qubit complement. Rotation signs are omitted from
the table; they are chosen using the Clifford sign condition of
Appendix~\ref{app:rotation_sign_selection}.

This example makes the general scaling transparent. The abstract
perfect-matching and complement templates of
Appendix~\ref{app:cab_templates} become ordinary physical Pauli strings
whose supports remain separated into four Jordan--Wigner rows. The
family therefore provides an explicit code--gate co-design example in
which CAB-index disjointness survives as physical disjointness while
the largest required Pauli generator grows only as $N/4$.

% ============================================================
\section{CAB-guided searches for physical gate realizations}
\label{app:hypercube_compilation}
% ============================================================

The CAB constructions of Sec.~\ref{sec:analytic_constructions} provide
analytic solutions to the syndrome-partition conditions at the level of
CAB indices. To obtain favorable physical Pauli realizations, we use
these analytic constructions to guide a bounded equal-syndrome
factorization search.

The numerical problem is closely related to the direct equal-syndrome
searches used for the qLDPC codes. There, we searched a finite set of
stabilizers for factorizations into commuting physical Pauli generators
with one common nonzero syndrome. Here the CAB construction supplies
additional information before the physical search begins by specifying
both the stabilizer product to be factorized and the common syndrome
that the physical Pauli generators should carry. The numerical search then
looks for low-weight, support-disjoint Pauli factorizations satisfying
those CAB-derived conditions. 

% ------------------------------------------------------------
\subsection{CAB-guided equal-syndrome factorization}
\label{app:cab_guided_compilation}
% ------------------------------------------------------------

A CAB template specifies a stabilizer product
\begin{align}
R
&=
\prod_{a\in\mathcal B}S_a
\label{eq:app_cab_guided_multiplier}
\end{align}
together with a common nonzero syndrome
$\boldsymbol\sigma_{\rm CAB}$ for the physical Pauli generators. The precise
choice depends on the analytic construction. For example, the matching
construction used for SWAP specifies the product of the selected
stabilizers and a syndrome that is nonzero on those same stabilizers.
For the nested CNOT and multi-source CNOT constructions, the stabilizer product is set by the control logical qubits while the common syndrome is carried
by the target stabilizer. The complement construction similarly fixes
the product of the selected stabilizers and the syndrome shared by the
four CAB Pauli generators.

Starting from this CAB-specified pair
$(R,\boldsymbol\sigma_{\rm CAB})$, we search for physical
factorizations
\begin{align}
R
&\doteq
G_1G_2\cdots G_q
\label{eq:app_cab_guided_factorization}
\end{align}
such that the physical Pauli generators $G_j$ commute, have pairwise
disjoint physical support, satisfy the chosen generator weight bound,
and all carry the full stabilizer syndrome
$\boldsymbol\sigma_{\rm CAB}$. Because the Pauli generator supports are disjoint and their product is fixed
to be $R$, the search can be organized as an exact-cover problem on
the physical support of $R$. Candidate Pauli generators are formed from
subsets of that support, filtered by the required weight and syndrome,
and combined until they cover the support of $R$ exactly.

% ------------------------------------------------------------
\subsection{Bounded stabilizer dressing}
\label{app:stabilizer_dressing}
% ------------------------------------------------------------

The CAB-specified stabilizer product $R$ may have a physical Pauli
pattern that does not admit a sufficiently low-weight or
support-disjoint factorization. We therefore enlarge the physical
search by multiplying $R$ by additional stabilizers, $R'=RD$, with $D \in \mathcal S$. The resulting operator $R'$ is again a stabilizer and can have a
substantially different physical Pauli support from $R$. We search
each allowed $R'$ from the beginning for a disjoint equal-syndrome
factorization with the same CAB-specified syndrome.

For the hypercube searches reported in this work, the dressing
operator $D$ is restricted to products of at most two additional
stabilizer generators. Stabilizers already used in the CAB construction
are excluded from this dressing set. This bounded dressing step preserves the CAB guidance while enlarging
the set of physical Pauli patterns available to the factorization
search. It is the mechanism used to obtain several of the additional
addressed logical gates reported for the hypercube codes. The complete CAB-guided factorization search is summarized in
Algorithm~\ref{alg:cab_guided_search}.

\begin{algorithm}[t]
\caption{CAB-guided search for Pauli-rotation-depth-one Clifford LGAs}
\label{alg:cab_guided_search}
\begin{algorithmic}[1]
\Require $(R,\boldsymbol{\sigma}_{\rm CAB})$,
Pauli generator weight bound $w_{\max}$,
stabilizer-dressing bound $r_{\max}$
\Ensure Codespace-preserving depth-one LGAs and logical Clifford actions

\State Initialize $\mathcal A=\varnothing$

\For{each allowed stabilizer dressing $D$}
    \State $R' \leftarrow RD$
    \State $\mathcal G \leftarrow
\{\,R'|_Q:\ Q\subseteq\operatorname{supp}(R')\,\}$

\State $\mathcal G \leftarrow
\{\,G\in\mathcal G:
\mathrm{wt}(G)\leq w_{\max},
\ \boldsymbol{\sigma}_{\boldsymbol S}(G)
=\boldsymbol{\sigma}_{\rm CAB}\,\}$

    \For{each even-cardinality, support-disjoint
    $\{G_j\}\subseteq\mathcal G$}
        \If{$[G_i,G_j]=0$ for all $i,j$ and
        $\prod_j G_j\doteq R'$}
            \State Determine rotation signs
            \If{the signed LGA preserves the codespace}
                \State Compute logical Clifford action $L$
                \State Add $(\{G_j\},L)$ to $\mathcal A$
            \EndIf
        \EndIf
    \EndFor
\EndFor
\State \Return $\mathcal A$
\end{algorithmic}
\end{algorithm}

% ============================================================
\section{Recursive hypercube construction and bounded gate catalogs}
\label{app:hypercube_catalogs}
% ============================================================

This appendix gives the recursive hypercube construction and the
bounded physical searches underlying the Clifford-gate results of
Sec.~\ref{sec:hypercube_results}. We first prove the recursion for the
code parameters. We then discuss the CAB-guided searches used to
improve physical realizations and to identify additional logical
Clifford actions beyond those analytically found in the main text. Finally, we summarize the resulting catalogs for the
$[[16,4,4]]$ and $[[64,8,8]]$ codes.

% ------------------------------------------------------------
\subsection{Recursive code construction}
\label{app:hypercube_recursion}
% ------------------------------------------------------------

The recursion begins with the $[[4,2,2]]$ seed code
$\mathcal H^{(0)}$. Suppose the previous level is an
$[[n_{\ell-1},k_{\ell-1},d_{\ell-1}]]$ code
$\mathcal H^{(\ell-1)}$. The level-$\ell$ four-copy code is formed from four
independent copies,
\begin{align}
\mathcal K^{(\ell)}
&=
\left(
\mathcal H^{(\ell-1)}
\right)^{\otimes4}.
\label{eq:app_hypercube_scaffold}
\end{align}
Before imposing the outer-code checks, this code has parameters
$[[4n_{\ell-1},4k_{\ell-1},d_{\ell-1}]]$.

For each logical qubit $a$ of the previous code, let
$\overline X_a^{(s)}$ and $\overline Z_a^{(s)}$ denote its logical
Pauli representatives in inner-code copy $s=1,\ldots,4$. We add the two
outer-code checks
\begin{align}
T_a^X
&=
-\overline X_a^{(1)}
\overline X_a^{(2)}
\overline X_a^{(3)}
\overline X_a^{(4)},
&
T_a^Z
&=
-\overline Z_a^{(1)}
\overline Z_a^{(2)}
\overline Z_a^{(3)}
\overline Z_a^{(4)}.
\label{eq:app_hypercube_refinement_checks}
\end{align}
Thus the four copies of each previous-level logical qubit are
constrained by the same $[[4,2,2]]$ structure as the seed code.

The new checks commute with the inherited stabilizers. For fixed $a$,
$T_a^X$ and $T_a^Z$ anticommute once in each inner-code copy and
therefore commute overall. Checks associated with different logical
qubits commute by the logical Pauli relations of the previous level.
For each of the $k_{\ell-1}$ logical qubits we add two independent
checks, so
\begin{align}
k_\ell
&=
4k_{\ell-1}-2k_{\ell-1}
=
2k_{\ell-1}.
\end{align}

The distance also doubles. Reduce a physical Pauli in the normalizer of
the concatenated code modulo the inherited stabilizers of the four scaffold
copies. It then acts as a Pauli operator on the
$4k_{\ell-1}$ inner-code logical qubits. A nontrivial logical Pauli of
the concatenated code must act nontrivially on at least one of the outer
$[[4,2,2]]$ logical families. Every nontrivial logical Pauli of that
outer code occupies at least two of its four positions. The original
physical operator must therefore induce nontrivial inner logical
operators in at least two disjoint inner-code copies. Each contributes
physical weight at least $d_{\ell-1}$, giving
$d_\ell\geq2d_{\ell-1}$.

Conversely, let $P$ be any nontrivial minimum-weight logical Pauli of
the inner code. Apply the same representative in two inner-code copies,
giving $P^{(1)}P^{(2)}$. For each outer-code check, the two copies
contribute the same commutation sign, so their product commutes with
all outer-code checks. Its outer Pauli class is nontrivial: in every
logical family on which $P$ acts, it occupies two of the four copies.
The resulting logical operator has weight $2d_{\ell-1}$, proving
$d_\ell\le2d_{\ell-1}$ and hence $d_\ell=2d_{\ell-1}$.
The recursion is therefore
\begin{align}
[[n_{\ell-1},k_{\ell-1},d_{\ell-1}]]
\longrightarrow
[[4n_{\ell-1},2k_{\ell-1},2d_{\ell-1}]].
\label{eq:app_hypercube_parameter_recursion}
\end{align}
Starting from $[[4,2,2]]$, the first two error-correcting levels are
\begin{align}
[[4,2,2]]
&\longrightarrow
[[16,4,4]]
\longrightarrow
[[64,8,8]].
\end{align}

\begin{table*}[t]
\centering
\footnotesize
\renewcommand{\arraystretch}{1.10}
\begin{tabular*}{\textwidth}
{@{\extracolsep{\fill}}lcccccc@{}}
\toprule
Search
& CAB targets
& $w_{\max}$
& additional stabilizers
& verified realizations
& logical actions
& additional actions \\
\midrule

$[[16,4,4]]$ final-code search
& 35
& 6
& $\leq2$
& 1380
& 60
& 43 \\

$[[64,8,8]]$ final-code search
& 631
& 6
& 0
& 33
& 15
& 0 \\

$[[64,8,8]]$ pair search
& 84
& 8
& 0
& 12
& 12
& 0 \\

$[[64,8,8]]$ dressed-pair search
& 72
& 8
& $\leq2$
& 448
& 72
& 48 \\

\bottomrule
\end{tabular*}
\caption{Bounded physical searches used to supplement the analytic CAB
constructions. ``CAB targets'' counts the logical CAB constructions
used to initialize each search. ``Additional stabilizers'' gives the
maximum number of stabilizer generators multiplied into the
CAB-specified stabilizer product. ``verified realizations'' counts the
code-preserving physical realizations returned by the search. ``Logical
actions'' counts the distinct Clifford transformations
implemented by those realizations, and ``additional actions'' counts the
actions added to the existing hypercube catalog.}
\label{tab:app_action_unbiased_search_scope}
\end{table*}

% ------------------------------------------------------------
\subsection{Bounded physical search domains}
\label{app:hypercube_search_domains}
% ------------------------------------------------------------

The hypercube catalogs combine the analytic CAB constructions with
physical searches guided by those constructions. We include CAB gates
constructed directly on the final code, gates constructed on the
four-copy inner-code that also preserve the outer-code checks, and
lower-level gates promoted through the recursion. The CAB-guided
factorization procedure of Appendix~\ref{app:hypercube_compilation} is then
used to find alternative physical realizations with lower Pauli generator weight or Pauli-rotation depth.

For the $[[16,4,4]]$ code, the CAB search on the final code starts from
$35$ logical targets: six SWAPs, twelve directed CNOTs, one
four-qubit complement, and sixteen multi-source CNOTs with two or three controls.
The physical search uses cumulative generator weight cutoffs through
$w_{\max}=6$ and allows the CAB-specified stabilizer product to be
multiplied by products of at most two additional stabilizer
generators.

For the $[[64,8,8]]$ final code, the CAB search starts from all
$28$ SWAPs, all $56$ directed CNOTs, all $99$ complements on
subsets of size four, six, or eight, and all $448$ multi-source CNOTs with two or
three controls, giving $631$ final-code targets. The baseline search
uses Pauli generator weights through $w_{\max}=6$ without stabilizer
dressing. A focused pair search extends the cutoff to
$w_{\max}=8$ for all $28$ SWAPs and $56$ directed CNOTs. A further search covers the remaining 24 logical-qubit pairs, using the same weight-eight cutoff and allowing products of at most two additional stabilizer generators. 

The recursive construction also supplies CAB gates on the
sixteen-logical-qubit four-copy code. We include all $120$ SWAPs on that four-copy code
and all $240$ directed CNOTs. The larger complement and multi-source CNOT spaces
are sampled with fixed deterministic bounds in the accompanying
implementation.

To keep each physical search finite, we impose fixed deterministic caps: at most \(5000\) stabilizer dressings and \(50{,}000\) candidate Pauli generators are examined, each dressed stabilizer contributes at most eight exact-cover factorizations, and the search retains at most \(32\) physical partitions for further analysis, with up to four taken through final signed codespace verification.
These bounds define the finite searches summarized in Table~\ref{tab:app_action_unbiased_search_scope}.

% ------------------------------------------------------------
\subsection{The \texorpdfstring{$[[16,4,4]]$}{[[16,4,4]]} catalog}
\label{app:l1_hypercube_catalog}
% ------------------------------------------------------------

The CAB constructions, constructions on the four-copy code, promoted gates, and
their directly targeted physical searches provide $125$
physical realizations with Pauli-rotation depth one implementing $23$ distinct logical Clifford actions. Classifying the remaining
verified realizations returned by the bounded search identifies
$43$ additional logical actions. We retain one physical realization for each
of these additional actions, giving a final set of $168$ LGAs implementing $66$ distinct logical
Clifford gates. All $43$ additional actions already occur at the weight-four search
stage and use four physical Pauli generators with weight profile
$(2,2,4,4)$. Their logical-action families are included in
Table~\ref{tab:app_hypercube_family_counts}.

The addressed two-qubit gates illustrate how the CAB-guided search
extends the analytic construction. The nested CAB construction gives
both CNOT orientations on the natural logical pairs $(1,2)$ and
$(3,4)$ using four weight-two Pauli generators. These LGAs saturate the
equal-syndrome bound $w_{\max}=d/2$. The bounded search also
finds both CNOT orientations on $(1,4)$ and $(2,3)$ with Pauli generator
profile $(2,2,4,4)$. The resulting CNOT edges form a connected graph
on the four logical qubits. The matching construction gives SWAP on $(1,2)$ and $(3,4)$ with
Pauli generator profile $(3,3,3,3)$. The bounded search also gives SWAP on
$(1,4)$ and $(2,3)$ with profile $(2,2,4,4)$, producing a
connected set of SWAPs.

\begin{table*}[t]
\centering
\footnotesize
\renewcommand{\arraystretch}{1.10}
\begin{tabular*}{\textwidth}
{@{\extracolsep{\fill}}lrrrrrrr@{}}
\toprule
Code
& CNOT
& SWAP
& multi-souce CNOT
& complement
& composite
& general
& total \\
\midrule

$[[16,4,4]]$
& 8
& 4
& 32
& 4
& 9
& 9
& 66 \\

$[[64,8,8]]$
& 32
& 4
& 48
& 11
& 23
& 0
& 118 \\

\bottomrule
\end{tabular*}
\caption{Logical Clifford actions in the final hypercube
catalogs, grouped by logical-action family. ``Composite'' denotes an
action for which an explicit Clifford circuit decomposition was found.
``General'' denotes the remaining Clifford transformations
in the catalog.}
\label{tab:app_hypercube_family_counts}
\end{table*}

% ------------------------------------------------------------
\subsection{The \texorpdfstring{$[[64,8,8]]$}{[[64,8,8]]} catalog}
\label{app:l2_hypercube_catalog}
% ------------------------------------------------------------

The CAB constructions, constructions on the four-copy code, promoted gates, and
their directly targeted physical searches provide $198$
physical realizations with Pauli-rotation depth one implementing $70$ distinct
nonidentity logical Clifford actions. Classifying the additional LGAs returned by the bounded searches identifies
$48$ more logical actions. Retaining one physical realization for each gives
a final catalog of $246$ LGAs implementing $118$ distinct logical Clifford gates. All $48$ additional actions use four physical Pauli generators with weight
profile $(4,4,8,8)$ up to ordering. Their logical actions lie in the
local-Clifford family of multi-source CNOT transformations.

The nested CAB construction gives both CNOT orientations on the four
natural logical pairs $(1,2)$, $(3,4)$, $(5,6)$, and $(7,8)$ using
four weight-four Pauli generators. These LGAs again saturate
$w_{\max}=d/2$. The searches add both CNOT orientations on the twelve
additional logical pairs listed in
Sec.~\ref{subsec:hypercube_analytic_summary}. Every additional CNOT found in this search uses four Pauli generators with weight profile
$(4,4,8,8)$ up to ordering. Together with the four analytic pairs,
these gates give individually addressed CNOTs on sixteen of the
twenty-eight logical pairs and form a connected CNOT graph on all
eight logical qubits.

% ============================================================
\section{Middle Reed--Muller hierarchy}
\label{app:middle_rm_hierarchy}
% ============================================================

This appendix proves the two properties of the middle quantum
Reed--Muller construction used in
Sec.~\ref{subsec:middle_rm_nonclifford}: the physical Pauli generators have
one common nonzero stabilizer syndrome, and their relative logical
operators generate the required family of commuting logical $Z$
operators. The general equal-syndrome sign construction of
Appendix~\ref{app:logical_sign_construction} then gives the logical
$C^rZ$ gate.

The relevant code family is
\begin{align}
\mathcal Q_r
&=
\left[
\left[
2^{2r},
\binom{2r}{r},
2^r
\right]
\right],
\qquad r\geq1.
\label{eq:app_middle_rm_parameters}
\end{align}
The $r=1$ member is the $[[4,2,2]]$ code and gives the first member of
the controlled-phase hierarchy.

% ------------------------------------------------------------
\subsection{Physical Pauli generators and their common syndrome}
\label{app:middle_rm_polynomial_code}
\label{app:rm_common_syndrome}
% ------------------------------------------------------------

Label the $2^{2r}$ physical qubits by binary strings
$\boldsymbol x=(x_1,\ldots,x_{2r})\in\mathbb F_2^{2r}$. For a binary
polynomial $f(\boldsymbol x)$, let $Z(f)$ denote the Pauli operator
whose support is the set of points on which $f=1$, and define $X(f)$
similarly.

For the middle Reed--Muller code, the $X$- and $Z$-type stabilizers
are generated by polynomials of degree at most $r-1$. Degree-$r$
monomials determine logical Pauli classes. To see this for a
degree-$r$ monomial $b$, note that its overlap with any
degree-$(r-1)$ stabilizer polynomial has even parity: every monomial
appearing in their product has degree at most $2r-1$ and therefore has
even support on $\mathbb F_2^{2r}$. Thus $Z(b)$ commutes with every
$X$-type stabilizer. Since $b$ lies outside the degree-$(r-1)$
stabilizer space, it represents a nontrivial logical $Z$. The
$\binom{2r}{r}$ degree-$r$ monomials give the expected number of
independent logical $Z$ classes.

Choose a set $C$ of $r+1$ coordinates and let $W$ contain the remaining
$r-1$ coordinates. For each
$\boldsymbol u\in\mathbb F_2^{r+1}$, fix the coordinates in $C$ to
$\boldsymbol u$ and allow the coordinates in $W$ to vary freely. The
corresponding physical Pauli generator is
\begin{align}
G_{\boldsymbol u}
&=
\prod_{\boldsymbol x:\,\boldsymbol x_C=\boldsymbol u}
Z_{\boldsymbol x}.
\label{eq:app_rm_slice_Pauli generator}
\end{align}
There are $2^{r+1}$ such Pauli generators. Their supports partition the
physical qubits, so they are mutually support-disjoint, and each
Pauli generator has weight $2^{r-1}$. Their total product is the all-qubit
operator $Z(1)$, which is a $Z$-type stabilizer.

It remains to show that all of the Pauli generators carry the same nonzero
syndrome. Since each $G_{\boldsymbol u}$ is $Z$ type, it commutes with
every $Z$-type stabilizer. Consider an $X$-type stabilizer $X(f)$,
where $\deg f\leq r-1$. Its commutation with
$G_{\boldsymbol u}$ is determined by the parity of the support of $f$
after the coordinates in $C$ are fixed to $\boldsymbol u$.

Consider one monomial appearing in $f$. If it omits any coordinate in
$W$, then that unused coordinate can be flipped without changing the
monomial. Its nonzero values on the Pauli generator therefore occur in pairs
and contribute even overlap. Since $|W|=r-1$, the only monomial of
degree at most $r-1$ that contains every coordinate in $W$ is
$m_W=\prod_{j\in W}x_j$. This monomial is nonzero at exactly one point
of every Pauli generator, independent of $\boldsymbol u$.

The overlap parity with $X(f)$ therefore depends only on the
coefficient of $m_W$ in $f$ and is the same for every
$G_{\boldsymbol u}$. Hence
\begin{align}
\boldsymbol{\sigma}_{\boldsymbol S}(G_{\boldsymbol u})
&=
\boldsymbol v
\qquad
\forall\boldsymbol u.
\label{eq:app_rm_common_syndrome}
\end{align}
The common syndrome is nonzero because the stabilizer $X(m_W)$
anticommutes with every Pauli generator. For $r=1$, the set $W$ is empty and
$m_W=1$, so the same argument applies to the $[[4,2,2]]$ base case.

% ------------------------------------------------------------
\subsection{Relative logical operators}
\label{app:rm_relative_logicals}
% ------------------------------------------------------------

Take $G_{\boldsymbol 0}$ as the reference Pauli generator. Since every
$G_{\boldsymbol u}$ has the same syndrome, the product
$G_{\boldsymbol 0}G_{\boldsymbol u}$ has zero syndrome and therefore
acts as a logical Pauli operator on the codespace.

The Pauli generator with fixed coordinate values
$\boldsymbol u$ has indicator polynomial
\begin{align}
\delta_{\boldsymbol u}(\boldsymbol x_C)
&=
\prod_{i\in C:u_i=1}x_i
\prod_{i\in C:u_i=0}(1+x_i).
\label{eq:app_rm_slice_indicator}
\end{align}
Thus $G_{\boldsymbol u}=Z(\delta_{\boldsymbol u})$, and the relative
operator $G_{\boldsymbol 0}G_{\boldsymbol u}$ is represented by
$\delta_{\boldsymbol 0}+\delta_{\boldsymbol u}$ over $\mathbb F_2$.

Both indicator polynomials contain the same degree-$(r+1)$ monomial
$\prod_{j\in C}x_j$, so this highest-degree term cancels. All remaining
terms have degree at most $r$. Terms of degree at most $r-1$ are
$Z$-type stabilizers, so the logical action is determined entirely by
the degree-$r$ terms.

For each $i\in C$, define
$b_i=\prod_{j\in C\setminus\{i\}}x_j$ and let
$B_i=Z(b_i)$. Each $B_i$ is a logical $Z$ operator. The monomial $b_i$
appears in $\delta_{\boldsymbol 0}$ for every $i$. In
$\delta_{\boldsymbol u}$, it appears precisely when $u_i=0$.
Consequently, it survives in
$\delta_{\boldsymbol 0}+\delta_{\boldsymbol u}$ precisely when
$u_i=1$. Therefore
\begin{align}
G_{\boldsymbol 0}G_{\boldsymbol u}
&\sim
\prod_{i\in C}B_i^{u_i},
\label{eq:app_rm_relative_logical_identity}
\end{align}
where $\sim$ denotes equality up to multiplication by stabilizers.

The $r+1$ operators $B_i$ correspond to distinct degree-$r$
monomials and give independent commuting logical $Z$ classes. As
$\boldsymbol u$ runs over $\mathbb F_2^{r+1}$, the Pauli generators therefore
realize every product of these $r+1$ logical operators exactly once.

% ------------------------------------------------------------
\subsection{Logical controlled phase}
\label{app:rm_controlled_phase}
% ------------------------------------------------------------

The Pauli generators now satisfy the hypotheses of the higher-level
equal-syndrome construction of
Appendix~\ref{app:logical_sign_construction}: they share one nonzero
syndrome, while their relative logical operators run through all
products of $r+1$ independent commuting logical $Z$ operators.

Assign the rotation direction
$\epsilon_{\boldsymbol u}=(-1)^{|\boldsymbol u|}$ and use physical
rotation angle $\pi/2^{r+1}$. The sign construction of
Appendix~\ref{app:logical_sign_construction} then gives
\begin{align}
\overline U_r
&=
\overline{C^rZ}.
\label{eq:app_rm_crz_action}
\end{align}
This yields $CZ$ for $r=1$, $CCZ$ for $r=2$, $C^3Z$ for $r=3$, and
continues through the controlled-phase hierarchy.

The physical supports of the $2^{r+1}$ Pauli generators partition the full
code block, so the construction has Pauli-rotation depth one. Each Pauli generator
has weight $2^{r-1}$, while the code distance is $d=2^r$. Hence every
member of the family has Pauli generator weight $d/2$ and saturates the
equal-syndrome weight bound.

\end{document}